\documentclass{book}
\usepackage{roboto}

\usepackage[english]{babel}
\usepackage[letterpaper,top=2cm,bottom=2cm,left=3cm,right=3cm,marginparwidth=1.75cm]{geometry}

\usepackage{pdflscape}
\usepackage{rotating}
\usepackage{longtable}
\usepackage{array}
\usepackage{float}
\usepackage{amsmath}
\usepackage{graphicx}
\usepackage{inconsolata}
\usepackage[colorlinks=true, allcolors=blue]{hyperref}
\usepackage{enumitem}
\usepackage{tikz} 
\usepackage{listings}
\usepackage{xcolor}
\usepackage[T1]{fontenc}
\usepackage{IEEEtrantools}  
\usepackage{epigraph}  

\definecolor{bgcolor}{rgb}{0.97,0.97,0.97}
\definecolor{codeblue}{rgb}{0.1,0.1,0.8}
\definecolor{codegreen}{rgb}{0,0.4,0}
\definecolor{codegray}{rgb}{0.4,0.4,0.4}
\definecolor{codepurple}{rgb}{0.5,0,0.5}
\definecolor{codered}{rgb}{0.6,0.2,0.2}
\definecolor{lightgray}{rgb}{0.9,0.9,0.9}
\definecolor{darkgray}{rgb}{0.6,0.6,0.6} 

\makeatletter
\renewcommand{\paragraph}{%
  \@startsection{paragraph}{4}{\z@}{1ex}{-1em}{\normalfont\normalsize\bfseries\color{gray}}}
\makeatother

\lstdefinestyle{python}{
    language=Python,
    basicstyle=\ttfamily\small\color{black}\usefont{T1}{zi4}{m}{n},  
    keywordstyle=\bfseries\color{codeblue},  
    stringstyle=\color{codegreen},  
    commentstyle=\slshape\color{codegray},  
    showstringspaces=false,
    numbers=left,
    numberstyle=\tiny\color{codegray},  
    stepnumber=1,
    numbersep=8pt,
    frame=single,
    rulecolor=\color{darkgray},  
    breaklines=true,
    backgroundcolor=\color{bgcolor},
    tabsize=4,
    captionpos=b,
    morekeywords={self}, 
}

\lstdefinestyle{cmd}{
    language=bash,
    basicstyle=\ttfamily\small\color{black}\usefont{T1}{zi4}{m}{n},  
    keywordstyle=\bfseries\color{blue},
    stringstyle=\color{codegreen},
    commentstyle=\itshape\color{gray},
    showstringspaces=false,
    numbers=none,
    frame=single,
    rulecolor=\color{darkgray},  
    breaklines=true,
    backgroundcolor=\color{bgcolor},
    tabsize=4,
    captionpos=b,
}

\title{Deep Learning and Machine Learning: Advancing Big Data Analytics and Management with Design Patterns}
\author{
    Keyu Chen\textsuperscript{*} \\ 
    \textit{Georgia Institute of Technology} \\
    kchen637@gatech.edu
    \and
    Ziqian Bi\textsuperscript{*$\dagger$}\\
    \textit{Indiana University} \\
    bizi@iu.edu
    \and
    Tianyang Wang\textsuperscript{*} \\ 
    \textit{Xi’an Jiaotong-Liverpool University} \\
    Tianyang.Wang21@student.xjtlu.edu.cn
    \and
    Yizhu Wen \\
    \textit{University of Hawaii} \\
    yizhuw@hawaii.edu
    \and
    Pohsun Feng \\
    \textit{National Taiwan Normal University} \\
    41075018h@ntnu.edu.tw
    \and
    Qian Niu \\ 
    \textit{Kyoto University} \\
    niu.qian.f44@kyoto-u.jp
    \and
    Junyu Liu \\ 
    \textit{Kyoto University} \\
    liu.junyu.82w@st.kyoto-u.ac.jp
    \and
    Benji Peng \\ 
    \textit{AppCubic} \\
    benji@appcubic.com
    \and
    Sen Zhang \\ 
    \textit{Rutgers University} \\
    sen.z@rutgers.edu
    \and
    Ming Li \\ 
    \textit{Georgia Institute of Technology} \\
    mli694@gatech.edu
    \and
    Xuanhe Pan \\ 
    \textit{University of Wisconsin-Madison} \\
    xpan73@wisc.edu
    \and
    Jiawei Xu \\ 
    \textit{Purdue University} \\
    xu1644@purdue.edu
    \and
    Jinlang Wang \\ 
    \textit{University of Wisconsin-Madison} \\
    jinlang.wang@wisc.edu
    \and
    Xinyuan Song \\ 
    \textit{Emory University} \\
    xinyuan.song@emory.edu
    \and
    Ming Liu\textsuperscript{$\dagger$} \\ 
    \textit{Purdue University} \\
    liu3183@purdue.edu
}

\date{} 

\begin{document}

\maketitle

\begingroup
\renewcommand\thefootnote{}\footnote{
    \textsuperscript{*} Equal contribution \\
    \textsuperscript{$\dagger$} Corresponding author
}
\addtocounter{footnote}{0}
\endgroup

\epigraph{"Simple things should be simple, complex things should be possible."}{\textit{Alan Kay}}

\epigraph{"A design that doesn't take change into account risks major redesign in the future"}{\textit{Erich Gamma}}

\epigraph{"A pattern is an idea that has been useful in one practical context and will probably be useful in others."}{\textit{Martin Fowler}}

\epigraph{"Design patterns help you learn from others’ successes instead of your own failures."}{\textit{Mark Johnson}}

\epigraph{"A change in perspective is worth 80 IQ points."}{\textit{Alan Kay}}

\tableofcontents  

\chapter*{Introduction to the Series: \\ Deep Learning and Machine Learning: Advancing Big Data Analytics Management}

\noindent
The series, \textit{Deep Learning and Machine Learning: Advancing Big Data Analytics Management}, is a comprehensive collection designed to guide readers through the complexities of modern data analytics, machine learning, and deep learning techniques. With the exponential growth of big data, the need for advanced analytics and sophisticated methodologies has become paramount. Each book in this series tackles a key aspect of this expansive domain, offering in-depth discussions, practical examples, and strategic insights to help readers effectively manage and leverage big data in today's fast-evolving digital landscape.

The series begins with \textit{Handy Appetizer}, which sets the stage by providing a foundational understanding of key concepts and techniques in machine learning\cite{peng2024deeplearningmachinelearning} and deep learning\cite{chen2024deeplearningmachinelearning}. As readers progress through the subsequent volumes, they are introduced to more specialized topics, such as the intricacies of classifiers in \textit{Classifier}, the use of TensorFlow's pretrained models in \textit{TF Pretrained}, and critical frameworks like PyTorch in \textit{PyTorch}\cite{li2024deeplearningmachinelearning}. 

Each book in the series builds on the last, creating a cohesive and layered understanding of the field, culminating in the exploration of crucial advanced topics, including \textit{GPGPU}, \textit{NLP}, and the security aspects of models in \textit{Model Security}\cite{peng2024securinglargelanguagemodels}. Furthermore, the series expands into specialized areas like \textit{Diffusion}, offering insights into cutting-edge developments in deep learning research.

\textit{Design Pattern}, the fifth book in the series, is a particularly important volume. It emphasizes the architectural principles and reusable solutions necessary to create scalable, maintainable, and efficient deep learning and machine learning systems. As data complexity grows, so does the challenge of managing code structure and ensuring flexibility across various application scenarios. This book provides an in-depth examination of key design patterns that are essential for building robust systems, such as the Singleton, Factory, Observer, and Strategy patterns. These patterns, derived from classical software engineering, are adapted to meet the unique demands of machine learning model management and deployment. The book also explores how these patterns facilitate better collaboration between teams and improve the maintainability of codebases, which is crucial in large-scale data analytics projects\cite{niu2024large}.

As a reader, you will gain valuable insights into structuring your machine learning applications for scalability and performance. The examples provided in \textit{Design Pattern} not only illustrate each pattern's utility but also show how to implement them in practical machine learning and deep learning pipelines. This volume is critical for those looking to enhance their technical skills while developing a deeper understanding of system design within the context of big data analytics.

Together, the books in this series provide a structured, comprehensive learning path for practitioners, researchers, and enthusiasts aiming to advance their expertise in machine learning\cite{Peng_2024}, deep learning, and big data analytics management.

\bigskip
\noindent
Enjoy your journey through the transformative world of data-driven innovation.

\setcounter{part}{3}
\part{Becoming a Master}

\setcounter{chapter}{50}
\chapter{Design Patterns}

Today, artificial intelligence (AI) has deeply impacted various industries, from healthcare and finance to autonomous driving, and is nearly ubiquitous \cite{shaheen2021applications,jan2023artificial,cao2022ai,grigorescu2020survey}. Building efficient and scalable system architectures is crucial when developing AI models and applications. Design patterns offer a proven, structured approach to achieving this goal \cite{martin2000design,zhang2011we}. These patterns simplify the design of deep learning frameworks and enhance the code's readability and maintainability. By mastering design patterns, developers can better manage and understand the complexities within AI systems, ensuring flexibility and stability even as requirements evolve \cite{agerbo1998preserve,mcnatt2001coupling,ampatzoglou2013building}. Thus, design patterns are not only core knowledge for traditional software development but are also essential foundations for understanding and designing AI systems.

\section{Introduction}

\subsection{What Are Design Patterns?}

Design patterns are proven solutions to common software design problems. They represent best practices that have evolved and serve as templates for solving specific design challenges. Each pattern describes a problem, the solution, and the consequences of applying that solution \cite{wedyan2020impact}. 

In software engineering, particularly in object-oriented programming, design patterns provide a way to structure your code to make it more reusable, maintainable, and flexible \cite{garcia2005modularizing}. Instead of reinventing the wheel whenever you face a problem, you can apply a tried-and-tested design pattern, thus improving your development efficiency and code quality.

\subsection{History of Design Patterns}

The concept of design patterns was originally introduced in the field of architecture by Christopher Alexander in the 1970s. His book \textit{A Pattern Language} has described how common patterns of architecture could be used to solve design problems in building construction \cite{alexander2018pattern}.

In the context of software development, the term "design patterns" gained prominence in 1995 when four computer scientists—Erich Gamma, Richard Helm, Ralph Johnson, and John Vlissides—published a book titled \textit{Design Patterns: Elements of Reusable Object-Oriented Software} \cite{gamma1995elements}. These authors are often referred to as the "Gang of Four" (GoF). Their book documented 23 fundamental design patterns that have since become a cornerstone of software engineering.

\subsection{Why Do We Need Design Patterns?}

As you gain experience in software development, you will start to encounter recurring problems. These problems are not unique to you; developers worldwide face similar challenges. Design patterns provide generalized solutions to these problems, allowing you to write better code, faster.

Here are a few reasons why design patterns are important:

\begin{itemize}
    \item \textbf{Code Reusability}: Design patterns encourage reusing code, which saves time and effort. Instead of building new solutions from scratch, developers can apply a known pattern that solves a particular problem.
    \item \textbf{Improved Communication}: Patterns provide a shared language for developers. When you mention a specific design pattern, such as the "Singleton" or "Observer," other developers will immediately understand the structure of your solution without needing a lengthy explanation.
    \item \textbf{Maintainability}: Following design patterns helps make your code easier to maintain and extend. They promote separation of concerns, meaning that each part of your application has a clear, focused responsibility.
    \item \textbf{Scalability}: Design patterns help structure your code in a way that is more adaptable to change, especially as your application grows.
\end{itemize}

\subsection{Drawbacks of Design Patterns}

While design patterns offer many advantages, they also come with some potential downsides:

\begin{itemize}
    \item \textbf{Over-Engineering}: Applying a design pattern where it is not needed can make your code unnecessarily complex. Beginners sometimes use patterns just for the sake of it, resulting in over-engineering.
    \item \textbf{Learning Curve}: For new developers, learning design patterns may feel overwhelming at first. It takes time to fully understand when and where to apply a particular pattern.
    \item \textbf{Inflexibility}: Relying too heavily on predefined patterns may limit your creativity in finding unique solutions. Not all problems neatly fit into a specific pattern.
\end{itemize}

\subsection{Types of Design Patterns}

The Gang of Four categorized design patterns into three main types:

\begin{itemize}
    \item \textbf{Creational Patterns}: These patterns deal with object creation mechanisms, trying to create objects in a way that is suitable to the situation. They help make the system independent of how objects are created, composed, and represented. Some common examples are:
    \begin{itemize}
        \item \textbf{Singleton Pattern}: Ensures a class has only one instance and provides a global point of access to that instance.
        \item \textbf{Factory Method Pattern}: Provides an interface for creating objects, but lets subclasses alter the type of objects that will be created.
        \item \textbf{Builder Pattern}: Separates the construction of a complex object from its representation, allowing the same construction process to create different representations.
    \end{itemize}
    
    \item \textbf{Structural Patterns}: These patterns focus on composing classes and objects to form larger structures. Structural patterns ease the design by identifying a simple way to realize relationships between entities. Examples include:
    \begin{itemize}
        \item \textbf{Adapter Pattern}: Allows incompatible interfaces to work together by acting as a bridge between them.
        \item \textbf{Decorator Pattern}: Allows behavior to be added to individual objects, dynamically, without affecting the behavior of other objects from the same class.
        \item \textbf{Facade Pattern}: Provides a simplified interface to a complex subsystem.
    \end{itemize}
    
    \item \textbf{Behavioral Patterns}: These patterns are concerned with algorithms and the assignment of responsibilities between objects. They help in better interaction and communication between objects. Examples include:
    \begin{itemize}
        \item \textbf{Observer Pattern}: Defines a one-to-many dependency between objects so that when one object changes state, all its dependents are notified and updated automatically.
        \item \textbf{Strategy Pattern}: Allows a family of algorithms to be defined, encapsulated, and made interchangeable.
        \item \textbf{Command Pattern}: Encapsulates a request as an object, thereby allowing for parameterization of clients with different requests.
    \end{itemize}
\end{itemize}

\subsection{How Design Patterns Differ}

It’s important to understand the differences between the three types of design patterns:

\begin{itemize}
    \item \textbf{Creational vs. Structural}: Creational patterns focus on the process of object creation, whereas structural patterns deal with how objects and classes are combined to form larger structures.
    \item \textbf{Structural vs. Behavioral}: While structural patterns are concerned with how objects are organized, behavioral patterns define how objects interact with one another.
    \item \textbf{Creational vs. Behavioral}: Creational patterns aim to separate the concern of object creation, while behavioral patterns focus on communication and interaction between objects.
\end{itemize}

\subsection{An Example of a Simple Design Pattern: Singleton}

Let’s look at a simple example to demonstrate a design pattern. The Singleton pattern \cite{freeman2015singleton} is one of the most commonly used patterns in Python. It ensures that a class has only one instance and provides a global point of access to that instance.

\begin{lstlisting}[style=python]
class Singleton:
    _instance = None

    def __new__(cls):
        if cls._instance is None:
            cls._instance = super(Singleton, cls).__new__(cls)
        return cls._instance

# Test the Singleton pattern
singleton1 = Singleton()
singleton2 = Singleton()

print(singleton1 is singleton2)  # Outputs: True
\end{lstlisting}

In this example, the \texttt{\_\_new\_\_} method ensures that only one instance of the \texttt{Singleton} class is created. Even though we create two variables, \texttt{singleton1} and \texttt{singleton2}, they both point to the same instance.

\subsection{Conclusion}

Design patterns are essential tools in software engineering, particularly in object-oriented programming. They allow developers to follow best practices, solve common problems efficiently, and write code that is maintainable, scalable, and easy to understand. 

By understanding the different types of patterns—creational, structural, and behavioral—you will be better equipped to decide which pattern is most appropriate for solving a specific problem. In the upcoming chapters, we will dive into each of these patterns, providing practical Python examples to solidify your understanding.

\section{Creational Patterns}
\subsection{Introduction}

In object-oriented programming, Creational Design Patterns \cite{ng2008identification} deal with the process of object creation. Rather than creating objects directly in a straightforward way, these patterns provide various ways to create objects that improve flexibility and reuse of existing code. The creational patterns help to make the system independent of how its objects are created, composed, and represented. 

In Python, several creational patterns can be implemented easily thanks to its dynamic and high-level nature. However, understanding these patterns can help beginners and new developers write better, more flexible code that can be extended without much modification. Let’s start by diving into why these patterns are needed and how they solve common issues.

\subsubsection{Why Creational Patterns are Necessary}

In software development, there is often a need to create objects during runtime. The naive approach would be to instantiate these objects directly using their constructors. However, this can lead to a situation where:

\begin{itemize}
    \item The object creation process becomes tightly coupled with the classes that use them.
    \item Changes in the way objects are created force you to modify existing code.
    \item You need to manage the lifecycle of objects manually.
\end{itemize}

Creational patterns abstract the object instantiation process and allow the system to delegate the responsibility of creating instances to specific factories or builders. This helps in scenarios where:

\begin{itemize}
    \item You want to create different types or families of objects without knowing the exact class of object that will be created.
    \item The creation logic is complex and needs to be centralized or made more flexible.
    \item You want to enforce a specific way of creating objects (e.g., singleton objects that only allow one instance).
\end{itemize}

Let’s look at a simple example to illustrate this. Suppose we have a program that needs to create different shapes, such as circles and squares. Without a creational pattern, we might write something like this:

\begin{lstlisting}[style=python]
class Circle:
    def draw(self):
        print("Drawing a Circle")

class Square:
    def draw(self):
        print("Drawing a Square")

# Direct instantiation
circle = Circle()
square = Square()

circle.draw()
square.draw()
\end{lstlisting}

In this example, we directly instantiate the objects using the class names, which works but is not very flexible. If we need to change the way we create these objects or manage different kinds of objects, we would need to modify the existing code. Creational patterns solve this problem by abstracting the creation process.

\subsubsection{Common Creational Patterns}

There are five main creational design patterns that are commonly used:

\begin{enumerate}
    \item \textbf{Factory Method Pattern}: Provides an interface for creating objects in a superclass, but allows subclasses to alter the type of objects that will be created.
    \item \textbf{Abstract Factory Pattern}: Encapsulates a group of individual factories that have a common theme without specifying their concrete classes.
    \item \textbf{Builder Pattern}: Separates the construction of a complex object from its representation so that the same construction process can create different representations.
    \item \textbf{Prototype Pattern}: Allows for the creation of new objects by copying existing objects (prototypes), which can be more efficient in some cases.
    \item \textbf{Singleton Pattern}: Ensures that a class has only one instance and provides a global point of access to it.
\end{enumerate}

In the following sections, we will go over each of these patterns in detail with clear Python examples, so you can understand how and when to use them in your code.

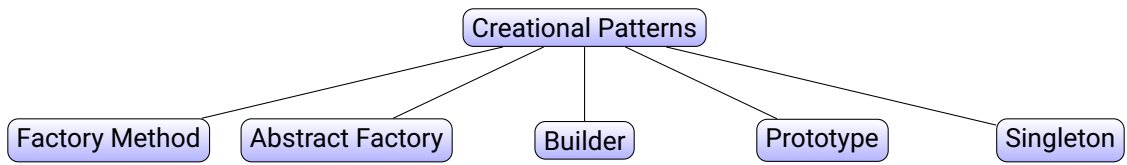
\begin{tikzpicture}[sibling distance=9em,
  every node/.style = {shape=rectangle, rounded corners,
    draw, align=center, top color=white, bottom color=blue!30}]]
  \node {Creational Patterns}
    child { node {Factory Method} }
    child { node {Abstract Factory} }
    child { node {Builder} }
    child { node {Prototype} }
    child { node {Singleton} };
\end{tikzpicture}

By the end of this chapter, you will have a solid understanding of how to use each of these patterns to improve the structure and flexibility of your Python code.

\subsection{Singleton Pattern}

The Singleton Pattern \cite{freeman2015singleton} is one of the simplest and most widely used design patterns in software engineering. It ensures that a class has only one instance and provides a global point of access to that instance.

\subsubsection{Motivation}

The motivation for using the Singleton Pattern comes from scenarios where it is necessary to have exactly one instance of a class. Common use cases include situations where:

\begin{itemize}
    \item Managing access to shared resources, such as databases or file systems.
    \item Centralized control over certain application components, such as configuration settings.
    \item Implementing logging, where multiple instances could lead to inconsistencies.
\end{itemize}

For example, imagine an application that connects to a database. If multiple instances of the database connection class are created, it could lead to unnecessary resource consumption and potential connection errors. The Singleton Pattern ensures that only one connection is made and reused throughout the application.

\subsubsection{Structure}

The structure of the Singleton Pattern consists of a class that restricts the creation of its instances and provides a method to access the single instance.

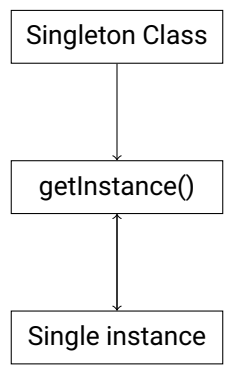
\begin{center}
\begin{tikzpicture}[node distance=2cm, auto]
  \node (Class) [draw, rectangle, minimum width=8em, minimum height=2em, text centered] {Singleton Class};
  \node (Method) [below of=Class, draw, rectangle, minimum width=8em, minimum height=2em, text centered] {getInstance()};
  \node (Instance) [below of=Method, draw, rectangle, minimum width=8em, minimum height=2em, text centered] {Single instance};

  \draw[->] (Class) -- (Method);
  \draw[->] (Method) -- (Instance);
  \draw[->] (Instance) -- (Method);
\end{tikzpicture}
\end{center}

In Python, we typically use a class variable to store the single instance and control access to it through a static or class method.

\subsubsection{Participants}

There is only one key participant in the Singleton Pattern:

\begin{itemize}
    \item \textbf{Singleton}: This class defines a static method for accessing a single instance. It ensures that only one instance of the class is created.
\end{itemize}

\subsubsection{How It Works}

The Singleton works by checking if an instance of the class already exists. If it does not, it creates the instance and stores it. If it does exist, it simply returns the already created instance. This ensures that there is only one instance of the class throughout the program’s lifecycle.

Here is a step-by-step explanation of how a Singleton is typically implemented in Python:

\begin{enumerate}
    \item A class attribute is used to store the instance.
    \item A class method (often named \texttt{getInstance()}) checks if the instance exists.
    \item If the instance does not exist, the method creates a new one and stores it in the class attribute.
    \item The method then returns the instance.
\end{enumerate}

\subsubsection{Advantages and Disadvantages}

\paragraph{Advantages:}

\begin{itemize}
    \item \textbf{Controlled access to a single instance:} Ensures that only one instance of a class exists, which is useful for managing global resources like database connections.
    \item \textbf{Reduced resource usage:} Since only one instance is created, memory and system resources are used more efficiently.
    \item \textbf{Simplifies debugging:} Easier to track the state of a single instance, making debugging and maintenance simpler.
\end{itemize}

\paragraph{Disadvantages:}

\begin{itemize}
    \item \textbf{Global state:} Singleton introduces a global state into the application, which can lead to unexpected behaviors if the state is modified from different parts of the program.
    \item \textbf{Difficult to test:} Singleton classes can be difficult to test, as their global nature can make it hard to isolate different parts of the code for testing.
    \item \textbf{Concurrency issues:} In multithreaded applications, ensuring that only one instance is created requires additional synchronization, which can introduce complexity.
\end{itemize}

\subsubsection{Use Cases}

The Singleton Pattern is used in scenarios where only a single instance of a class is needed. Some examples include:

\begin{itemize}
    \item \textbf{Database connections:} Only one connection object is needed to manage access to the database.
    \item \textbf{Configuration managers:} Centralizing configuration settings ensures that all parts of the application refer to the same configuration.
    \item \textbf{Logging:} A single logging object can manage logs across the entire application.
    \item \textbf{Cache management:} A single instance can manage cached data to optimize performance across different parts of the application.
\end{itemize}

\subsubsection{Code Example}

Here’s a basic Python implementation of the Singleton Pattern:

\begin{lstlisting}[style=python]
class Singleton:
    _instance = None
    
    @classmethod
    def getInstance(cls):
        if cls._instance is None:
            cls._instance = Singleton()
        return cls._instance

# Usage
s1 = Singleton.getInstance()
s2 = Singleton.getInstance()

print(s1 is s2)  # Output: True
\end{lstlisting}

In this example, the \texttt{getInstance()} method checks if the instance (\texttt{\_instance}) exists. If not, it creates a new instance of the \texttt{Singleton} class and stores it in \texttt{\_instance}. If the instance already exists, it simply returns it. This ensures that \texttt{s1} and \texttt{s2} refer to the same instance.

\subsubsection{Pattern Extensions}

There are some variations of the Singleton Pattern, such as:

\paragraph{Thread-safe Singleton:}

In a multithreaded environment, it is necessary to ensure that the Singleton instance is created in a thread-safe manner. Here’s an example of how to implement a thread-safe Singleton in Python:

\begin{lstlisting}[style=python]
import threading

class Singleton:
    _instance = None
    _lock = threading.Lock()  # A lock to synchronize threads
    
    @classmethod
    def getInstance(cls):
        with cls._lock:
            if cls._instance is None:
                cls._instance = Singleton()
        return cls._instance

# Usage in a multithreaded environment
s1 = Singleton.getInstance()
s2 = Singleton.getInstance()

print(s1 is s2)  # Output: True
\end{lstlisting}

In this version, the \texttt{getInstance()} method uses a lock to synchronize access and ensure that only one thread can create the instance at a time.

\paragraph{Lazy Initialization Singleton:}

In some cases, you may want to delay the creation of the Singleton instance until it is needed. This is known as lazy initialization, and it can be achieved as shown:

\begin{lstlisting}[style=python]
class Singleton:
    _instance = None
    
    def __new__(cls):
        if cls._instance is None:
            cls._instance = super(Singleton, cls).__new__(cls)
        return cls._instance

# Usage
s1 = Singleton()
s2 = Singleton()

print(s1 is s2)  # Output: True
\end{lstlisting}

Here, the \texttt{\_\_new\_\_()} method ensures that only one instance of the class is created, deferring its creation until the class is first instantiated.

\subsection{Factory Method Pattern}

The \textbf{Factory Method Pattern} \cite{chung2011factory} is a creational design pattern that provides an interface for creating objects in a super-class but allows subclasses to alter the type of objects that will be created. The key idea is that subclasses will override a specific method in the superclass to create different types of objects.

This pattern is useful when you want to delegate the instantiation of objects to subclasses while maintaining control over which specific types of objects are created.

\subsubsection{Motivation}

Imagine you are building a system where you need to create various types of products, such as different types of user interfaces, vehicles, or database connections. A naive approach would be to create these objects directly using constructors in multiple parts of your code. This leads to tight coupling between the classes that create the objects and the classes that use them.

The \textbf{Factory Method Pattern} solves this problem by allowing the creation logic to be placed in a method that subclasses can override, providing flexibility and ensuring loose coupling. With this approach, the client code only interacts with the abstract product and is unaware of which concrete product class will be instantiated.

\textbf{Example:} Consider a case where you have a class that needs to generate buttons for a GUI, but the type of button (e.g., Windows or Mac) depends on the environment. Using the Factory Method Pattern, the code that generates the button is separated from the code that uses it.

\subsubsection{Structure}

The basic structure of the Factory Method Pattern involves the following components:



 
\begin{itemize}
    \item \textbf{Creator (Abstract Class or Interface)}: Declares the factory method, which returns an object of type \texttt{Product}. The creator may also contain default logic that uses this factory method.
    \item \textbf{Concrete Creator (Subclass)}: Implements the factory method to return a specific concrete product.
    \item \textbf{Product (Abstract Class or Interface)}: Defines the interface for the objects the factory method creates.
    \item \textbf{Concrete Product (Subclass)}: A specific implementation of the product that the factory method returns.
\end{itemize}

The structure can be illustrated using a class diagram: \\
 
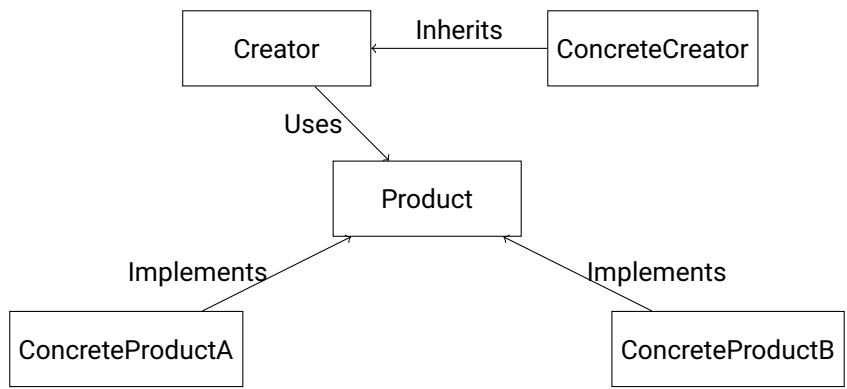
\begin{tikzpicture}
\node (Creator) at (0, 0) [draw, rectangle, minimum width=2.5cm, minimum height=1cm, text centered] {Creator};
\node (ConcreteCreator) at (5, 0) [draw, rectangle, minimum width=2.5cm, minimum height=1cm, text centered] {ConcreteCreator};
\node (Product) at (2, -2) [draw, rectangle, minimum width=2.5cm, minimum height=1cm, text centered] {Product};
\node (ConcreteProductA) at (-2, -4) [draw, rectangle, minimum width=2.5cm, minimum height=1cm, text centered] {ConcreteProductA};
\node (ConcreteProductB) at (6, -4) [draw, rectangle, minimum width=2.5cm, minimum height=1cm, text centered] {ConcreteProductB};

\draw[->] (ConcreteCreator) -- (Creator) node[midway, above] {Inherits};
\draw[->] (ConcreteProductA) -- (Product) node[midway, left] {Implements};
\draw[->] (ConcreteProductB) -- (Product) node[midway, right] {Implements};
\draw[->] (Creator) -- (Product) node[midway, left] {Uses};

\end{tikzpicture}

\subsubsection{Participants}

\begin{itemize}
    \item \textbf{Creator}: Defines the factory method that returns an instance of the \texttt{Product}. It may also provide a default implementation for the factory method, which returns a default product.
    \item \textbf{Concrete Creator}: Overrides the factory method to return an instance of a specific \texttt{Concrete Product}.
    \item \textbf{Product}: Declares an interface for objects that are created by the factory method.
    \item \textbf{Concrete Product}: Implements the \texttt{Product} interface.
\end{itemize}

\subsubsection{How It Works}

The Factory Method Pattern works by delegating the responsibility of creating objects to subclasses. The base class defines a method (the factory method) that the subclasses can override to instantiate their specific types of products. The client interacts with the base class and relies on the factory method to create the product without knowing the exact class of the product.

\textbf{Steps:}
\begin{enumerate}
    \item The client calls the factory method on the creator (abstract class or base class).
    \item The base class defers the object creation process to the subclass by calling the factory method.
    \item The subclass overrides the factory method and creates a specific instance of the product.
    \item The client receives a product that adheres to the \texttt{Product} interface without knowing the specific type of product.
\end{enumerate}

\subsubsection{Advantages and Disadvantages}

\textbf{Advantages:}
\begin{itemize}
    \item \textbf{Decoupling}: The Factory Method Pattern promotes loose coupling between the client and the product creation logic.
    \item \textbf{Flexibility}: The code is open for extension since you can easily introduce new products without modifying the existing client code.
    \item \textbf{Single Responsibility Principle}: The responsibility for creating products is isolated to the factory methods.
\end{itemize}

\textbf{Disadvantages:}
\begin{itemize}
    \item \textbf{Complexity}: It adds extra complexity by introducing additional subclasses to handle the creation process.
    \item \textbf{Overhead}: For simple applications, the overhead of subclassing and factory methods might be unnecessary.
\end{itemize}

\subsubsection{Use Cases}

The Factory Method Pattern is commonly used in the following scenarios:

\begin{itemize}
    \item When the exact type of object to be created is not known until runtime.
    \item When the creation of objects involves a complex process that is not simply handled by constructors.
    \item When you want to provide a way for different subclasses to create different products without changing the client code.
\end{itemize}

\textbf{Example Use Cases:}
\begin{itemize}
    \item GUI libraries that create platform-specific components (Windows/Mac/Linux buttons, menus, etc.).
    \item Logging frameworks that need to create different types of loggers (e.g., file, console, database).
    \item Database connection libraries that need to create different types of connections depending on the database type (e.g., MySQL, PostgreSQL).
\end{itemize}

\subsubsection{Code Example}

Here is a Python code example to illustrate the Factory Method Pattern.

\begin{lstlisting}[style=python]
from abc import ABC, abstractmethod

# Product Interface
class Button(ABC):
    @abstractmethod
    def render(self):
        pass

# Concrete Product A
class WindowsButton(Button):
    def render(self):
        print("Rendering a Windows button.")

# Concrete Product B
class MacButton(Button):
    def render(self):
        print("Rendering a Mac button.")

# Creator (Factory)
class Dialog(ABC):
    @abstractmethod
    def create_button(self) -> Button:
        pass
    
    def render_dialog(self):
        button = self.create_button()
        button.render()

# Concrete Creator A
class WindowsDialog(Dialog):
    def create_button(self) -> Button:
        return WindowsButton()

# Concrete Creator B
class MacDialog(Dialog):
    def create_button(self) -> Button:
        return MacButton()

# Client code
def client_code(dialog: Dialog):
    dialog.render_dialog()

# Example Usage
if __name__ == "__main__":
    platform = input("Enter your platform (Windows/Mac): ").strip().lower()
    if platform == "windows":
        dialog = WindowsDialog()
    elif platform == "mac":
        dialog = MacDialog()
    else:
        print("Unsupported platform.")
        exit(1)

    client_code(dialog)
\end{lstlisting}

\textbf{Explanation:}
\begin{itemize}
    \item The \texttt{Button} class is an abstract class (product interface) that defines the \texttt{render} method.
    \item \texttt{WindowsButton} and \texttt{MacButton} are concrete products that implement the \texttt{Button} interface.
    \item The \texttt{Dialog} class is the abstract creator, and its \texttt{create\_button} method is the factory method.
    \item \texttt{WindowsDialog} and \texttt{MacDialog} are concrete creators that override \texttt{create\_button} to return the correct type of button.
\end{itemize}

\subsubsection{Pattern Extensions}

There are several ways to extend the Factory Method Pattern:

\begin{itemize}
    \item \textbf{Parameterization}: You can pass parameters to the factory method to allow for more dynamic product creation. For example, you could pass configurations or types to determine which product to create.
    \item \textbf{Abstract Factory Pattern}: If you need to create families of related products (e.g., different types of buttons, windows, and dialogs), you can extend this pattern into the Abstract Factory Pattern.
    \item \textbf{Lazy Initialization}: In cases where product creation is expensive, you can implement lazy initialization to create products only when they are needed.
\end{itemize}

\subsection{Abstract Factory Pattern}
The Abstract Factory Pattern \cite{bulajic2012approach} is one of the creational design patterns, focusing on creating families of related objects without specifying their concrete classes. This pattern is especially useful when a system needs to be independent of how its objects are created, composed, and represented. The pattern provides a way to encapsulate a group of individual factories with a common interface.

\subsubsection{Motivation}
Imagine you're building a GUI toolkit that can support different operating systems such as Windows, Mac, and Linux. Each operating system has its own unique set of GUI components (buttons, checkboxes, etc.), but you want to create a unified API that allows developers to use these components without worrying about the specifics of the operating system they are working on. The Abstract Factory Pattern allows you to define a set of abstract classes for these components (e.g., \texttt{Button}, \texttt{Checkbox}), and then create concrete factories to instantiate these components for each operating system (e.g., \texttt{WindowsFactory}, \texttt{MacFactory}, \texttt{LinuxFactory}). 

\subsubsection{Structure}
The Abstract Factory Pattern typically consists of the following components:
\begin{itemize}
    \item \textbf{AbstractFactory}: Declares a set of creation methods for abstract product types (e.g., \texttt{createButton()}, \texttt{createCheckbox()}).
    \item \textbf{ConcreteFactory}: Implements the creation methods for specific product families (e.g., \texttt{WindowsFactory}, \texttt{MacFactory}).
    \item \textbf{AbstractProduct}: Declares an interface for a type of product (e.g., \texttt{Button}, \texttt{Checkbox}).
    \item \textbf{ConcreteProduct}: Implements the abstract product interfaces (e.g., \texttt{WindowsButton}, \texttt{MacCheckbox}).
    \item \textbf{Client}: Uses only the interfaces declared by the abstract factory and abstract product classes. The client is unaware of the specific classes used to create the concrete objects.
\end{itemize}

Here's a simple diagram representing the structure: \\

  
\usetikzlibrary{positioning}
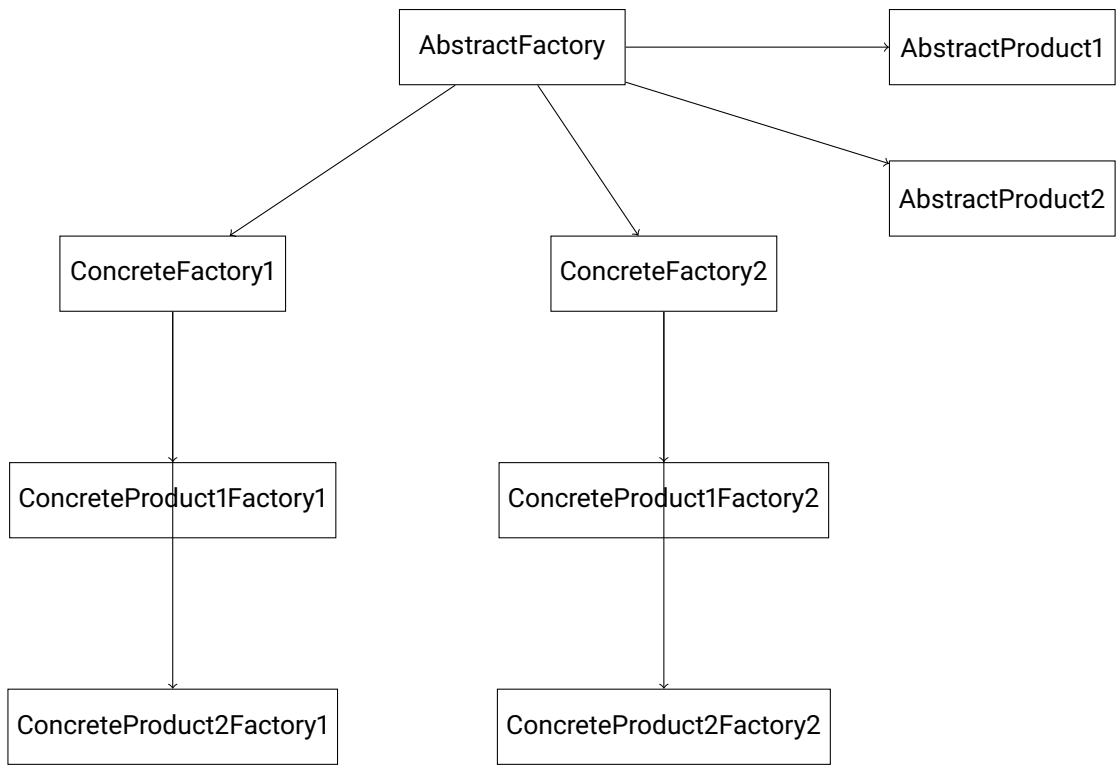
\begin{tikzpicture}
  \node (AF) [draw, rectangle, minimum width=3cm, minimum height=1cm] {AbstractFactory};
  \node (CF1) [below left=2cm and 1.5cm of AF, draw, rectangle, minimum width=3cm, minimum height=1cm] {ConcreteFactory1};
  \node (CF2) [below right=2cm and -1cm of AF, draw, rectangle, minimum width=3cm, minimum height=1cm] {ConcreteFactory2};
  \node (AP1) [right=3.5cm of AF, draw, rectangle, minimum width=3cm, minimum height=1cm] {AbstractProduct1};
  \node (AP2) [below=1cm of AP1, draw, rectangle, minimum width=3cm, minimum height=1cm] {AbstractProduct2};
  \node (CP1F1) [below=2cm of CF1, draw, rectangle, minimum width=3cm, minimum height=1cm] {ConcreteProduct1Factory1};
  \node (CP2F1) [below=2cm of CP1F1, draw, rectangle, minimum width=3cm, minimum height=1cm] {ConcreteProduct2Factory1};
  \node (CP1F2) [below=2cm of CF2, draw, rectangle, minimum width=3cm, minimum height=1cm] {ConcreteProduct1Factory2};
  \node (CP2F2) [below=2cm of CP1F2, draw, rectangle, minimum width=3cm, minimum height=1cm] {ConcreteProduct2Factory2};

  \draw[->] (AF) -- (CF1);
  \draw[->] (AF) -- (CF2);
  \draw[->] (CF1) -- (CP1F1);
  \draw[->] (CF1) -- (CP2F1);
  \draw[->] (CF2) -- (CP1F2);
  \draw[->] (CF2) -- (CP2F2);
  \draw[->] (AF) -- (AP1);
  \draw[->] (AF) -- (AP2);
\end{tikzpicture}

\subsubsection{Participants}
Here are the main participants in the Abstract Factory Pattern:

\begin{itemize}
    \item \textbf{AbstractFactory} (e.g., \texttt{GUIFactory}): Declares the creation methods for different types of products. 
    \item \textbf{ConcreteFactory} (e.g., \texttt{WindowsFactory}, \texttt{MacFactory}): Implements the creation methods to produce products that are specific to a particular family.
    \item \textbf{AbstractProduct} (e.g., \texttt{Button}, \texttt{Checkbox}): Declares the interface for the different products that can be created.
    \item \textbf{ConcreteProduct} (e.g., \texttt{WindowsButton}, \texttt{MacCheckbox}): Implements the abstract product interfaces.
    \item \textbf{Client}: Uses the factory to create instances of the products and works only with their abstract interfaces.
\end{itemize}

\subsubsection{How It Works}
1. The client code interacts with an abstract factory (\texttt{GUIFactory}), which provides a method for creating each type of product.
2. The factory method calls return abstract products (\texttt{Button}, \texttt{Checkbox}), which the client can use without needing to know the specific class of the product.
3. Depending on the specific factory implementation (\texttt{WindowsFactory}, \texttt{MacFactory}), the appropriate product implementation is created (\texttt{WindowsButton}, \texttt{MacButton}).
4. This process abstracts the creation logic from the client, allowing the same client code to work with different types of product families.

\subsubsection{Advantages and Disadvantages}
\textbf{Advantages}:
\begin{itemize}
    \item \textbf{Flexibility}: The client code can work with any product family without being coupled to its specific classes.
    \item \textbf{Consistency}: Ensures that a family of related products is used together (e.g., Windows GUI elements are consistent in style and functionality).
    \item \textbf{Scalability}: It's easy to add new families of products without changing existing code, just by introducing new factories and concrete products.
\end{itemize}

\textbf{Disadvantages}:
\begin{itemize}
    \item \textbf{Complexity}: The pattern introduces additional complexity by requiring a new set of abstract classes and interfaces.
    \item \textbf{Difficult to extend products individually}: If you want to introduce a new product into an existing family, you must modify every factory to support it.
\end{itemize}

\subsubsection{Use Cases}
The Abstract Factory Pattern is most useful in the following scenarios:
\begin{itemize}
    \item When a system needs to be independent of how its products are created.
    \item When you want to ensure that a family of related products is used together (e.g., all GUI elements should follow the same style and behavior).
    \item When you need to provide a library of products that work well together but vary between different contexts (e.g., different platforms or configurations).
    \item When a system must support different themes or configurations that involve distinct product families.
\end{itemize}

\subsubsection{Code Example}
Here's an example of how the Abstract Factory Pattern can be implemented in Python for a GUI toolkit that supports multiple platforms (Windows and Mac).

\begin{lstlisting}[style=python]
from abc import ABC, abstractmethod

# Abstract Product: Button
class Button(ABC):
    @abstractmethod
    def click(self) -> None:
        pass

# Concrete Product: Windows Button
class WindowsButton(Button):
    def click(self) -> None:
        print("Windows Button clicked!")

# Concrete Product: Mac Button
class MacButton(Button):
    def click(self) -> None:
        print("Mac Button clicked!")

# Abstract Product: Checkbox
class Checkbox(ABC):
    @abstractmethod
    def check(self) -> None:
        pass

# Concrete Product: Windows Checkbox
class WindowsCheckbox(Checkbox):
    def check(self) -> None:
        print("Windows Checkbox checked!")

# Concrete Product: Mac Checkbox
class MacCheckbox(Checkbox):
    def check(self) -> None:
        print("Mac Checkbox checked!")

# Abstract Factory
class GUIFactory(ABC):
    @abstractmethod
    def create_button(self) -> Button:
        pass

    @abstractmethod
    def create_checkbox(self) -> Checkbox:
        pass

# Concrete Factory: Windows Factory
class WindowsFactory(GUIFactory):
    def create_button(self) -> Button:
        return WindowsButton()

    def create_checkbox(self) -> Checkbox:
        return WindowsCheckbox()

# Concrete Factory: Mac Factory
class MacFactory(GUIFactory):
    def create_button(self) -> Button:
        return MacButton()

    def create_checkbox(self) -> Checkbox:
        return MacCheckbox()

# Client code
def client(factory: GUIFactory) -> None:
    button = factory.create_button()
    checkbox = factory.create_checkbox()

    button.click()
    checkbox.check()

# Test Windows factory
print("Testing Windows factory:")
client(WindowsFactory())

# Test Mac factory
print("\nTesting Mac factory:")
client(MacFactory())
\end{lstlisting}

In this example:
\begin{itemize}
    \item \texttt{Button} and \texttt{Checkbox} are abstract products.
    \item \texttt{WindowsButton}, \texttt{MacButton}, \texttt{WindowsCheckbox}, and \texttt{MacCheckbox} are concrete products.
    \item \texttt{GUIFactory} is the abstract factory, while \texttt{WindowsFactory} and \texttt{MacFactory} are concrete factories.
    \item The \texttt{client} function works with the abstract \texttt{GUIFactory} and uses its methods to create buttons and checkboxes, regardless of the specific platform.
\end{itemize}

\subsubsection{Pattern Extensions}
The Abstract Factory Pattern can be extended or modified in several ways:
\begin{itemize}
    \item \textbf{Factory as a Singleton}: If your application only needs one instance of each factory, consider implementing the factory as a Singleton.
    \item \textbf{Product Variations}: You can extend the product hierarchy to include more variations of products (e.g., different styles or themes).
    \item \textbf{Dynamic Factories}: In some scenarios, you may want to select a factory dynamically at runtime based on external configurations or user input.
\end{itemize}

By abstracting the creation process and hiding specific implementation details, the Abstract Factory Pattern provides a powerful tool for building flexible, extensible systems, particularly when multiple product families or variations are involved.

\subsection{Builder Pattern}
The Builder pattern \cite{chung2011builder} is a design pattern that is used to create complex objects step by step. Unlike other creational patterns that return the object immediately, the Builder pattern allows you to create an object by specifying its parts or attributes in a controlled and step-by-step manner.

\subsubsection{Motivation}
In some cases, creating a complex object may require a step-by-step approach, especially when the object can have multiple configurations or involves various parts that need to be built in a certain order. The Builder pattern is particularly useful when:
\begin{itemize}
    \item The process of object construction is complex.
    \item Different representations of the object can be created using the same building process.
    \item You want to decouple the object creation logic from the actual object itself.
\end{itemize}
A good real-world analogy is the construction of a house. A house can be built with different configurations (number of rooms, type of flooring, type of roof), but the process of building it always follows the same steps.

\subsubsection{Structure}
The Builder pattern typically consists of the following components:
\begin{itemize}
    \item \textbf{Builder}: An interface or abstract class that defines the steps required to build the object.
    \item \textbf{ConcreteBuilder}: Implements the Builder interface and defines the actual building process.
    \item \textbf{Director}: Responsible for controlling the building process and ensuring that steps are executed in the right order.
    \item \textbf{Product}: The final object that is being built.
\end{itemize}

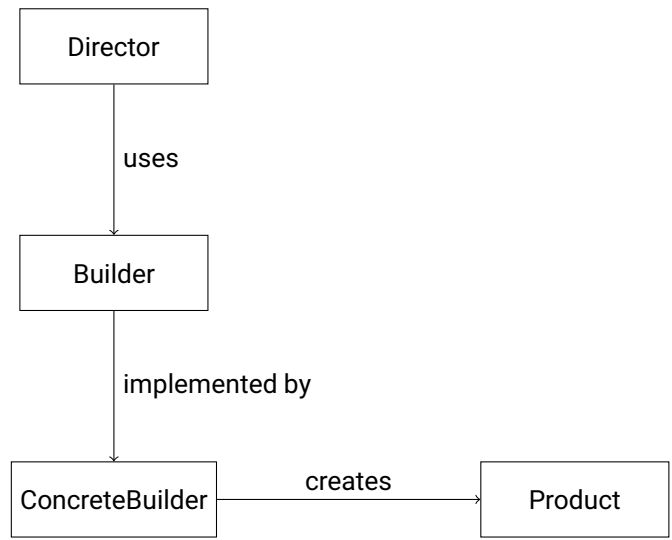
\begin{center}
\begin{tikzpicture}[scale=1.0, every node/.style={scale=1.0}]
  \node (director) [draw, rectangle, minimum width=2.5cm, minimum height=1cm] {Director};
  \node (builder) [draw, rectangle, below=2cm of director, minimum width=2.5cm, minimum height=1cm] {Builder};
  \node (concretebuilder) [draw, rectangle, below=2cm of builder, minimum width=2.5cm, minimum height=1cm] {ConcreteBuilder};
  \node (product) [draw, rectangle, right=3.5cm of concretebuilder, minimum width=2.5cm, minimum height=1cm] {Product};

  \draw[->] (director) -- (builder) node[midway,right] {uses};
  \draw[->] (builder) -- (concretebuilder) node[midway,right] {implemented by};
  \draw[->] (concretebuilder) -- (product) node[midway,above] {creates};
\end{tikzpicture}
\end{center}

\subsubsection{Participants}
\begin{itemize}
    \item \textbf{Builder}: Specifies an abstract interface for creating parts of a product object.
    \item \textbf{ConcreteBuilder}: Implements the Builder interface to construct and assemble parts of the product.
    \item \textbf{Director}: Constructs the object using the Builder interface. It is responsible for executing the steps in the correct sequence.
    \item \textbf{Product}: The object that is being built.
\end{itemize}

\subsubsection{How It Works}
The Builder pattern works by separating the construction of a complex object from its representation. This allows the same construction process to create different representations. The client directs the construction using a Director, which uses a Builder to assemble the final object step-by-step. The Builder hides the complexity of the construction process and the product is returned only when the building process is complete.

The steps are:
\begin{enumerate}
    \item The Director controls the building process.
    \item The Builder interface specifies the steps to be implemented.
    \item A ConcreteBuilder implements these steps.
    \item The final product is returned after the steps are completed.
\end{enumerate}

\subsubsection{Advantages and Disadvantages}
\textbf{Advantages}:
\begin{itemize}
    \item Provides control over the construction process.
    \item Allows for different representations of the object using the same construction process.
    \item Decouples the construction process from the product's internal representation.
\end{itemize}

\textbf{Disadvantages}:
\begin{itemize}
    \item Can increase complexity when the number of steps or configurations increases.
    \item Requires creating a new ConcreteBuilder for each representation of the object.
\end{itemize}

\subsubsection{Use Cases}
The Builder pattern is ideal for the following scenarios:
\begin{itemize}
    \item Creating objects with many optional parts or configurations.
    \item When the construction process involves several steps or phases.
    \item For objects that need to be assembled in a specific order.
    \item For creating objects that can be reused or customized.
\end{itemize}

Some common examples in software development:
\begin{itemize}
    \item Building complex UI components where each part needs to be customized.
    \item Constructing objects that require validation or conditional creation of components.
    \item Creating database query builders, where different SQL clauses are combined in a step-by-step fashion.
\end{itemize}

\subsubsection{Code Example}
Here’s a Python implementation of the Builder pattern:

\begin{lstlisting}[style=python]
class Product:
    def __init__(self):
        self.parts = []
    
    def add(self, part):
        self.parts.append(part)
    
    def show(self):
        print("Product parts:", ', '.join(self.parts))

# Builder interface
class Builder:
    def build_part_a(self):
        pass
    
    def build_part_b(self):
        pass
    
    def get_product(self):
        pass

# Concrete Builder
class ConcreteBuilder(Builder):
    def __init__(self):
        self.product = Product()
    
    def build_part_a(self):
        self.product.add("Part A")
    
    def build_part_b(self):
        self.product.add("Part B")
    
    def get_product(self):
        return self.product

# Director
class Director:
    def __init__(self, builder):
        self._builder = builder
    
    def construct(self):
        self._builder.build_part_a()
        self._builder.build_part_b()

# Client code
builder = ConcreteBuilder()
director = Director(builder)
director.construct()

product = builder.get_product()
product.show()
\end{lstlisting}

In this example:
\begin{itemize}
    \item \texttt{Product} is the final object that is being built.
    \item \texttt{Builder} defines the interface for creating parts of the product.
    \item \texttt{ConcreteBuilder} implements the methods to build different parts of the product.
    \item \texttt{Director} controls the building process.
\end{itemize}
The Director ensures that the parts of the product are constructed in the correct order. Finally, the client retrieves the product after the building process is complete.

\subsubsection{Pattern Extensions}
There are several ways to extend the Builder pattern:
\begin{itemize}
    \item \textbf{Fluent Interface}: The Builder pattern can be extended with a fluent interface, where methods return the builder itself, allowing method chaining for more readable code.
    \begin{lstlisting}[style=python]
class FluentBuilder:
    def __init__(self):
        self.product = Product()
    
    def build_part_a(self):
        self.product.add("Part A")
        return self
    
    def build_part_b(self):
        self.product.add("Part B")
        return self
    
    def get_product(self):
        return self.product

# Fluent Builder usage
builder = FluentBuilder()
product = builder.build_part_a().build_part_b().get_product()
product.show()
    \end{lstlisting}
    
    \item \textbf{Multiple Builders}: If a product has many possible configurations, you can define multiple ConcreteBuilders to handle different variations. For example, you could have separate builders for different types of products, each building the product in a different way.

    \item \textbf{Composite Builder}: In cases where you are building a complex object that contains other complex objects, you might use a composite builder. For example, you could build a car that consists of a chassis, engine, and wheels, each of which can be built with its builder.
\end{itemize}

\subsection{Prototype Pattern}

The Prototype Pattern is a creational design pattern used to instantiate new objects by copying or cloning an existing object, known as the prototype. It is particularly useful when object creation is costly in terms of resources or time, and you want to avoid repeating the construction process.

\subsubsection{Motivation}
The primary motivation behind the Prototype Pattern \cite{chung2011prototype} is to create new objects by cloning an existing object rather than constructing one from scratch. This pattern is useful when:
\begin{itemize}
    \item The process of object creation is expensive (e.g., involves complex computations or costly I/O operations).
    \item The structure of the object is complicated or involves many configurations that are difficult to set up.
    \item You need to create multiple instances of a class with only slight differences in their state.
\end{itemize}

For example, imagine a scenario where you need to create different types of vehicles (cars, trucks, motorcycles) with different attributes such as color, engine type, and so on. Instead of creating each vehicle from scratch, you can create a prototype of each type of vehicle and clone it as needed, modifying only the necessary attributes.

\subsubsection{Structure}
The Prototype Pattern consists of several key components:

\begin{itemize}
    \item \textbf{Prototype:} An interface or abstract class that declares the method for cloning itself.
    \item \textbf{ConcretePrototype:} A class that implements the prototype interface and defines how to clone itself.
    \item \textbf{Client:} The client class that requests the creation of new objects using prototypes.
\end{itemize}

Here's a diagram to illustrate the structure:

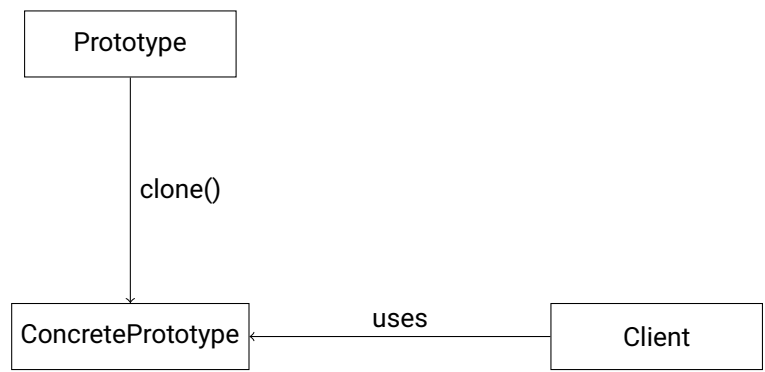
\begin{center}
\begin{tikzpicture}
    \node[draw, rectangle, minimum width=8em, minimum height=2.5em, text centered] (prototype) {Prototype};
    \node[draw, rectangle, below=3cm of prototype, minimum width=8em, minimum height=2.5em, text centered] (concretePrototype) {ConcretePrototype};
    \node[draw, rectangle, right=4cm of concretePrototype, minimum width=8em, minimum height=2.5em, text centered] (client) {Client};
    
    \draw[->] (prototype.south) -- (concretePrototype.north) node[midway, right] {clone()};
    \draw[->] (client.west) -- (concretePrototype.east) node[midway, above] {uses};
    
\end{tikzpicture}
\end{center}

\subsubsection{Participants}
\begin{itemize}
    \item \textbf{Prototype:} Declares the cloning method (in Python, this could be implemented via the `copy` method or a custom method).
    \item \textbf{ConcretePrototype:} Implements the cloning method. This class holds the actual object data and defines how to create a duplicate of itself.
    \item \textbf{Client:} Uses the prototype's clone method to create a new object, rather than using a constructor.
\end{itemize}

\subsubsection{How It Works}
The Prototype Pattern works by creating a "template" or "prototype" object, which can then be cloned whenever a new object is needed. In Python, this can be achieved using the `copy` module or by defining a custom method for copying the object.

Here is how it works step by step:
\begin{enumerate}
    \item A class implements a `clone` method (or uses Python's `copy` module).
    \item The `clone` method creates and returns a copy of the object.
    \item The client can create new objects by calling the `clone` method of an existing object.
    \item Optionally, after cloning, the client can modify the cloned object as necessary.
\end{enumerate}

\subsubsection{Advantages and Disadvantages}
\textbf{Advantages:}
\begin{itemize}
    \item Reduces the need for repetitive and costly object creation.
    \item Provides a flexible way to instantiate objects with complex configurations.
    \item Allows for adding or removing prototypes at runtime, making it highly adaptable.
\end{itemize}

\textbf{Disadvantages:}
\begin{itemize}
    \item Cloning objects can be complex if they involve deep copies (objects with references to other objects).
    \item If not implemented correctly, it can lead to unexpected behavior when dealing with mutable objects.
    \item Managing object prototypes and ensuring that they are up to date can introduce additional complexity.
\end{itemize}

\subsubsection{Use Cases}
The Prototype Pattern is particularly useful in scenarios such as:
\begin{itemize}
    \item When object creation is expensive (e.g., connecting to a database or making network requests).
    \item When a system uses many objects that share the same structure or configuration but differ in a few attributes.
    \item When you want to keep track of object states or versions.
    \item Video games, where you need to create many similar characters or assets.
\end{itemize}

\subsubsection{Code Example}
Let's look at an example of how the Prototype Pattern might be implemented in Python:

\begin{lstlisting}[style=python]
import copy

class Prototype:
    def clone(self):
        pass

class ConcretePrototype(Prototype):
    def __init__(self, name, color):
        self.name = name
        self.color = color

    def __str__(self):
        return f"{self.name} ({self.color})"

    def clone(self):
        # Here we can use a deep copy to ensure all attributes are copied properly
        return copy.deepcopy(self)

# Client code
if __name__ == "__main__":
    # Create a prototype object
    prototype_vehicle = ConcretePrototype("Car", "Red")
    print(f"Original Prototype: {prototype_vehicle}")

    # Clone the prototype
    cloned_vehicle = prototype_vehicle.clone()
    print(f"Cloned Vehicle: {cloned_vehicle}")

    # Modify the clone
    cloned_vehicle.color = "Blue"
    print(f"Modified Cloned Vehicle: {cloned_vehicle}")
    print(f"Original Prototype after modification: {prototype_vehicle}")
\end{lstlisting}

In this example, we have a `ConcretePrototype` class that can clone itself. The `clone` method uses Python's `copy.deepcopy` function to ensure that all attributes are copied correctly, even if they contain references to other objects.

When we run this code, we get the following output:

\begin{lstlisting}[style=cmd]
Original Prototype: Car (Red)
Cloned Vehicle: Car (Red)
Modified Cloned Vehicle: Car (Blue)
Original Prototype after modification: Car (Red)
\end{lstlisting}

Notice how modifying the cloned object does not affect the original prototype.

\subsubsection{Pattern Extensions}
There are several ways to extend the Prototype Pattern:
\begin{itemize}
    \item \textbf{Prototype Registry:} You can maintain a registry of prototypes, which allows for more dynamic and flexible object creation. Instead of directly instantiating prototypes, the client can request an object from the registry.
    \item \textbf{Deep and Shallow Cloning:} Depending on the situation, you may need to implement both deep and shallow copies in the prototype. Deep copying duplicates all referenced objects, while shallow copying duplicates only the top-level object.
    \item \textbf{Versioning Prototypes:} In some cases, it might be useful to store different versions of a prototype to capture different states or configurations over time.
\end{itemize}

The Prototype Pattern is a versatile and efficient way to manage object creation, especially in cases where performance or configuration complexity is a concern.

\section{Structural Patterns}
\subsection{Introduction}

In software design, structural patterns \cite{syrquin1988patterns} are concerned with how objects and classes are composed to form larger structures. These patterns simplify the design by identifying a simple way to realize relationships between different objects or entities. The main goal of structural patterns is to facilitate the design of complex systems by organizing objects and classes to ensure that the system remains flexible and efficient.

In Python, structural patterns help you organize your code into larger, reusable components. The focus is on how these components are structured and interact with one another. These patterns promote the idea of combining objects and classes in different ways to build more complex functionality without being locked into rigid or hard-coded relationships.

\subsubsection{Key Concepts in Structural Patterns}
There are a few key ideas to understand when working with structural design patterns in Python:
\begin{itemize}
    \item \textbf{Composition over Inheritance:} One of the fundamental principles behind structural patterns is favoring composition over inheritance. This means that rather than creating large class hierarchies using inheritance, we combine objects to build more flexible systems.
    \item \textbf{Decoupling Components:} Structural patterns often aim to reduce the dependency between components, making it easier to extend or modify parts of a system without breaking the entire application.
    \item \textbf{Flexibility and Reusability:} By thoughtfully organizing objects, structural patterns promote code that is easier to reuse and extend, which leads to more maintainable systems.
\end{itemize}

\subsubsection{Common Structural Patterns}
Several structural patterns are frequently used in Python programming. Some of the most commonly known patterns include:
\begin{itemize}
    \item \textbf{Adapter Pattern:} This pattern allows objects with incompatible interfaces to work together by introducing an adapter that makes their interfaces compatible.
    \item \textbf{Decorator Pattern:} This pattern is used to dynamically add behavior to objects without modifying their class, by wrapping them with decorator objects.
    \item \textbf{Facade Pattern:} This pattern provides a simplified interface to a complex subsystem, making it easier for clients to interact with the subsystem without needing to understand its complexities.
    \item \textbf{Proxy Pattern:} The proxy pattern introduces a surrogate or placeholder object that controls access to another object, providing controlled or delayed access.
    \item \textbf{Composite Pattern:} This pattern is used to treat individual objects and compositions of objects uniformly, often by organizing objects into tree-like structures.
\end{itemize}

Each of these patterns solves a specific design problem related to object composition and interactions. In the following sections, we will explore each pattern in detail, with Python examples to illustrate how they can be implemented.

To help explain the structural patterns, we'll use clear and concrete examples in Python, keeping things simple and beginner-friendly. Let's start by exploring these patterns in the upcoming sections.

\subsection{Adapter Pattern}

The Adapter Pattern \cite{hummel2009managed} is a structural design pattern that allows incompatible interfaces to work together. Essentially, it acts as a bridge between two incompatible classes, allowing them to collaborate. This pattern is often used in situations where we have an existing class and need it to interact with another class with a different interface, without modifying the original class.

\subsubsection{Motivation}

The need for an adapter arises when we have a class or function that expects an object with a specific interface, but we are given a different object that cannot be used directly. Instead of modifying the existing classes (which may not always be possible or desirable), we use an adapter to convert the interface of the existing class into the interface expected by the client. 

For example, suppose we are working with a payment processing library that expects objects with a method `process\_payment()`, but we have an existing class that performs a similar function with a method `make\_payment()`. In this case, we can use an adapter to convert the interface of `make\_payment()` into `process\_payment()`.

\subsubsection{Structure}

The Adapter Pattern involves the following key components:

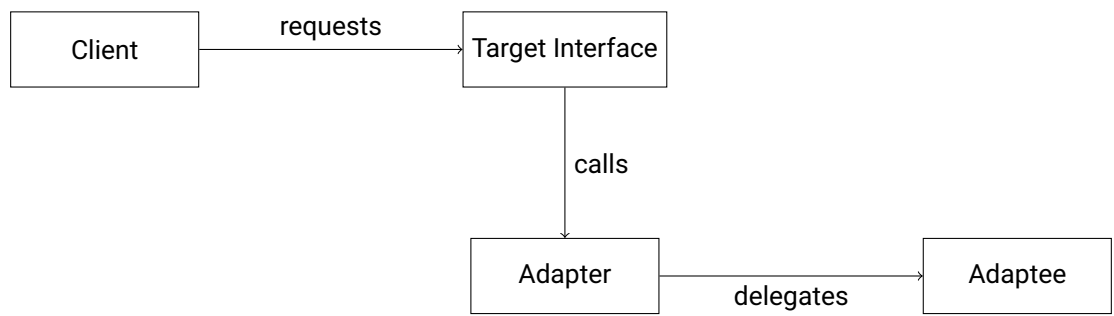
\begin{center}
\begin{tikzpicture}[scale=1]
\node[draw, minimum width=2.5cm, minimum height=1cm] (Client) {Client};
\node[draw, right=3.5cm of Client, minimum width=2.5cm, minimum height=1cm] (Target) {Target Interface};
\node[draw, below=2cm of Target, minimum width=2.5cm, minimum height=1cm] (Adapter) {Adapter};
\node[draw, right=3.5cm of Adapter, minimum width=2.5cm, minimum height=1cm] (Adaptee) {Adaptee};

\draw[->] (Client) -- node[above] {requests} (Target);
\draw[->] (Target) -- node[right] {calls} (Adapter);
\draw[->] (Adapter) -- node[below] {delegates} (Adaptee);
\end{tikzpicture}
\end{center}

\begin{itemize}
    \item \textbf{Client}: The object that requires the functionality of another class, but expects it to follow a specific interface.
    \item \textbf{Target Interface}: The interface that the client expects.
    \item \textbf{Adapter}: The object that implements the `Target Interface` and translates requests from the client into calls to the `Adaptee`.
    \item \textbf{Adaptee}: The existing object with an incompatible interface, which contains the actual functionality needed by the client.
\end{itemize}

\subsubsection{Participants}

\begin{itemize}
    \item \textbf{Client}: The class or object that interacts with the system, making requests for specific services.
    \item \textbf{Adapter}: The class that adapts the interface of the `Adaptee` to the `Target Interface` required by the `Client`. It implements the `Target Interface` and delegates calls to the `Adaptee`.
    \item \textbf{Adaptee}: The class with an existing, incompatible interface that needs to be adapted.
    \item \textbf{Target Interface}: The interface that defines the expected methods for the client.
\end{itemize}

\subsubsection{How It Works}

The Adapter Pattern works by wrapping the `Adaptee` with a class that translates its interface to match the `Target Interface`. The `Adapter` class implements the `Target Interface`, and its methods internally call the corresponding methods of the `Adaptee`.

In Python, we can achieve this by using composition: the `Adapter` holds an instance of the `Adaptee` and translates method calls as necessary.

\begin{lstlisting}[style=python]
# Target Interface
class PaymentProcessor:
    def process_payment(self, amount):
        pass

# Adaptee (incompatible interface)
class OldPaymentSystem:
    def make_payment(self, amount):
        print(f"Processing payment of ${amount} in the old system")

# Adapter that makes OldPaymentSystem compatible with PaymentProcessor
class PaymentAdapter(PaymentProcessor):
    def __init__(self, old_system):
        self.old_system = old_system

    def process_payment(self, amount):
        # Translate the call to the Adaptee's method
        self.old_system.make_payment(amount)

# Client code
def complete_payment(payment_processor, amount):
    payment_processor.process_payment(amount)

old_payment_system = OldPaymentSystem()
adapter = PaymentAdapter(old_payment_system)
complete_payment(adapter, 100)
\end{lstlisting}

In this example:
\begin{itemize}
    \item `PaymentProcessor` is the `Target Interface` that the `Client` expects.
    \item `OldPaymentSystem` is the `Adaptee`, which has an incompatible `make\_payment()` method.
    \item `PaymentAdapter` is the `Adapter` that wraps the `OldPaymentSystem` and translates `process\_payment()` calls into `make\_payment()` calls.
    \item The `Client` code uses the `Target Interface` (`process\_payment()`), and the adapter ensures that the `OldPaymentSystem` works seamlessly with the `Client`.
\end{itemize}

\subsubsection{Advantages and Disadvantages}

\textbf{Advantages}:
\begin{itemize}
    \item \textbf{Reuse of existing functionality}: You can adapt existing classes without modifying their source code, making them reusable in new systems.
    \item \textbf{Improves code flexibility}: By decoupling the client from the adaptee, the code becomes more flexible and can evolve without requiring changes to existing components.
    \item \textbf{Open/Closed Principle}: You can extend the behavior of classes without changing their code.
\end{itemize}

\textbf{Disadvantages}:
\begin{itemize}
    \item \textbf{Increased complexity}: Adding an adapter adds another layer of abstraction, which can increase the complexity of the system.
    \item \textbf{Performance overhead}: Depending on the complexity of the adaptation, the adapter may introduce performance overhead due to additional method calls and object creation.
\end{itemize}

\subsubsection{Use Cases}

The Adapter Pattern is useful in the following scenarios:
\begin{itemize}
    \item When you need to integrate a legacy system with a modern interface.
    \item When you want to use a third-party library that does not follow the interface required by your codebase.
    \item When you want to reuse existing classes, but their interfaces are not compatible with the system you are building.
    \item In situations where a class cannot be modified directly due to external ownership or the risk of breaking other code.
\end{itemize}

\textbf{Example:} Suppose you're building an application that needs to log messages, but your existing logging system has a method called `write\_log()`, whereas your new system expects a `log\_message()` method. An adapter can bridge this gap and allow you to use the old logging system without changing its implementation.

\begin{lstlisting}[style=python]
# Target Interface
class Logger:
    def log_message(self, message):
        pass

# Adaptee (incompatible interface)
class OldLogger:
    def write_log(self, message):
        print(f"Logging message: {message}")

# Adapter
class LoggerAdapter(Logger):
    def __init__(self, old_logger):
        self.old_logger = old_logger

    def log_message(self, message):
        # Delegate the call to the old logger
        self.old_logger.write_log(message)

# Client code
def log_activity(logger, message):
    logger.log_message(message)

old_logger = OldLogger()
adapter = LoggerAdapter(old_logger)
log_activity(adapter, "Adapter Pattern in action!")
\end{lstlisting}

\subsubsection{Code Example}

Here's another Python example demonstrating how the Adapter Pattern can be used to integrate incompatible interfaces:

\begin{lstlisting}[style=python]
# Target Interface
class MediaPlayer:
    def play(self, audio_type, file_name):
        pass

# Adaptee (with incompatible interface)
class AdvancedMediaPlayer:
    def play_vlc(self, file_name):
        print(f"Playing vlc file: {file_name}")
    
    def play_mp4(self, file_name):
        print(f"Playing mp4 file: {file_name}")

# Adapter class
class MediaAdapter(MediaPlayer):
    def __init__(self, audio_type):
        self.advanced_media_player = AdvancedMediaPlayer()

    def play(self, audio_type, file_name):
        if audio_type == "vlc":
            self.advanced_media_player.play_vlc(file_name)
        elif audio_type == "mp4":
            self.advanced_media_player.play_mp4(file_name)

# Client code
class AudioPlayer(MediaPlayer):
    def play(self, audio_type, file_name):
        if audio_type in ["vlc", "mp4"]:
            adapter = MediaAdapter(audio_type)
            adapter.play(audio_type, file_name)
        else:
            print(f"Invalid media: {audio_type} format not supported")

# Usage
player = AudioPlayer()
player.play("mp4", "song.mp4")
player.play("vlc", "movie.vlc")
\end{lstlisting}

In this example, the `AudioPlayer` class can play media files in different formats, but it uses the `MediaAdapter` to translate the play request to the `AdvancedMediaPlayer` methods.

\subsubsection{Pattern Extensions}

The Adapter Pattern can be extended in several ways:
\begin{itemize}
    \item \textbf{Two-Way Adapter}: An adapter that allows two classes with incompatible interfaces to communicate in both directions. It adapts both interfaces so that either can be used as the client.
    \item \textbf{Class Adapter}: In some languages, an adapter can be created using inheritance rather than composition, allowing the adapter to inherit behaviors from both the client and the adaptee.
    \item \textbf{Object Adapter}: This is the most common form of the Adapter Pattern in Python, where the adapter uses composition and contains an instance of the adaptee to forward requests.
\end{itemize}
  
\subsection{Bridge Pattern}
The \textbf{Bridge Pattern} \cite{harmes2008bridge} is a structural design pattern that aims to decouple an abstraction from its implementation so that the two can vary independently. This pattern is especially useful when both the abstraction and the implementation may evolve, or when an object needs to support multiple implementations dynamically.

The Bridge Pattern is often confused with the Adapter Pattern, but they serve different purposes. While the Adapter is used to make two incompatible interfaces work together, the Bridge Pattern is used to separate the abstraction from the implementation so they can evolve separately.

\subsubsection{Motivation}
Consider the case of building a graphic system where you have different types of shapes, such as circles and squares, and you want to be able to render them in different ways, such as using a vector or raster rendering engine. Without the Bridge Pattern, you might end up with a class hierarchy that multiplies classes unnecessarily: `VectorCircle', `RasterCircle', `VectorSquare', `RasterSquare', and so on. As the number of shapes and rendering techniques grows, this approach becomes unmanageable.

The Bridge Pattern allows you to separate the abstraction (in this case, the shapes) from their implementation (the rendering engines). This way, you can add new shapes or rendering techniques without changing existing code, promoting better scalability and flexibility.

\subsubsection{Structure}
The structure of the Bridge Pattern can be broken down as follows:

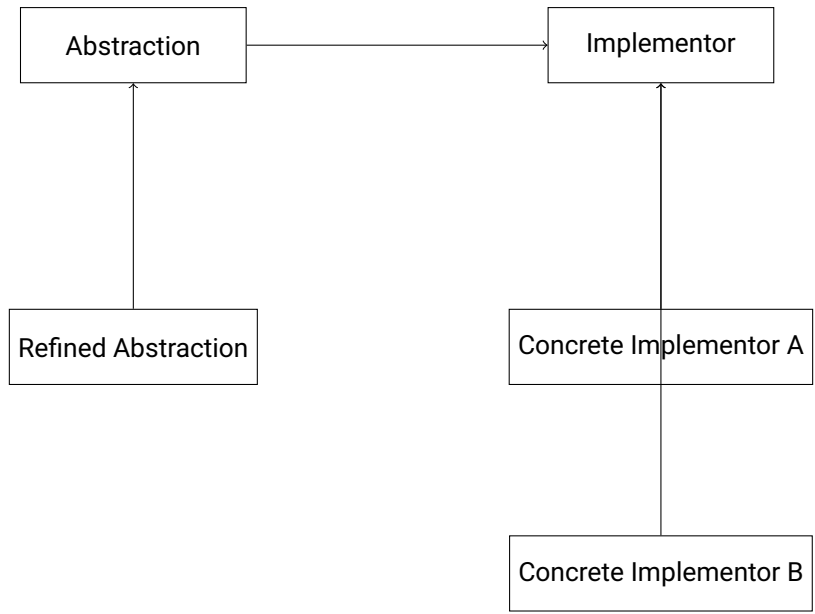
\begin{center}
\begin{tikzpicture}
  \node (abstraction) [draw, rectangle, minimum width=3cm, minimum height=1cm] {Abstraction};
  \node (implementor) [right=4cm of abstraction, draw, rectangle, minimum width=3cm, minimum height=1cm] {Implementor};
  \node (refinedAbstraction) [below=3cm of abstraction, draw, rectangle, minimum width=3cm, minimum height=1cm] {Refined Abstraction};
  \node (concreteImplementorA) [below=3cm of implementor, draw, rectangle, minimum width=3cm, minimum height=1cm] {Concrete Implementor A};
  \node (concreteImplementorB) [below=2cm of concreteImplementorA, draw, rectangle, minimum width=3cm, minimum height=1cm] {Concrete Implementor B};

  \draw[->] (abstraction) -- (implementor);
  \draw[->] (refinedAbstraction) -- (abstraction);
  \draw[->] (concreteImplementorA) -- (implementor);
  \draw[->] (concreteImplementorB) -- (implementor);
\end{tikzpicture}
\end{center}

\begin{itemize}
    \item \textbf{Abstraction}: Defines the interface for the abstraction and maintains a reference to an object of type \textit{Implementor}.
    \item \textbf{Refined Abstraction}: Extends the interface defined by \textit{Abstraction}.
    \item \textbf{Implementor}: Defines the interface for the implementation classes.
    \item \textbf{Concrete Implementors}: Implement the interface defined by the \textit{Implementor}.
\end{itemize}

In Python, this structure is commonly achieved using composition, where the abstraction holds a reference to the implementor.

\subsubsection{Participants}

\begin{itemize}
    \item \textbf{Abstraction}: The high-level abstraction that defines the interface for client interactions. It holds a reference to the implementor object.
    \item \textbf{Refined Abstraction}: A subclass of the \textit{Abstraction} that extends the base functionality while using the implementor for low-level operations.
    \item \textbf{Implementor}: Defines the interface that will be implemented by concrete classes to provide the specific low-level functionality.
    \item \textbf{Concrete Implementors}: Implement the interface defined by the \textit{Implementor}, providing different specific implementations.
\end{itemize}

\subsubsection{How It Works}
The Abstraction defines the interface that the client uses. The Abstraction, however, does not implement all the functionality directly. Instead, it delegates some work to the Implementor, which provides the specific implementation of certain methods. This allows the abstraction and implementation to vary independently.

For example, consider an application that renders different shapes. You can define an abstract class for shapes (the Abstraction), and this class can hold a reference to a rendering engine (the Implementor). Concrete shapes like Circle and Square (Refined Abstraction) can then use the rendering engine to perform the actual rendering, allowing you to change the rendering strategy without modifying the shapes.

\subsubsection{Advantages and Disadvantages}
\textbf{Advantages:}
\begin{itemize}
    \item \textbf{Separation of concerns}: The Bridge Pattern allows you to separate abstraction and implementation, making your code easier to understand and maintain.
    \item \textbf{Flexibility}: You can change either the abstraction or the implementation without affecting the other.
    \item \textbf{Scalability}: Adding new implementations or new abstractions is straightforward, avoiding the exponential growth of subclasses.
\end{itemize}

\textbf{Disadvantages:}
\begin{itemize}
    \item \textbf{Increased complexity}: Introducing additional abstraction layers can make the system more complex and harder to understand at first glance.
    \item \textbf{More classes}: As with most design patterns, the Bridge Pattern introduces more classes into the system, which could add to the overall codebase.
\end{itemize}

\subsubsection{Use Cases}
The Bridge Pattern is useful in the following situations:

\begin{itemize}
    \item When you need to switch implementations at runtime.
    \item When you want to decouple abstraction from implementation so that they can evolve independently.
    \item When you have a class explosion problem due to combining multiple independent factors (such as different shapes and rendering methods).
\end{itemize}

\textbf{Example:}

\begin{enumerate}
    \item A graphical user interface (GUI) framework where different platforms (Windows, Linux, macOS) use different rendering techniques but need a consistent API.
    \item A payment processing system where various payment methods (credit card, PayPal) need different back-end implementations but the client should interact with a unified interface.
\end{enumerate}

\subsubsection{Code Example}
Here is a Python example of the Bridge Pattern:

\begin{lstlisting}[style=python]
# Implementor
class Renderer:
    def render_circle(self, radius):
        pass

# Concrete Implementors
class VectorRenderer(Renderer):
    def render_circle(self, radius):
        print(f"Drawing a circle of radius {radius} using vector rendering.")

class RasterRenderer(Renderer):
    def render_circle(self, radius):
        print(f"Drawing pixels for a circle of radius {radius} using raster rendering.")

# Abstraction
class Shape:
    def __init__(self, renderer):
        self.renderer = renderer

    def draw(self):
        pass

    def resize(self, factor):
        pass

# Refined Abstraction
class Circle(Shape):
    def __init__(self, renderer, radius):
        super().__init__(renderer)
        self.radius = radius

    def draw(self):
        self.renderer.render_circle(self.radius)

    def resize(self, factor):
        self.radius *= factor

# Client Code
vector_renderer = VectorRenderer()
raster_renderer = RasterRenderer()

circle1 = Circle(vector_renderer, 5)
circle1.draw()  # Drawing a circle of radius 5 using vector rendering.
circle1.resize(2)
circle1.draw()  # Drawing a circle of radius 10 using vector rendering.

circle2 = Circle(raster_renderer, 3)
circle2.draw()  # Drawing pixels for a circle of radius 3 using raster rendering.
\end{lstlisting}

In this example:
\begin{itemize}
    \item \texttt{Renderer} is the implementor, providing an interface for rendering operations.
    \item \texttt{VectorRenderer} and \texttt{RasterRenderer} are concrete implementors that provide specific rendering strategies.
    \item \texttt{Shape} is the abstraction, which holds a reference to the \texttt{Renderer}.
    \item \texttt{Circle} is the refined abstraction that extends \texttt{Shape} and uses a renderer to perform its operations.
\end{itemize}

\subsubsection{Pattern Extensions}
The Bridge Pattern can be extended to support even more complex hierarchies. For example, you could introduce more complex shapes, like squares or polygons, while still supporting various rendering engines. The same principle of separating the abstraction (shape) from the implementation (rendering engine) would apply.

You can also combine the Bridge Pattern with other design patterns. For example, combining it with the Factory Pattern can help create appropriate implementations of the bridge dynamically at runtime based on certain conditions.

\subsection{Composite Pattern}
The Composite Pattern \cite{riehle1997composite} is a structural design pattern that allows you to compose objects into tree-like structures, where individual objects and groups of objects are treated uniformly. This pattern helps to work with complex hierarchical structures by simplifying their management through a unified interface.

\subsubsection{Motivation}
In many applications, you might encounter objects that are organized hierarchically. A common example is a file system, where directories can contain both files and other directories. From a user’s perspective, a directory containing files and another directory is still treated as a "directory," regardless of whether it contains individual files or more directories.

This leads to the need for treating both individual objects (like files) and composite objects (like directories containing other files or directories) in a uniform way. The Composite Pattern helps to handle these cases by providing a common interface for both single objects and composite objects.

\subsubsection{Structure}
The structure of the Composite Pattern consists of the following core components:

\begin{itemize}
    \item \textbf{Component:} Declares the common interface for both individual objects and composites. This interface defines operations that can be implemented by either single objects or groups of objects.
    \item \textbf{Leaf:} Represents the individual objects that do not have any children. Implements the Component interface.
    \item \textbf{Composite:} Represents the composite objects, which are containers that can hold leaf elements or other composite objects. Implements the Component interface.
    \item \textbf{Client:} Interacts with objects through the Component interface, which means the client can work with both individual objects and composite objects uniformly.
\end{itemize}

A typical structure for a Composite pattern looks like this:

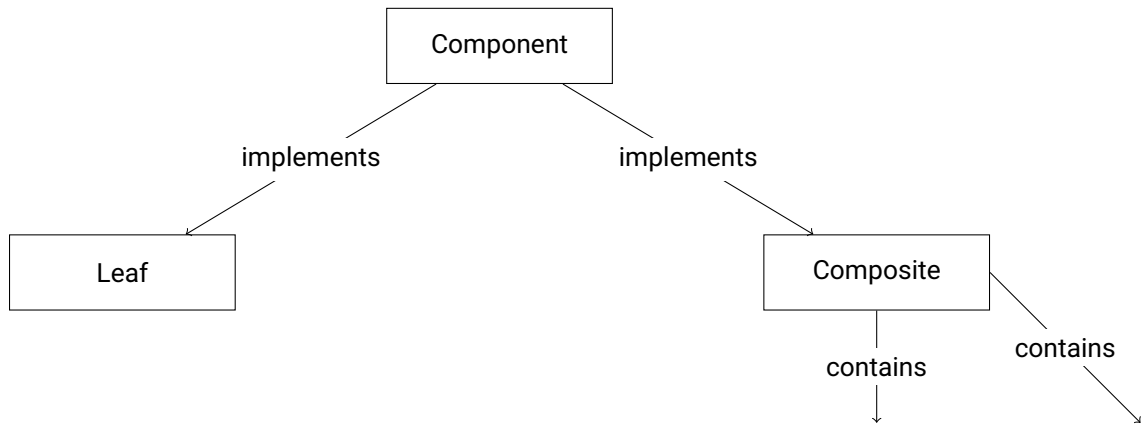
\begin{center}
\begin{tikzpicture}
  \node [draw, rectangle, minimum width=3cm, minimum height=1cm] (component) {Component};
  \node [below left=2cm and 2cm of component, draw, rectangle, minimum width=3cm, minimum height=1cm] (leaf) {Leaf};
  \node [below right=2cm and 2cm of component, draw, rectangle, minimum width=3cm, minimum height=1cm] (composite) {Composite};

  \draw [->] (component) -- (leaf) node[midway, fill=white] {implements};
  \draw [->] (component) -- (composite) node[midway, fill=white] {implements};
  \draw [->] (composite) -- +(0,-2) node[midway, fill=white] {contains};
  \draw [->] (composite.east) -- +(2,-2) node[midway, fill=white] {contains};
\end{tikzpicture}
\end{center}

\subsubsection{Participants}
The participants in the Composite Pattern are:

\begin{itemize}
    \item \textbf{Component:} Declares the common interface for objects in the composition. In Python, this is typically represented by an abstract base class or an interface.
    \item \textbf{Leaf:} Implements the Component interface and represents the individual, non-composite objects.
    \item \textbf{Composite:} Implements the Component interface and stores child components. These children can be either Leafs or other Composites.
    \item \textbf{Client:} Works with components through the Component interface.
\end{itemize}

\subsubsection{How It Works}
In the Composite Pattern, the key idea is that the Client interacts with the Component interface, not worrying about whether it's working with individual objects or groups of objects. The Composite object holds references to other Components, and it can invoke operations on them. This allows for recursively treating individual objects and compositions of objects in the same way.

When a Client invokes an operation on a Composite object, the Composite forwards the operation to its child components, which may themselves be individual objects or other Composites. This recursive behavior allows clients to handle a complex tree structure of objects as if it were a single object.

\subsubsection{Advantages and Disadvantages}
\textbf{Advantages:}

\begin{itemize}
    \item \textbf{Uniformity:} You can treat individual objects and compositions of objects uniformly, simplifying the client’s code.
    \item \textbf{Extensibility:} You can easily add new types of components (either Leafs or Composites) without modifying the existing code.
    \item \textbf{Hierarchical structure:} Naturally supports working with tree structures, such as file systems or organizational charts.
\end{itemize}

\textbf{Disadvantages:}

\begin{itemize}
    \item \textbf{Complexity:} The pattern can introduce additional complexity, especially when the structure becomes deeply nested.
    \item \textbf{Overhead:} Managing components and maintaining a tree-like structure can incur overhead.
\end{itemize}

\subsubsection{Use Cases}
The Composite Pattern is commonly used in scenarios where objects form a tree structure. Examples include:

\begin{itemize}
    \item \textbf{File systems:} Directories can contain both files and other directories, forming a recursive structure.
    \item \textbf{Graphical user interfaces:} Windows, panels, and buttons can be treated as part of a composite structure where each container can hold other UI elements.
    \item \textbf{Organization charts:} In organizational charts, managers can have subordinates, some of whom might be managers themselves.
\end{itemize}

\subsubsection{Code Example}

Here is a Python implementation of the Composite Pattern:

\begin{lstlisting}[style=python]
from abc import ABC, abstractmethod

# Component: Declares the interface for objects in the composition.
class Component(ABC):
    @abstractmethod
    def operation(self):
        pass

# Leaf: Represents leaf objects in the composition. It has no children.
class Leaf(Component):
    def __init__(self, name):
        self.name = name

    def operation(self):
        print(f"Leaf {self.name} operation executed.")

# Composite: Represents a composite object (a container for children).
class Composite(Component):
    def __init__(self, name):
        self.name = name
        self.children = []

    def add(self, component):
        self.children.append(component)

    def remove(self, component):
        self.children.remove(component)

    def operation(self):
        print(f"Composite {self.name} operation executed.")
        for child in self.children:
            child.operation()

# Client code:
if __name__ == "__main__":
    # Create leaf objects
    leaf1 = Leaf("A")
    leaf2 = Leaf("B")

    # Create composite objects
    composite1 = Composite("Composite 1")
    composite2 = Composite("Composite 2")

    # Build the structure
    composite1.add(leaf1)
    composite1.add(composite2)
    composite2.add(leaf2)

    # Execute operations
    composite1.operation()
\end{lstlisting}

In this example, we have a `Component` interface, a `Leaf` class representing individual elements, and a `Composite` class that can hold both Leafs and other Composites. The client interacts with all these objects through the `Component` interface.

\subsubsection{Pattern Extensions}
The Composite Pattern can be extended in several ways to handle more specific use cases:

\begin{itemize}
    \item \textbf{Transparency vs Safety:} In some implementations, the `Composite` class exposes methods like `add` and `remove`, while the `Leaf` class does not. This is called a "safe" implementation. In a "transparent" implementation, both `Leaf` and `Composite` implement these methods, but the `Leaf` throws an exception if these methods are called.
    \item \textbf{Shared Components:} Components can be shared across multiple composites to reduce memory usage. This is similar to the Flyweight pattern.
    \item \textbf{Decorator Pattern:} You can combine the Composite pattern with the Decorator pattern to add responsibilities to components dynamically while maintaining a hierarchical structure.
\end{itemize}

\subsection{Decorator Pattern}
The Decorator Pattern \cite{gardner2007decorator} is a structural design pattern that allows you to dynamically attach additional responsibilities to an object. It is a flexible alternative to subclassing for extending functionality. By using decorators, you can avoid monolithic classes and instead compose behavior in a more modular way.

\subsubsection{Motivation}
The primary motivation for using the decorator pattern is to add functionalities to objects without modifying their class. In real-world situations, we often need to enhance the behavior of objects in a dynamic, runtime manner without creating an extensive inheritance tree. For example, you might want to add logging, caching, or validation behavior to a function or an object without modifying its original code.

Consider an example where we have a simple `Coffee` class that provides a basic coffee beverage. Over time, we want to enhance the coffee by adding milk, sugar, or other ingredients. Instead of modifying the original `Coffee` class or creating subclasses like `MilkCoffee`, `SugarMilkCoffee`, and so on, we can dynamically add these enhancements using decorators.

\subsubsection{Structure}
The Decorator Pattern typically involves the following structure:

\begin{itemize}
  \item \textbf{Component}: Defines an interface for objects that can have responsibilities added to them.
  \item \textbf{ConcreteComponent}: A class that implements the Component interface. This is the original object to which decorators are applied.
  \item \textbf{Decorator}: A base class that implements the Component interface and holds a reference to a Component object. It delegates all operations to the component object and allows subclasses to add extra functionality.
  \item \textbf{ConcreteDecorator}: Subclasses of the Decorator class that add responsibilities or behavior to the component.
\end{itemize}

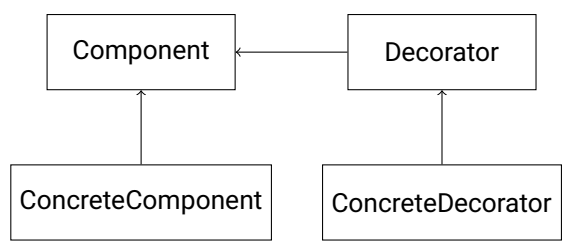
\begin{center}
\begin{tikzpicture}
  [class/.style={rectangle, draw=black, fill=white, text centered, minimum height=1cm, minimum width=2.5cm}]
  
  \node (Component) at (0,0) [class] {Component};
  \node (ConcreteComponent) at (0,-2) [class] {ConcreteComponent};
  \node (Decorator) at (4,0) [class] {Decorator};
  \node (ConcreteDecorator) at (4,-2) [class] {ConcreteDecorator};

  \draw[->] (ConcreteComponent) -- (Component);
  \draw[->] (Decorator) -- (Component);
  \draw[->] (ConcreteDecorator) -- (Decorator);
\end{tikzpicture}
\end{center}

\subsubsection{Participants}
\begin{itemize}
  \item \textbf{Component}: Defines the interface for objects that can have behavior dynamically added to them.
  \item \textbf{ConcreteComponent}: Implements the Component interface and provides the base behavior.
  \item \textbf{Decorator}: Maintains a reference to a Component object and defines an interface that conforms to the Component interface.
  \item \textbf{ConcreteDecorator}: Adds extra behavior to the component.
\end{itemize}

\subsubsection{How It Works}
The Decorator Pattern works by wrapping an object in a decorator class. Each decorator class adds its specific behavior and delegates the original task to the wrapped object. You can use multiple decorators to compose various functionalities dynamically.

When a method is called on the decorated object, the call is forwarded from the outermost decorator to the original component, with each decorator potentially adding its functionality along the way.

For example, suppose we have a `Coffee` class, and we want to add "milk" and "sugar" as decorators:

\begin{lstlisting}[style=python]
class Coffee:
    def cost(self):
        return 5

class MilkDecorator:
    def __init__(self, coffee):
        self._coffee = coffee
    
    def cost(self):
        return self._coffee.cost() + 1.5

class SugarDecorator:
    def __init__(self, coffee):
        self._coffee = coffee
    
    def cost(self):
        return self._coffee.cost() + 0.5

# Usage
coffee = Coffee()
milk_coffee = MilkDecorator(coffee)
milk_sugar_coffee = SugarDecorator(milk_coffee)

print(milk_sugar_coffee.cost())  # Outputs: 7.0
\end{lstlisting}

In this example:
1. We start with a `Coffee` object.
2. Then we wrap it with a `MilkDecorator` to add the cost of milk.
3. Finally, we wrap it with a `SugarDecorator` to add the cost of sugar.

Each decorator adds its functionality to the base object without modifying its original code.

\subsubsection{Advantages and Disadvantages}

\textbf{Advantages:}
\begin{itemize}
  \item \textbf{Flexibility}: You can dynamically add or remove responsibilities to an object without modifying its class.
  \item \textbf{No Large Inheritance Trees}: The Decorator Pattern allows you to avoid creating a large number of subclasses to accommodate different combinations of behavior.
  \item \textbf{Composability}: You can combine multiple decorators to create complex behavior without modifying existing code.
\end{itemize}

\textbf{Disadvantages:}
\begin{itemize}
  \item \textbf{Complexity}: Decorators can introduce complexity in the system because you have to track which objects are decorated and in what order.
  \item \textbf{Overhead}: Each decorator introduces an additional layer of abstraction, which can result in performance overhead and make debugging more difficult.
  \item \textbf{Harder to Understand}: Since decorators are applied dynamically, the flow of the program may become harder to follow, especially when many decorators are applied.
\end{itemize}

\subsubsection{Use Cases}
The Decorator Pattern is often used in scenarios where:
\begin{itemize}
  \item You need to add responsibilities to an object dynamically at runtime.
  \item You want to avoid subclassing to extend behavior.
  \item Different combinations of responsibilities need to be applied to an object.
  \item Examples: 
    \begin{itemize}
      \item Logging: Adding logging to method calls dynamically.
      \item I/O Streams: Wrapping input/output streams to add functionality like buffering, compression, or encryption.
      \item UI Components: Adding borders, scrollbars, and shadows to UI elements without modifying their base classes.
    \end{itemize}
\end{itemize}

\subsubsection{Code Example}
Here is a complete example illustrating the Decorator Pattern in Python:

\begin{lstlisting}[style=python]
# Basic interface for a coffee
class Coffee:
    def cost(self):
        raise NotImplementedError

# Basic Coffee class
class SimpleCoffee(Coffee):
    def cost(self):
        return 5

# Decorator base class
class CoffeeDecorator(Coffee):
    def __init__(self, coffee):
        self._coffee = coffee
    
    def cost(self):
        return self._coffee.cost()

# Concrete Decorator: Milk
class MilkDecorator(CoffeeDecorator):
    def cost(self):
        return self._coffee.cost() + 1.5

# Concrete Decorator: Sugar
class SugarDecorator(CoffeeDecorator):
    def cost(self):
        return self._coffee.cost() + 0.5

# Usage
coffee = SimpleCoffee()
milk_coffee = MilkDecorator(coffee)
milk_sugar_coffee = SugarDecorator(milk_coffee)

print(f"Cost of milk sugar coffee: {milk_sugar_coffee.cost()}")
\end{lstlisting}

In this example, we defined a base `Coffee` interface, a `SimpleCoffee` class, and two decorators: `MilkDecorator` and `SugarDecorator`. The decorators add functionality (increased cost) without altering the `SimpleCoffee` class itself.

\subsubsection{Pattern Extensions}
Some possible extensions of the Decorator Pattern:
\begin{itemize}
  \item \textbf{Decorator Chaining}: You can chain multiple decorators to apply multiple enhancements in sequence.
  \item \textbf{Mutability}: Decorators can be designed to allow adding or removing behavior dynamically, making them more flexible.
  \item \textbf{Custom Behavior}: Decorators can modify not only the method output but also the input, adding even more flexibility.
  \item \textbf{Decorator Factories}: You can create factories that return decorators dynamically, based on some conditions, for more control over which decorators get applied.
\end{itemize}

\subsection{Facade Pattern}
The Facade Pattern \cite{jiang2011design} is a structural design pattern that provides a simplified interface to a complex system of classes, libraries, or frameworks. It hides the complexities of the system behind a facade, which is an easy-to-use interface, allowing the client to interact with the system without needing to understand its inner workings.

\subsubsection{Motivation}
In software development, complex systems often consist of multiple subsystems with many interdependent classes. Interacting directly with these systems can become cumbersome, especially for beginners, or when the complexity increases over time. The Facade Pattern solves this by offering a simplified, higher-level interface to interact with the system, hiding the underlying complexity. This makes the system easier to use and reduces the learning curve for users who only need access to specific functionality.

For example, imagine a system that handles multiple services such as file processing, email sending, and logging. Without a facade, the client would need to manage all these services directly, which can be confusing and error-prone. By introducing a facade, the client only interacts with a single class, which simplifies the usage of the system.

\subsubsection{Structure}
The structure of the Facade Pattern is relatively simple. It consists of three main components:

\begin{itemize}
    \item \textbf{Facade:} The Facade class provides a simple interface to the client by interacting with one or more subsystems. It abstracts the complexity of the system by wrapping the interactions with the subsystems.
    \item \textbf{Subsystems:} These are the complex parts of the system. They contain detailed implementation and provide real functionality. The subsystems don’t have any knowledge of the facade; they work independently.
    \item \textbf{Client:} The client interacts with the facade rather than directly with the subsystems.
\end{itemize}

Here’s a diagram that illustrates the structure of the Facade Pattern:

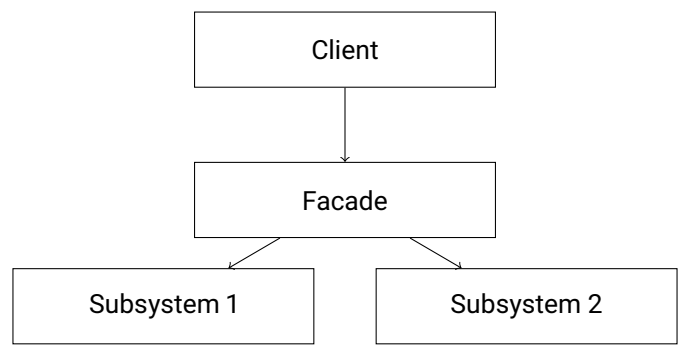
\begin{center}
\begin{tikzpicture}[node distance=2cm, every node/.style={draw, rectangle, minimum width=4cm, minimum height=1cm, text centered}]
    \node (client) {Client};
    \node (facade) [below of=client] {Facade};
    \node (subsystem1) [below left of=facade, xshift=-1cm] {Subsystem 1};
    \node (subsystem2) [below right of=facade, xshift=1cm] {Subsystem 2};

    \draw[->] (client) -- (facade);
    \draw[->] (facade) -- (subsystem1);
    \draw[->] (facade) -- (subsystem2);
\end{tikzpicture}
\end{center}

\subsubsection{Participants}
\begin{itemize}
    \item \textbf{Facade:} The class that provides the simplified interface and delegates requests to the appropriate subsystem classes.
    \item \textbf{Subsystem classes:} These classes handle the detailed work requested by the facade. They do not know about the facade and operate independently.
    \item \textbf{Client:} The client interacts only with the facade to perform high-level operations without dealing with the underlying complexity.
\end{itemize}

\subsubsection{How It Works}
The client interacts with the Facade class, which in turn interacts with one or more subsystems. The facade shields the client from the complexities of the subsystems by providing an easy-to-use interface.

Here’s a step-by-step explanation of how the Facade Pattern works:

\begin{enumerate}
    \item The client sends a request to the facade.
    \item The facade receives the request and processes it. Depending on the request, the facade calls methods on one or more subsystems.
    \item The subsystems execute the actual logic and return the result to the facade.
    \item The facade returns the simplified result to the client.
\end{enumerate}

The client is completely unaware of how the subsystems work. It only knows that it can interact with the facade to get its task done.

\subsubsection{Advantages and Disadvantages}

\textbf{Advantages:}
\begin{itemize}
    \item \textbf{Simplified interface:} The facade provides a simple interface to interact with a complex system, making it easier to use.
    \item \textbf{Loose coupling:} The facade reduces dependencies between the client and the subsystems. The client only depends on the facade, not the individual subsystems.
    \item \textbf{Improved maintainability:} Since the client does not interact directly with the subsystems, changes to the subsystems do not affect the client.
    \item \textbf{Code readability:} The facade helps organize and structure complex code by grouping related operations behind a single class.
\end{itemize}

\textbf{Disadvantages:}
\begin{itemize}
    \item \textbf{Potential over-simplification:} In some cases, the facade may overly simplify the system, limiting the client’s ability to use advanced features of the subsystems.
    \item \textbf{Additional layer of abstraction:} The facade adds another layer of abstraction, which may not always be necessary, especially in simple systems.
\end{itemize}

\subsubsection{Use Cases}
The Facade Pattern is useful in the following situations:
\begin{itemize}
    \item When working with complex libraries or frameworks and you want to expose only specific functionality to the client.
    \item When you want to reduce dependencies between different parts of your code.
    \item When you are refactoring legacy systems and need to provide a simplified interface to existing code.
    \item When different parts of the system need to be decoupled to improve maintainability and flexibility.
\end{itemize}

\subsubsection{Code Example}
Here’s a Python example that demonstrates the Facade Pattern in action.

\begin{lstlisting}[style=python]
# Subsystem classes
class FileHandler:
    def read_file(self, file_name):
        return f"Reading data from {file_name}"
    
    def write_file(self, file_name, data):
        return f"Writing {data} to {file_name}"

class EmailSender:
    def send_email(self, recipient, content):
        return f"Sending email to {recipient}: {content}"

class Logger:
    def log(self, message):
        return f"Log entry: {message}"

# Facade class
class SystemFacade:
    def __init__(self):
        self.file_handler = FileHandler()
        self.email_sender = EmailSender()
        self.logger = Logger()
    
    def process_and_notify(self, file_name, data, recipient):
        # Reading and writing to a file
        read_result = self.file_handler.read_file(file_name)
        self.logger.log(read_result)
        
        write_result = self.file_handler.write_file(file_name, data)
        self.logger.log(write_result)
        
        # Sending an email notification
        email_result = self.email_sender.send_email(recipient, "File processed")
        self.logger.log(email_result)
        
        return f"File {file_name} processed and notification sent to {recipient}"

# Client code
if __name__ == "__main__":
    facade = SystemFacade()
    result = facade.process_and_notify("example.txt", "Some important data", "user@example.com")
    print(result)
\end{lstlisting}

In this example:
- \texttt{FileHandler}, \texttt{EmailSender}, and \texttt{Logger} are the subsystems.
- \texttt{SystemFacade} is the facade that simplifies the interaction with these subsystems.
- The client only interacts with \texttt{SystemFacade}, without needing to know how the file handling, email sending, and logging work behind the scenes.

\subsubsection{Pattern Extensions}
The Facade Pattern can be extended or combined with other patterns to increase flexibility and functionality.

\textbf{Abstract Facade:} You can create an abstract class or interface for the facade, allowing for different implementations of the facade to handle different subsystems or behaviors.

\textbf{Facade with Dependency Injection:} Instead of hardcoding the subsystems within the facade, you can pass the subsystems as dependencies through the constructor. This allows for greater flexibility and testability.

\begin{lstlisting}[style=python]
class SystemFacade:
    def __init__(self, file_handler, email_sender, logger):
        self.file_handler = file_handler
        self.email_sender = email_sender
        self.logger = logger
\end{lstlisting}

This version of the facade allows the client to inject different subsystems, making the facade more modular and adaptable to changes in the underlying system.

By combining the Facade Pattern with other patterns, such as Dependency Injection, you can create a flexible and scalable architecture suitable for a wide variety of use cases.

\subsection{Flyweight Pattern}
The Flyweight Pattern \cite{rana2019impact} is a structural design pattern that helps reduce memory usage by sharing common parts of objects between multiple instances. Instead of creating new instances of a class each time, the Flyweight Pattern reuses existing objects when appropriate. This is particularly useful when many similar objects are created in an application, such as graphical elements in a GUI or characters in a text editor. The Flyweight Pattern can drastically reduce memory overhead, especially when dealing with a large number of objects that share similar data.

\subsubsection{Motivation}
In certain applications, you may need to create a large number of similar objects, and each object may contain a mix of unique and shared information. For example, if you are designing a text editor that displays each letter as an object, it would be inefficient to create a separate object for each letter in a document, especially when many of them are the same letter, like "e". By using the Flyweight Pattern, you can create one shared object for each unique letter and reuse it multiple times, only storing the specific properties that differ between instances, such as position or color. This can lead to significant memory savings.

\begin{lstlisting}[style=python]
# Naive implementation without Flyweight
class Letter:
    def __init__(self, char, font, size, color):
        self.char = char
        self.font = font
        self.size = size
        self.color = color

# This will create a separate object for each letter, even if they are the same character.
letters = [Letter('a', 'Arial', 12, 'black') for _ in range(1000)]
\end{lstlisting}

The Flyweight Pattern solves this problem by creating a shared object (or flyweight) for each unique character, and reusing it across the program while maintaining the flexibility to handle different properties.

\subsubsection{Structure}
The Flyweight Pattern involves the following elements:

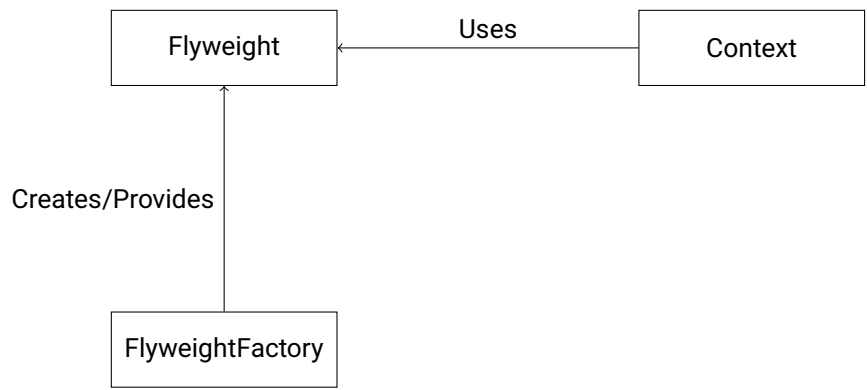
\begin{center}
\begin{tikzpicture}
  [node distance=3cm, auto]
  \node (Flyweight) [rectangle, draw, text centered, minimum width=3cm, minimum height=1cm] {Flyweight};
  \node (Context) [rectangle, draw, right=4cm of Flyweight, text centered, minimum width=3cm, minimum height=1cm] {Context};
  \node (FlyweightFactory) [rectangle, draw, below=3cm of Flyweight, text centered, minimum width=3cm, minimum height=1cm] {FlyweightFactory};

  \draw[->] (Context) -- (Flyweight) node[midway, above] {Uses};
  \draw[->] (FlyweightFactory) -- (Flyweight) node[midway, left] {Creates/Provides};
\end{tikzpicture}
\end{center}

\begin{itemize}
    \item \textbf{Flyweight}: The shared object that contains intrinsic (invariant) data.
    \item \textbf{Context}: The object that contains extrinsic (variant) data and works with the flyweight object.
    \item \textbf{FlyweightFactory}: A factory that manages flyweight objects and provides shared instances as needed.
\end{itemize}

\subsubsection{Participants}
\begin{itemize}
    \item \textbf{Flyweight}: The class that defines common shared data among multiple objects. It is designed to be reused by multiple clients.
    \item \textbf{ConcreteFlyweight}: The concrete implementation of the Flyweight that contains the intrinsic state.
    \item \textbf{Client (Context)}: The object that uses flyweight instances. It may contain additional information (extrinsic state) that is specific to the client.
    \item \textbf{FlyweightFactory}: This class is responsible for managing and returning flyweight instances.
\end{itemize}

\subsubsection{How It Works}
The Flyweight Pattern separates intrinsic (shared) data from extrinsic (unique) data. The intrinsic state is stored inside the flyweight object and is shared among multiple clients, while the extrinsic state is stored outside the flyweight and provided by clients during usage.

\begin{enumerate}
    \item \textbf{Intrinsic State}: This is the part of the object's state that can be shared and remains constant across all contexts where the flyweight is used. For example, the character in a text editor (e.g., 'a').
    \item \textbf{Extrinsic State}: This is the part of the object's state that varies across contexts and must be provided by the client. For example, the position of the character in the document.
\end{enumerate}

\begin{lstlisting}[style=python]
# Flyweight class
class LetterFlyweight:
    def __init__(self, char):
        self.char = char  # Intrinsic state (shared data)

    def render(self, font, size, color, position):
        print(f"Rendering letter '{self.char}' with font {font}, size {size}, color {color} at {position}")

# Flyweight Factory
class LetterFactory:
    _flyweights = {}

    @classmethod
    def get_flyweight(cls, char):
        if char not in cls._flyweights:
            cls._flyweights[char] = LetterFlyweight(char)
        return cls._flyweights[char]

# Usage
factory = LetterFactory()

letters = [
    ('a', 'Arial', 12, 'black', (10, 20)),
    ('b', 'Arial', 12, 'black', (20, 20)),
    ('a', 'Arial', 12, 'red', (30, 20)),
]

for char, font, size, color, position in letters:
    letter = factory.get_flyweight(char)
    letter.render(font, size, color, position)
\end{lstlisting}

In this example:

\begin{itemize}
    \item The character ('a' or 'b') is the intrinsic state shared between multiple contexts.
    \item The font, size, color, and position are the extrinsic state that changes depending on where and how the letter is used.
\end{itemize}

\subsubsection{Advantages and Disadvantages}

\paragraph{Advantages:}
\begin{itemize}
    \item \textbf{Memory Efficiency}: Reduces memory usage by sharing common data across multiple objects.
    \item \textbf{Performance Improvement}: Can improve performance when dealing with a large number of similar objects.
\end{itemize}

\paragraph{Disadvantages:}
\begin{itemize}
    \item \textbf{Complexity}: The pattern introduces complexity due to the separation of intrinsic and extrinsic data.
    \item \textbf{Context Management}: Extrinsic data needs to be carefully managed by the client, which can increase the burden on developers.
\end{itemize}

\subsubsection{Use Cases}
The Flyweight Pattern is suitable in scenarios where:
\begin{itemize}
    \item You need to handle a large number of similar objects.
    \item Memory optimization is critical.
    \item Shared data can be separated from unique data.
\end{itemize}

Typical examples include:
\begin{itemize}
    \item \textbf{Text Rendering}: Representing individual characters in a document, where each character can share font and size but differs in position or color.
    \item \textbf{GUI Elements}: Many graphical elements (like icons or buttons) share visual properties but differ in their location or state.
    \item \textbf{Game Development}: For example, in a game with many objects (like trees or cars), where the visual representation is shared but the position and state vary.
\end{itemize}

\subsubsection{Code Example}
Here’s an extended example using the Flyweight Pattern for managing graphical elements in a simple game:

\begin{lstlisting}[style=python]
# Flyweight class for trees
class TreeFlyweight:
    def __init__(self, type, color, texture):
        self.type = type  # Intrinsic state
        self.color = color  # Intrinsic state
        self.texture = texture  # Intrinsic state

    def render(self, x, y):
        print(f"Rendering a {self.type} tree with color {self.color} at position ({x}, {y})")

# Flyweight factory for trees
class TreeFactory:
    _flyweights = {}

    @classmethod
    def get_flyweight(cls, type, color, texture):
        key = (type, color, texture)
        if key not in cls._flyweights:
            cls._flyweights[key] = TreeFlyweight(type, color, texture)
        return cls._flyweights[key]

# Usage
factory = TreeFactory()

trees = [
    ('Oak', 'Green', 'Rough', (10, 20)),
    ('Pine', 'Dark Green', 'Smooth', (20, 30)),
    ('Oak', 'Green', 'Rough', (30, 40)),
]

for type, color, texture, position in trees:
    tree = factory.get_flyweight(type, color, texture)
    tree.render(*position)
\end{lstlisting}

\subsubsection{Pattern Extensions}
The Flyweight Pattern can be combined with other design patterns for more advanced functionality. For example:

\begin{itemize}
    \item \textbf{Composite + Flyweight}: You can use the Flyweight Pattern along with the Composite Pattern to create hierarchical structures (like documents) where both shared and unique elements can be efficiently managed.
    \item \textbf{Proxy + Flyweight}: The Proxy Pattern can be used to control access to flyweight objects, allowing for additional features like lazy initialization or access control.
\end{itemize}

By combining the Flyweight Pattern with others, you can create flexible, memory-efficient solutions for complex problems.

\subsection{Proxy Pattern}

The Proxy Pattern \cite{wuest2006proxy} is a structural design pattern. In this pattern, a class represents the functionality of another class. The proxy acts as an intermediary between the client and the object it wants to interact with, controlling access to that object.

\subsubsection{Motivation}

Imagine a scenario where you have an object that performs resource-intensive operations, such as connecting to a remote server or reading a large file from a disk. Every time the object is accessed, it incurs a significant cost. The Proxy Pattern can help by allowing the proxy object to control access to this expensive resource, ensuring that the actual object is only created or accessed when necessary.

For example, if an object requires expensive initialization, you might want to delay the object's creation until it is truly needed. The proxy will act as a stand-in and control when and how the real object is accessed. This can help improve performance or manage resources more efficiently.

\subsubsection{Structure}

The structure of the Proxy Pattern can be represented as follows:

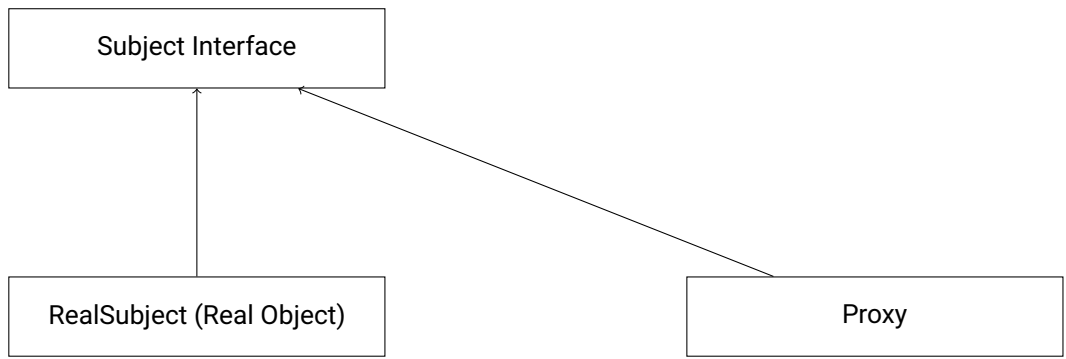
\begin{center}
\begin{tikzpicture}
  [class/.style={rectangle, draw=black, fill=white, text centered, minimum height=3em, minimum width=5cm}]
  
  \node[class] (Subject) {Subject Interface};
  \node[class, below=2.5cm of Subject] (RealSubject) {RealSubject (Real Object)};
  \node[class, right=4cm of RealSubject] (Proxy) {Proxy};

  \draw[->] (Proxy) -- (Subject);
  \draw[->] (RealSubject) -- (Subject);
  
\end{tikzpicture}
\end{center}

\begin{itemize}
    \item \textbf{Subject Interface}: This is a common interface that both the Real Subject and the Proxy implement. It declares the operations that clients can perform.
    \item \textbf{Real Subject}: This is the object that the Proxy controls access to. It performs the actual operations.
    \item \textbf{Proxy}: The Proxy contains a reference to the Real Subject and controls access to it. It may delay the creation or loading of the Real Subject, restrict its access, or perform other actions before forwarding the request to the Real Subject.
\end{itemize}

\subsubsection{Participants}

\begin{itemize}
    \item \textbf{Proxy}: The object that controls access to the Real Subject. It implements the same interface as the Real Subject and may forward requests to it. The proxy can also decide to handle requests itself or delay operations until the real object is needed.
    \item \textbf{Real Subject}: The actual object that is being proxied. It performs the real work and provides the core functionality requested by the client.
    \item \textbf{Client}: The object that interacts with the Proxy and indirectly with the Real Subject. The client usually doesn't know whether it is working with a Proxy or a Real Subject.
\end{itemize}

\subsubsection{How It Works}

In the Proxy Pattern, the proxy controls the access to a Real Subject. Let's break this down step by step:

1. The \textbf{Client} makes a request to the \textbf{Proxy}.
2. The \textbf{Proxy} checks or manages access to the \textbf{Real Subject}. This can include checking permissions, managing loading times, or adding extra functionality (such as logging).
3. Once the access control is done, the \textbf{Proxy} forwards the request to the \textbf{Real Subject}.
4. The \textbf{Real Subject} then performs the requested action and returns the result.
5. The \textbf{Proxy} may further modify the result before passing it back to the \textbf{Client}.

\subsubsection{Advantages and Disadvantages}

\textbf{Advantages}:
\begin{itemize}
    \item \textbf{Lazy Initialization}: The Proxy can delay the creation or initialization of expensive objects until they are needed.
    \item \textbf{Access Control}: The Proxy can control access to the Real Subject by verifying permissions or performing other checks.
    \item \textbf{Additional Functionality}: The Proxy can add extra functionality (such as logging or caching) to the operations without modifying the Real Subject.
    \item \textbf{Remote Proxy}: In some cases, the Proxy can be used to represent an object that resides in a different address space, such as a remote server.
\end{itemize}

\textbf{Disadvantages}:
\begin{itemize}
    \item \textbf{Increased Complexity}: The use of a proxy adds an extra layer of abstraction, which can make the code more complex.
    \item \textbf{Potential Performance Cost}: While proxies are often used to optimize performance, there is a potential performance cost due to the extra level of indirection.
\end{itemize}

\subsubsection{Use Cases}

The Proxy Pattern is particularly useful in scenarios such as:
\begin{itemize}
    \item \textbf{Lazy Loading}: If the Real Subject is resource-intensive to create, the Proxy can delay its creation until necessary.
    \item \textbf{Access Control}: Proxies can be used to control access to objects based on security rules or other conditions.
    \item \textbf{Logging and Auditing}: A Proxy can add logging or monitoring functionality without modifying the Real Subject.
    \item \textbf{Remote Proxy}: When the Real Subject is located on a different machine or server, the Proxy can provide a local representation that communicates with the remote object.
\end{itemize}

\subsubsection{Code Example}

Here is a simple example in Python that demonstrates the Proxy Pattern. We will create a `RealSubject` that performs a time-consuming task, and a `Proxy` that controls access to it.

\begin{lstlisting}[style=python]
# Subject Interface
class Subject:
    def request(self):
        pass

# RealSubject: Performs the actual operation
class RealSubject(Subject):
    def request(self):
        print("RealSubject: Handling the request...")

# Proxy: Controls access to the RealSubject
class Proxy(Subject):
    def __init__(self):
        self.real_subject = None

    def request(self):
        if self.real_subject is None:
            print("Proxy: Creating RealSubject.")
            self.real_subject = RealSubject()
        print("Proxy: Forwarding request to RealSubject.")
        self.real_subject.request()

# Client code
def client_code(subject: Subject):
    subject.request()

# Use the proxy to access the real object
print("Client: Working with Proxy:")
proxy = Proxy()
client_code(proxy)
\end{lstlisting}

In this example:
- The `RealSubject` performs the actual work of handling the request.
- The `Proxy` controls access to the `RealSubject`. It only creates the `RealSubject` when necessary and forwards the client's request to it.
- The `Client` interacts with the `Proxy`, which in turn forwards the request to the `RealSubject`.

\subsubsection{Pattern Extensions}

The Proxy Pattern has a few extensions and variations that are useful in specific contexts:
\begin{itemize}
    \item \textbf{Virtual Proxy}: This type of proxy is used to manage expensive resource objects. It delays the creation and initialization of the Real Subject until it is needed.
    \item \textbf{Remote Proxy}: This proxy is used to represent objects that exist in a different address space, such as on a remote server. The proxy provides a local interface to the remote object and manages the communication.
    \item \textbf{Protection Proxy}: This proxy controls access to the Real Subject based on access permissions or security rules.
    \item \textbf{Smart Proxy}: This proxy adds additional functionality, such as reference counting, caching, or logging, before forwarding the request to the Real Subject.
\end{itemize}

\section{Behavioral Patterns}
\subsection{Introduction}

Behavioral design patterns \cite{reimanis2019behavioral} are concerned with the interaction and responsibility between objects. Unlike structural patterns, which focus on how objects are organized and composed, behavioral patterns emphasize communication and the delegation of responsibility between different objects. These patterns are particularly useful in defining the flow of a program, ensuring that different parts of a system can interact efficiently while maintaining loose coupling between them.

In Python, behavioral patterns help developers manage complex interactions in a flexible and scalable way. By applying these patterns, we can create systems that are easier to maintain and extend as they grow in complexity. These patterns promote the distribution of responsibilities between objects, making the code more modular and adaptable.

\subsubsection{Key Concepts in Behavioral Patterns}
When learning about behavioral design patterns, there are some key principles and ideas that you should keep in mind:
\begin{itemize}
    \item \textbf{Encapsulating Behavior:} Behavioral patterns often focus on encapsulating how a certain action or operation is performed. This allows you to change how behavior is executed without altering the objects that use this behavior.
    \item \textbf{Delegating Responsibilities:} These patterns allow objects to delegate certain tasks or responsibilities to other objects, helping to distribute work across different parts of the system and preventing any single object from becoming too complex.
    \item \textbf{Communication Between Objects:} Many behavioral patterns are centered around how objects communicate with one another, ensuring that they can collaborate to perform tasks while keeping dependencies between them as low as possible.
\end{itemize}

\subsubsection{Common Behavioral Patterns}
Several behavioral design patterns are frequently used in Python to manage object interactions and responsibilities. Some of the most popular behavioral patterns include:
\begin{itemize}
    \item \textbf{Observer Pattern:} This pattern allows objects to subscribe to events and be notified when those events occur, promoting loose coupling between the objects that generate events and those that react to them.
    \item \textbf{Strategy Pattern:} The strategy pattern allows you to define a family of algorithms and make them interchangeable, enabling the algorithm used by an object to be selected dynamically at runtime.
    \item \textbf{Command Pattern:} This pattern encapsulates commands as objects, allowing you to parameterize objects with operations, queue operations, and support undoable actions.
    \item \textbf{Mediator Pattern:} This pattern centralizes communication between objects by introducing a mediator object that handles communication between other objects, preventing them from referring to each other directly.
    \item \textbf{Chain of Responsibility Pattern:} This pattern allows a request to pass through a chain of objects, where each object has a chance to handle the request, thus decoupling the sender from the receiver.
    \item \textbf{State Pattern:} This pattern allows an object to alter its behavior when its internal state changes, making the object appear as if it has changed its class.
\end{itemize}

Each of these patterns solves a specific challenge related to object behavior and interaction. They help you manage how objects communicate, delegate tasks, and encapsulate behavior. In the following sections, we will dive deeper into each of these patterns, providing clear and beginner-friendly Python examples to illustrate how they can be implemented.

By understanding behavioral patterns, you will gain the ability to design systems where objects can interact effectively without creating a tangled web of dependencies, ensuring your Python applications remain flexible and maintainable as they scale.

\subsection{Chain of Responsibility Pattern}
The Chain of Responsibility (CoR) pattern \cite{vinoski2002chain} is a behavioral design pattern that allows you to pass requests along a chain of handlers. The pattern decouples the sender of a request from its receiver by giving multiple objects a chance to handle the request. Handlers are organized in a chain, and each handler can either handle the request or pass it to the next handler in the chain.

\subsubsection{Motivation}
Sometimes, a request must be processed by more than one object. For example, in a software system, various components might be interested in handling an event, like logging a message, checking authentication, or validating user input. The Chain of Responsibility pattern helps us avoid coupling a request to any particular handler. Instead, we can build a chain of handlers where each one has a chance to process the request, creating a flexible and reusable flow.

For instance, imagine we are designing a customer service system where each request is either handled by a junior staff member, or a manager or escalated to the director. Each level of staff only handles specific kinds of requests, and if they can't handle it, they pass the request to the next level.

\subsubsection{Structure}
The structure of the Chain of Responsibility pattern can be described as follows:

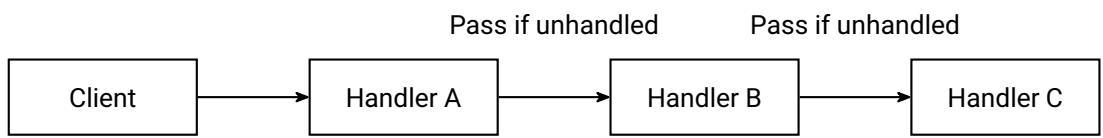
\usetikzlibrary{arrows.meta}

\begin{center}
\begin{tikzpicture}
  [node distance=4cm, auto, >={Stealth[round]}, thick]
  
  \node (Client) [draw, rectangle, minimum width=2.5cm, minimum height=1cm] {Client};
  \node (HandlerA) [right of=Client, draw, rectangle, minimum width=2.5cm, minimum height=1cm] {Handler A};
  \node (HandlerB) [right of=HandlerA, draw, rectangle, minimum width=2.5cm, minimum height=1cm] {Handler B};
  \node (HandlerC) [right of=HandlerB, draw, rectangle, minimum width=2.5cm, minimum height=1cm] {Handler C};

  \draw[->] (Client) -- (HandlerA);
  \draw[->] (HandlerA) -- node[above, yshift=0.7cm] {Pass if unhandled} (HandlerB);
  \draw[->] (HandlerB) -- node[above, yshift=0.7cm] {Pass if unhandled} (HandlerC);
  
\end{tikzpicture}
\end{center}

\textbf{Client}: The object that makes the request. \\
\textbf{Handler A, B, C}: The chain of handlers. Each one tries to handle the request or pass it to the next one.

\begin{itemize}
    \item \textbf{Client}: The object that makes the request.
    \item \textbf{Handler A, B, C}: The chain of handlers. Each one tries to handle the request or pass it to the next one.
\end{itemize}

\subsubsection{Participants}
The main participants in the Chain of Responsibility pattern are:

\begin{itemize}
    \item \textbf{Handler}: Defines an interface for handling requests. Optionally, it can reference the next handler in the chain.
    \item \textbf{ConcreteHandler}: Implements the handler's behavior and decides either to process the request or pass it along the chain.
    \item \textbf{Client}: The object that initiates the request.
\end{itemize}

\subsubsection{How It Works}
\begin{enumerate}
    \item \textbf{The client} sends a request to the first handler in the chain.
    \item \textbf{Each handler} examines the request and decides whether to handle it or pass it to the next handler.
    \item \textbf{If a handler can handle the request}, it processes it and the chain ends.
    \item \textbf{If a handler cannot handle the request}, it passes the request to the next handler in the chain.
    \item \textbf{The chain continues} until a handler processes the request or no handlers are left.
\end{enumerate}

\subsubsection{Advantages and Disadvantages}
\textbf{Advantages:}
\begin{itemize}
    \item \textbf{Decouples sender and receiver}: The sender of a request does not need to know which handler will process it.
    \item \textbf{Flexible and extensible}: New handlers can easily be added to the chain without modifying the existing code.
    \item \textbf{Reduces coupling}: The request does not need to be hard-coded to a specific handler, making the code more reusable and maintainable.
\end{itemize}

\textbf{Disadvantages:}
\begin{itemize}
    \item \textbf{No guarantee of handling}: If no handler can handle the request, it might go unprocessed.
    \item \textbf{Performance}: If the chain is long, the request may pass through many handlers before finding one that can handle it, which could introduce delays.
\end{itemize}

\subsubsection{Use Cases}
\begin{itemize}
    \item \textbf{Logging Systems}: Different log levels (debug, info, error) can be handled by different logging handlers.
    \item \textbf{UI Event Handling}: In a graphical user interface, events can be handled by different components like buttons, menus, or dialogs.
    \item \textbf{Authorization Chains}: Handling security requests where each component checks whether the user has the necessary permissions.
\end{itemize}

\subsubsection{Code Example}
Let’s walk through a Python example using the Chain of Responsibility pattern. Here, we will simulate a system where customer service requests are handled by different staff levels (Junior Staff, Manager, and Director).

\begin{lstlisting}[style=python]
class Handler:
    """The base handler interface."""
    def __init__(self, next_handler=None):
        self._next_handler = next_handler

    def handle(self, request):
        if self._next_handler:
            return self._next_handler.handle(request)
        return None

class JuniorStaff(Handler):
    """Concrete handler: Junior staff can handle simple requests."""
    def handle(self, request):
        if request == "simple":
            return "Junior staff handled the request."
        else:
            return super().handle(request)

class Manager(Handler):
    """Concrete handler: Manager handles moderate complexity requests."""
    def handle(self, request):
        if request == "moderate":
            return "Manager handled the request."
        else:
            return super().handle(request)

class Director(Handler):
    """Concrete handler: Director handles high complexity requests."""
    def handle(self, request):
        if request == "complex":
            return "Director handled the request."
        else:
            return super().handle(request)

# Client code
if __name__ == "__main__":
    # Setting up the chain
    chain = JuniorStaff(Manager(Director()))
    
    # Client makes requests
    print(chain.handle("simple"))   # Output: Junior staff handled the request.
    print(chain.handle("moderate")) # Output: Manager handled the request.
    print(chain.handle("complex"))  # Output: Director handled the request.
    print(chain.handle("unknown"))  # Output: None
\end{lstlisting}

In this code:
\begin{itemize}
    \item \textbf{Handler} is the base class that defines the interface for handling requests.
    \item \textbf{JuniorStaff}, \textbf{Manager}, and \textbf{Director} are concrete handlers that implement different levels of request handling.
    \item The client sets up the chain with \texttt{JuniorStaff}, \texttt{Manager}, and \texttt{Director} in order, so each one tries to handle the request based on its capabilities.
    \item If no handler can handle the request, the chain returns \texttt{None}.
\end{itemize}

\subsubsection{Pattern Extensions}
The Chain of Responsibility pattern can be extended in several ways:

\begin{itemize}
    \item \textbf{Dynamic Chains}: You can change the order of handlers dynamically at runtime based on application logic.
    \item \textbf{Circular Chains}: Handlers can pass the request around in a loop until it gets handled.
    \item \textbf{Multiple Requests}: The chain can be designed to handle multiple requests at once, processing each one until all are handled.
    \item \textbf{Logging Chains}: Handlers can log the request whether they handle it or pass it along, creating a trace of how the request was processed.
\end{itemize}

In conclusion, the Chain of Responsibility pattern provides a flexible way to process requests with multiple handlers. It allows for adding or modifying the chain without changing the client code, making it a powerful tool for many scenarios in Python.

\subsection{Command Pattern}

The Command Pattern \cite{harmes2008command} is a behavioral design pattern in which an object is used to encapsulate all the information needed to perform an action or trigger an event later. This information includes the method to call, the method’s arguments, and the object to which the method belongs. It is useful for separating concerns, allowing you to decouple the sender of a request from its receiver.

\subsubsection{Motivation}

Imagine a scenario where you have a GUI application that has multiple buttons, such as "Save," "Open," and "Close." Each button performs a different action, and if you want to allow undo/redo functionality, you must be able to track the actions executed by each button. You don’t want the button itself to know how to perform those actions. This is where the Command Pattern becomes very useful.

By encapsulating the action in a command object, you can decouple the invoker (the button) from the actual logic (such as saving or opening a file). This decoupling increases flexibility and allows additional functionalities like undo, redo, and logging.

\subsubsection{Structure}

The basic structure of the Command Pattern involves the following components:

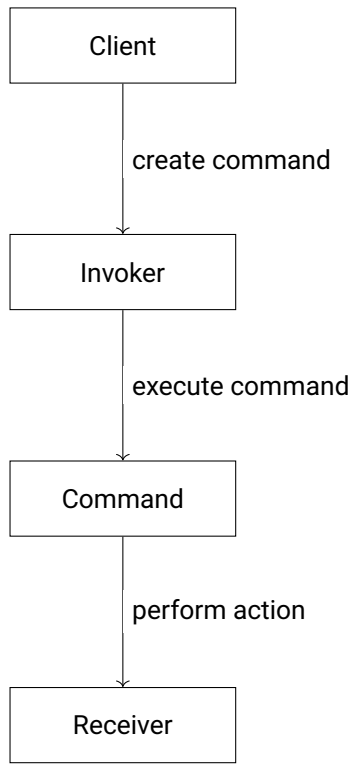
\begin{center}
\begin{tikzpicture}[node distance=2cm, every node/.style={fill=white, font=\sffamily}, align=center]
  
  \node (client) [draw, rectangle, minimum width=3cm, minimum height=1cm] {Client};
  \node (invoker) [draw, rectangle, below=of client, minimum width=3cm, minimum height=1cm] {Invoker};
  \node (command) [draw, rectangle, below=of invoker, minimum width=3cm, minimum height=1cm] {Command};
  \node (receiver) [draw, rectangle, below=of command, minimum width=3cm, minimum height=1cm] {Receiver};

  \draw [->] (client) -- (invoker) node[midway, right] {create command};
  \draw [->] (invoker) -- (command) node[midway, right] {execute command};
  \draw [->] (command) -- (receiver) node[midway, right] {perform action};

\end{tikzpicture}
\end{center}

\textbf{Client}: Creates a command object and associates it with the Invoker.

\textbf{Invoker}: Knows how to execute a command but does not know the details of the action.

\textbf{Command}: Declares an interface for executing operations. Concrete commands implement the execute method by invoking actions on the receiver.

\textbf{Receiver}: Contains the actual logic for performing the action.

\subsubsection{Participants}

\begin{itemize}
    \item \textbf{Command Interface:} Defines the structure for executing commands.
    \item \textbf{Concrete Command:} Implements the Command interface and calls the necessary methods on the receiver.
    \item \textbf{Invoker:} Triggers commands and knows nothing about the actions themselves.
    \item \textbf{Receiver:} Contains the core business logic that needs to be executed.
    \item \textbf{Client:} Initializes the concrete command and associates it with the invoker.
\end{itemize}

\subsubsection{How It Works}

The Command Pattern works by decoupling the object that invokes an operation from the object that knows how to perform it. The command object itself stores the details of the request, including the receiver and the method to invoke on the receiver. This allows for flexibility in command execution.

\begin{itemize}
    \item \textbf{Step 1:} The client creates a command object and passes it to the invoker.
    \item \textbf{Step 2:} The invoker executes the command when required.
    \item \textbf{Step 3:} The command itself calls the necessary methods on the receiver to perform the operation.
    \item \textbf{Step 4:} The receiver performs the actual work.
\end{itemize}

\subsubsection{Advantages and Disadvantages}

\textbf{Advantages:}
\begin{itemize}
    \item \textbf{Decoupling:} It decouples the sender and receiver, making the code more flexible and easier to maintain.
    \item \textbf{Extensibility:} You can easily extend the functionality by adding new commands without changing existing code.
    \item \textbf{Undo/Redo:} Commands can be stored for undo/redo functionality.
    \item \textbf{Logging:} You can log the commands for future reference or debugging purposes.
\end{itemize}

\textbf{Disadvantages:}
\begin{itemize}
    \item \textbf{Complexity:} Adds more classes and can make the code more complex.
    \item \textbf{Overhead:} Requires careful design to avoid excessive overhead in simple cases.
\end{itemize}

\subsubsection{Use Cases}

\begin{itemize}
    \item \textbf{GUI Applications:} Where each action (button click) corresponds to a different command.
    \item \textbf{Undo/Redo Operations:} Keeping track of commands executed to allow users to undo or redo actions.
    \item \textbf{Macro Recording:} Recording user actions as commands to be replayed later.
    \item \textbf{Task Scheduling:} Where tasks (commands) need to be delayed or executed at different times.
\end{itemize}

\subsubsection{Code Example}

Here is a Python example of the Command Pattern where we simulate a simple text editor with undo functionality:

\begin{lstlisting}[style=python]
# Command Interface
class Command:
    def execute(self):
        pass

    def undo(self):
        pass

# Concrete Command
class WriteCommand(Command):
    def __init__(self, document, text):
        self.document = document
        self.text = text

    def execute(self):
        self.document.write(self.text)

    def undo(self):
        self.document.erase(self.text)

# Receiver
class Document:
    def __init__(self):
        self.content = ""

    def write(self, text):
        self.content += text
        print(f"Document after writing: {self.content}")

    def erase(self, text):
        self.content = self.content.replace(text, "")
        print(f"Document after erasing: {self.content}")

# Invoker
class Editor:
    def __init__(self):
        self.history = []

    def execute_command(self, command):
        command.execute()
        self.history.append(command)

    def undo_last_command(self):
        if self.history:
            last_command = self.history.pop()
            last_command.undo()

# Client
document = Document()
editor = Editor()

write_hello = WriteCommand(document, "Hello ")
write_world = WriteCommand(document, "World!")

editor.execute_command(write_hello)  # Document after writing: Hello 
editor.execute_command(write_world)  # Document after writing: Hello World!
editor.undo_last_command()           # Document after erasing: Hello 
editor.undo_last_command()           # Document after erasing: 
\end{lstlisting}

In this example, the \texttt{Editor} is the invoker that triggers commands, while the \texttt{Document} is the receiver. Commands are encapsulated as objects, allowing the text editor to maintain a history of operations and undo them when needed.

\subsubsection{Pattern Extensions}

Some common extensions of the Command Pattern include:

\begin{itemize}
    \item \textbf{Composite Command:} A command that consists of multiple sub-commands.
    \item \textbf{Macro Command:} Used for executing a sequence of commands as a single command.
    \item \textbf{Queued Command:} Commands are queued and executed later or in sequence.
\end{itemize}
  
\subsection{Interpreter Pattern}
The Interpreter Pattern \cite{hills2011case} is a design pattern that specifies how to evaluate sentences in a language. This pattern is typically used to interpret expressions of a specific language or convert data from one format into another. It is especially useful when you have a set of rules or grammar that need to be evaluated repeatedly in different contexts.

\subsubsection{Motivation}
The main motivation behind the Interpreter Pattern is to interpret user input or another kind of language, be it formal or domain-specific. For example, in Python, you might create an interpreter for simple mathematical expressions like arithmetic operations, or you could create one for parsing a custom command language.

Imagine you need to process user-defined expressions such as "5 + 3 - 2", or you have a custom language for configuring some software. Manually writing logic to parse and evaluate these expressions can quickly become complex, error-prone, and hard to maintain. This is where the Interpreter Pattern comes in handy. It simplifies this problem by breaking it down into smaller pieces, where each piece represents a part of the language (such as an operand or operator) and is interpreted accordingly.

\subsubsection{Structure}
The Interpreter Pattern typically consists of the following structure:
\begin{itemize}
    \item \textbf{AbstractExpression} – Declares an abstract method for interpreting expressions.
    \item \textbf{TerminalExpression} – Implements the interpretation operation for terminal symbols (i.e., leaves in the expression tree).
    \item \textbf{NonTerminalExpression} – Represents operators or rules that define how to combine and interpret terminal expressions.
    \item \textbf{Context} – Contains information that is global to the interpreter, such as variable values or input data.
    \item \textbf{Client} – The part of the code that uses the interpreter to interpret or evaluate expressions.
\end{itemize}

Here's an example structure in a tree form:

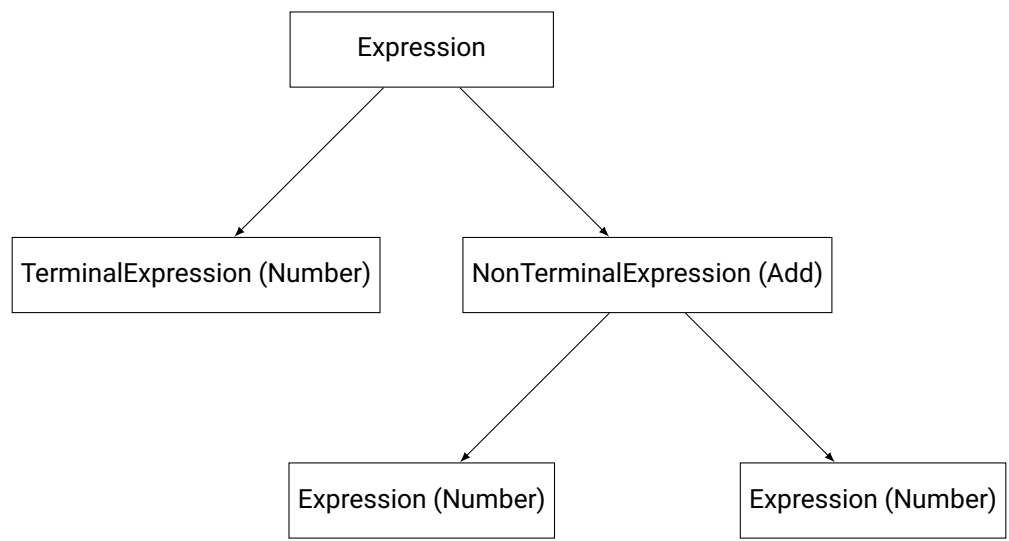
\begin{center}
\begin{tikzpicture}[
  level distance=3cm,   
  sibling distance=6cm,   
  edge from parent/.style={draw, -latex}, 
  every node/.style={draw, rectangle, minimum width=3.5cm, minimum height=1cm, align=center, font=\sffamily}] 
    \node {Expression}
        child { node {TerminalExpression (Number)} }
        child { node {NonTerminalExpression (Add)}
            child { node {Expression (Number)} }
            child { node {Expression (Number)} }
        };
\end{tikzpicture}
\end{center}

\subsubsection{Participants}
Here’s a breakdown of the roles each participant plays in the Interpreter Pattern:

\begin{itemize}
    \item \textbf{AbstractExpression} – This defines the abstract interface for all nodes in the expression tree. All concrete expression types (terminal or non-terminal) must implement this interface.
    \item \textbf{TerminalExpression} – These are leaf nodes in the expression tree. Each terminal expression represents an individual unit of the expression, like a number or variable.
    \item \textbf{NonTerminalExpression} – These nodes represent operations that combine terminal expressions, such as addition, subtraction, or other grammar rules.
    \item \textbf{Context} – This holds an external state that might be required during interpretation, such as variable bindings in the case of an expression involving variables.
    \item \textbf{Client} – This initiates the interpretation process by constructing the expression tree and invoking the interpretation method.
\end{itemize}

\subsubsection{How It Works}
The core of the Interpreter Pattern revolves around building a tree of expressions, where each node can be evaluated individually. The terminal nodes represent atomic values (such as numbers or variables), while the non-terminal nodes represent operations or rules that combine these values.

The process typically works as follows:

\begin{enumerate}
    \item The client builds an expression tree based on the input expression. For example, the arithmetic expression "5 + 3" will be represented as a tree with the "+" operator as a non-terminal node and "5" and "3" as terminal nodes.
    \item Once the expression tree is constructed, the client calls the \texttt{interpret} method on the root node (usually a non-terminal node).
    \item Each non-terminal node recursively evaluates its child nodes, combining their results to produce a final result.
    \item Terminal nodes directly return their values.
\end{enumerate}

\subsubsection{Advantages and Disadvantages}
\textbf{Advantages:}
\begin{itemize}
    \item It simplifies the implementation of grammar by breaking down expressions into smaller pieces.
    \item It’s easy to extend or modify the grammar by adding new terminal or non-terminal expressions.
    \item It can be reused in different contexts, such as parsing a language or evaluating mathematical expressions.
\end{itemize}

\textbf{Disadvantages:}
\begin{itemize}
    \item The Interpreter Pattern can become complex when the grammar becomes large. Each rule or operation in the grammar requires a new class.
    \item The pattern may be less efficient than alternatives, especially if the expression tree is large and needs to be evaluated repeatedly.
\end{itemize}

\subsubsection{Use Cases}
The Interpreter Pattern is commonly used in the following scenarios:

\begin{itemize}
    \item Parsing and evaluating mathematical expressions (e.g., calculators, parsers for arithmetic expressions).
    \item Implementing interpreters for domain-specific languages (DSLs), such as custom query languages or configuration languages.
    \item Syntax parsing for compilers or interpreters.
    \item Text processing where specific patterns need to be interpreted or transformed based on rules.
\end{itemize}

\subsubsection{Code Example}
Here is a Python example implementing the Interpreter Pattern to evaluate simple arithmetic expressions:

\begin{lstlisting}[style=python]
from abc import ABC, abstractmethod

# AbstractExpression
class Expression(ABC):
    @abstractmethod
    def interpret(self, context):
        pass

# TerminalExpression for numbers
class Number(Expression):
    def __init__(self, value):
        self.value = value

    def interpret(self, context):
        return self.value

# NonTerminalExpression for addition
class Add(Expression):
    def __init__(self, left, right):
        self.left = left
        self.right = right

    def interpret(self, context):
        return self.left.interpret(context) + self.right.interpret(context)

# NonTerminalExpression for subtraction
class Subtract(Expression):
    def __init__(self, left, right):
        self.left = left
        self.right = right

    def interpret(self, context):
        return self.left.interpret(context) - self.right.interpret(context)

# Context class (can be extended to hold more information)
class Context:
    pass

# Client code
if __name__ == "__main__":
    # Building the expression 5 + 3 - 2
    expr = Subtract(
        Add(Number(5), Number(3)),
        Number(2)
    )

    context = Context()
    result = expr.interpret(context)
    print(f"Result: {result}")  # Output: 6
\end{lstlisting}

\subsubsection{Pattern Extensions}
Here are a few possible extensions of the Interpreter Pattern:

\begin{itemize}
    \item \textbf{Adding more operators:} You can easily extend this pattern by adding new non-terminal expressions like multiplication (\texttt{Multiply}) or division (\texttt{Divide}).
    \item \textbf{Variables:} The context could be extended to include variable bindings. For example, you can add a \texttt{Variable} class that looks up values from the context.
    \item \textbf{Composite Pattern:} The Interpreter Pattern is often combined with the Composite Pattern, where complex expressions are made up of simpler ones.
    \item \textbf{Chaining Expressions:} You can extend the interpreter to support the chaining of expressions, such as adding parentheses for precedence or more complex evaluation rules.
\end{itemize}

\subsection{Iterator Pattern}
The Iterator Pattern \cite{gibbons2009essence} is a design pattern used to traverse elements of a collection without exposing the underlying structure of that collection. This pattern allows the client to sequentially access elements of a collection uniformly, irrespective of the internal structure.

\subsubsection{Motivation}
Imagine you have a collection of items like a list, set, or even a custom data structure in Python. As a programmer, you need a way to loop through or access these items one by one without worrying about how the collection is implemented internally. This is where the Iterator Pattern comes in.

The primary motivation for using the Iterator Pattern is to provide a consistent way to traverse a collection, allowing access to its elements one by one without exposing the underlying representation. The Iterator abstracts the navigation logic, so the client code remains clean and decoupled from the collection's internal details.

In Python, iterators are commonly used with collections like lists, tuples, dictionaries, and custom objects. By understanding the Iterator Pattern, you'll be able to work with collections more efficiently and even create your iterators for custom data structures.

\subsubsection{Structure}
The Iterator Pattern typically involves two primary participants: 
\begin{itemize}
  \item \textbf{Iterator}: Provides the interface for accessing and traversing elements. In Python, this is the object that implements the \texttt{\_\_iter\_\_()} and \texttt{\_\_next\_\_()} methods.
  \item \textbf{Iterable}: The collection object that implements the \texttt{\_\_iter\_\_()} method and returns an iterator.
\end{itemize}

Here’s a typical structure of the Iterator Pattern in Python:

\begin{lstlisting}[style=python]
class Iterator:
    def __init__(self, collection):
        self._collection = collection
        self._index = 0

    def __iter__(self):
        return self

    def __next__(self):
        if self._index < len(self._collection):
            result = self._collection[self._index]
            self._index += 1
            return result
        else:
            raise StopIteration
\end{lstlisting}

\subsubsection{Participants}
\begin{itemize}
  \item \textbf{Concrete Iterator}: The class responsible for implementing the iteration logic. It maintains the current position of the traversal.
  \item \textbf{Concrete Iterable}: The collection that returns an instance of the iterator when requested. It defines the \texttt{\_\_iter\_\_()} method to return the iterator object.
\end{itemize}

In Python, any object that implements the \texttt{\_\_iter\_\_()} method and the iterator object’s \texttt{\_\_next\_\_()} method can be used as an iterator.

\textbf{Concrete Iterator Example:}
\begin{lstlisting}[style=python]
class NumberIterator:
    def __init__(self, numbers):
        self._numbers = numbers
        self._index = 0

    def __iter__(self):
        return self

    def __next__(self):
        if self._index < len(self._numbers):
            result = self._numbers[self._index]
            self._index += 1
            return result
        else:
            raise StopIteration
\end{lstlisting}

\textbf{Concrete Iterable Example:}
\begin{lstlisting}[style=python]
class NumberCollection:
    def __init__(self):
        self._numbers = []

    def add_number(self, number):
        self._numbers.append(number)

    def __iter__(self):
        return NumberIterator(self._numbers)
\end{lstlisting}

\subsubsection{How It Works}
The Iterator Pattern separates the logic of traversing a collection from the actual collection itself. The collection (also known as the iterable) implements the \texttt{\_\_iter\_\_()} method, which returns an iterator object. The iterator object implements both the \texttt{\_\_iter\_\_()} and \texttt{\_\_next\_\_()} methods. 

When you use a \texttt{for} loop or call \texttt{next()} on an iterable in Python, it implicitly calls the \texttt{\_\_iter\_\_()} method to get an iterator and then calls the \texttt{\_\_next\_\_()} method to fetch elements one by one.

\textbf{Example of usage:}
\begin{lstlisting}[style=python]
collection = NumberCollection()
collection.add_number(1)
collection.add_number(2)
collection.add_number(3)

for number in collection:
    print(number)
\end{lstlisting}

This example prints:
\begin{lstlisting}[style=cmd]
1
2
3
\end{lstlisting}

\subsubsection{Advantages and Disadvantages}
\textbf{Advantages:}
\begin{itemize}
  \item \textbf{Simplifies traversal}: The Iterator Pattern provides a clean and consistent way to traverse different types of collections.
  \item \textbf{Encapsulation}: The internal structure of the collection is hidden from the client. The client only needs to know how to interact with the iterator.
  \item \textbf{Supports different traversal methods}: If needed, you can implement different iterators for different traversal strategies (e.g., forward, backward, filtered iteration).
\end{itemize}

\textbf{Disadvantages:}
\begin{itemize}
  \item \textbf{Overhead}: For simple collections, the Iterator Pattern may introduce unnecessary complexity by requiring additional classes.
  \item \textbf{State management}: Iterators maintain an internal state (e.g., the current index), which may make them harder to manage in certain scenarios, like parallel processing or when the underlying collection changes during iteration.
\end{itemize}

\subsubsection{Use Cases}
\begin{itemize}
  \item \textbf{Traversing custom collections}: When you create custom data structures and need a way to traverse them without exposing internal details.
  \item \textbf{Multiple traversal strategies}: When you need different ways to traverse a collection, such as forward, backward, or filtered iteration.
  \item \textbf{Lazy evaluation}: Iterators are useful when working with large datasets where it's inefficient to load everything into memory. Python generators are an example of this.
\end{itemize}

\subsubsection{Code Example}
Here’s a complete example that illustrates the use of the Iterator Pattern in Python.

\begin{lstlisting}[style=python]
class BookIterator:
    def __init__(self, books):
        self._books = books
        self._index = 0

    def __iter__(self):
        return self

    def __next__(self):
        if self._index < len(self._books):
            book = self._books[self._index]
            self._index += 1
            return book
        else:
            raise StopIteration

class Library:
    def __init__(self):
        self._books = []

    def add_book(self, book):
        self._books.append(book)

    def __iter__(self):
        return BookIterator(self._books)

# Usage:
library = Library()
library.add_book("Python Design Patterns")
library.add_book("Advanced Python Programming")

for book in library:
    print(book)
\end{lstlisting}

Output:
\begin{lstlisting}[style=cmd]
Python Design Patterns
Advanced Python Programming
\end{lstlisting}

\subsubsection{Pattern Extensions}
The Iterator Pattern can be extended to accommodate various needs:
\begin{itemize}
  \item \textbf{Bidirectional Iteration}: You can add support for reverse traversal by adding a method like \texttt{\_\_prev\_\_()} or by creating a separate reverse iterator.
  \item \textbf{Filtered Iteration}: You can create iterators that only return elements that meet specific criteria, such as filtering out \texttt{None} values or selecting even numbers from a list.
  \item \textbf{Iterator chaining}: In Python, you can chain multiple iterators together using tools like \texttt{itertools.chain()} to traverse multiple collections as if they were a single collection.
\end{itemize}

\textbf{Example of Bidirectional Iterator:}

\begin{lstlisting}[style=python]
class BidirectionalIterator:
    def __init__(self, collection):
        self._collection = collection
        self._index = 0

    def __iter__(self):
        return self

    def __next__(self):
        if self._index < len(self._collection):
            result = self._collection[self._index]
            self._index += 1
            return result
        else:
            raise StopIteration

    def previous(self):
        if self._index > 0:
            self._index -= 1
            return self._collection[self._index]
        else:
            raise StopIteration

# Usage:
collection = [10, 20, 30]
iterator = BidirectionalIterator(collection)

# Moving forward
print(next(iterator))  # 10
print(next(iterator))  # 20

# Moving backward
print(iterator.previous())  # 20
\end{lstlisting}

\subsection{Mediator Pattern}

\subsubsection{Motivation}
The Mediator Pattern \cite{chung2011mediator} is a behavioral design pattern that promotes loose coupling between interacting components in a system. It achieves this by introducing a mediator object, which controls and coordinates interactions between different objects. Instead of components communicating directly with each other, they communicate through the mediator. This reduces the dependencies between components, making the system easier to maintain and modify.

In real-world software, a common problem is that classes can become tightly coupled when they interact with many other classes. For example, in a chat application, different users (components) may need to communicate with each other. Without a mediator, each user would have to know how to send messages directly to every other user, creating a complex web of dependencies. The mediator pattern helps solve this by centralizing the communication through a single object — the mediator.

\subsubsection{Structure}
The key components of the Mediator pattern are:
\begin{itemize}
    \item \textbf{Mediator}: Defines an interface for communication between Colleague objects.
    \item \textbf{ConcreteMediator}: Implements the Mediator interface and coordinates communication between Colleagues.
    \item \textbf{Colleague}: Represents objects that are communicating with each other through the Mediator.
\end{itemize}

A simple example structure is illustrated as:

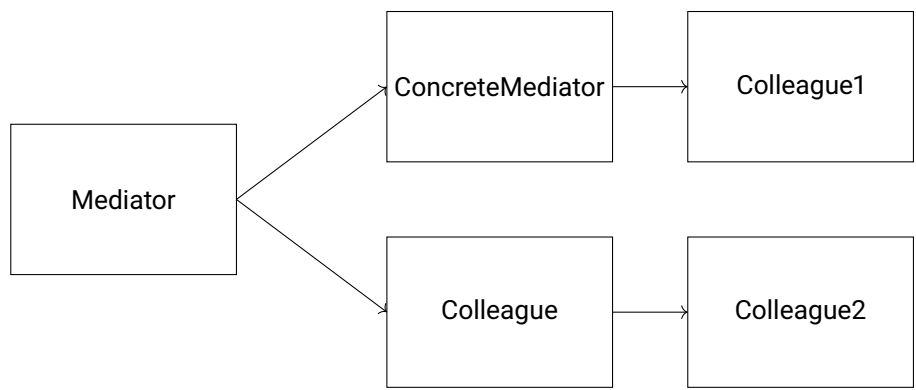
\begin{center}
\begin{tikzpicture}
  \draw (0,0) rectangle (3,2) node[midway] {Mediator};
  \draw (5,1.5) rectangle (8,3.5) node[midway] {ConcreteMediator};
  \draw (5,-1.5) rectangle (8,0.5) node[midway] {Colleague};
  \draw (9,1.5) rectangle (12,3.5) node[midway] {Colleague1};
  \draw (9,-1.5) rectangle (12,0.5) node[midway] {Colleague2};

  \draw[->] (3,1) -- (5,2.5);     
  \draw[->] (3,1) -- (5,-0.5);    
  \draw[->] (8,2.5) -- (9,2.5);    
  \draw[->] (8,-0.5) -- (9,-0.5);  
\end{tikzpicture}
\end{center}

In this diagram, the Mediator object knows the Colleague objects and manages the interaction between them, but the Colleague objects do not know each other directly. They only know how to communicate with the Mediator.

\subsubsection{Participants}
The participants in the Mediator pattern are:

\begin{itemize}
    \item \textbf{Mediator}: The interface that defines the contract for communication between Colleague objects. It has methods to send and receive messages from colleagues.
    \item \textbf{ConcreteMediator}: Implements the Mediator interface and holds references to the Colleagues. It defines how the communication is managed between them.
    \item \textbf{Colleague}: Represents individual components or classes that need to communicate. They send messages through the mediator rather than directly to each other.
\end{itemize}

\subsubsection{How It Works}
1. Colleagues interact with the Mediator rather than with each other directly.
2. A ConcreteMediator maintains references to the Colleagues and routes messages between them.
3. When one Colleague wants to communicate with another, it sends the message to the Mediator.
4. The Mediator then forwards the message to the appropriate Colleague or takes some action based on the message.

This setup ensures that the Colleagues do not need to be aware of each other’s details, thus promoting loose coupling.

\subsubsection{Advantages and Disadvantages}
\textbf{Advantages:}
\begin{itemize}
    \item \textbf{Reduces coupling between components}: Since Colleagues communicate through the Mediator, they are not tightly bound to each other.
    \item \textbf{Centralizes control}: The mediator centralizes communication logic, making the system more manageable.
    \item \textbf{Simplifies object interaction}: Complex webs of interaction between objects are simplified by handling all communication in one place.
\end{itemize}

\textbf{Disadvantages:}
\begin{itemize}
    \item \textbf{Mediator can become complex}: If many objects are involved, the Mediator itself can become a complex and bloated class.
    \item \textbf{Potential performance bottleneck}: Since all communication goes through the mediator, it can become a bottleneck in performance-critical applications.
\end{itemize}

\subsubsection{Use Cases}
The Mediator pattern is useful in scenarios where:
\begin{itemize}
    \item A system involves many components that interact in complex ways, and reducing direct dependencies between them is necessary.
    \item Communication logic between objects is becoming too complex and needs to be centralized.
    \item You need to decouple classes that depend heavily on each other, such as in a chat system, traffic control system, or GUI elements in an application.
\end{itemize}

\subsubsection{Code Example}
Let's consider a simple chat room system where users send messages to each other via a chat mediator. Instead of sending messages directly to each other, users send them through the mediator.

\begin{lstlisting}[style=python]
class ChatRoomMediator:
    def show_message(self, user, message):
        print(f"[{user}] {message}")

class User:
    def __init__(self, name, mediator):
        self.name = name
        self.mediator = mediator

    def send_message(self, message):
        self.mediator.show_message(self.name, message)

# Create a mediator
mediator = ChatRoomMediator()

# Create users
john = User("John", mediator)
jane = User("Jane", mediator)

# Users send messages through the mediator
john.send_message("Hello, Jane!")
jane.send_message("Hi, John!")
\end{lstlisting}

In this example, the `ChatRoomMediator` acts as the Mediator, and the `User` is the Colleague. The users don’t communicate with each other directly, but through the mediator, which centralizes the message delivery.

\subsubsection{Pattern Extensions}
There are several ways to extend the basic Mediator pattern:
\begin{itemize}
    \item \textbf{Enhanced Mediator Logic}: The mediator can be made more intelligent by adding business logic to manage or filter communications between colleagues.
    \item \textbf{Hierarchical Mediators}: In large systems, a hierarchy of mediators can be used to organize communications better. For instance, a higher-level mediator could handle communication between different modules, while lower-level mediators handle communication within the module.
    \item \textbf{Event-Driven Mediator}: The mediator can be modified to handle asynchronous events or to work with event-based systems, where communication happens in response to events rather than direct method calls.
\end{itemize}

\subsection{Memento Pattern}

\subsubsection{Motivation}
The Memento Pattern \cite{chung2011memento} is a behavioral design pattern that provides a way to capture and store the current state of an object without exposing its internal structure so that the object can be restored to this state later. The main motivation behind this pattern is to give an external entity (like a caretaker) the ability to undo or roll back changes to an object’s state while keeping the implementation details of the object hidden.

For example, in a text editor, the user might want to undo the last few typing operations. The Memento Pattern would allow the editor to save its current state at key points and then restore to any saved state when requested, without revealing how the editor works internally.

\subsubsection{Structure}
The Memento Pattern consists of three main components:
\begin{itemize}
    \item \textbf{Originator}: This is the object whose state needs to be saved. It creates a memento containing a snapshot of its current state and uses it to restore its previous state.
    \item \textbf{Memento}: This is a data structure that stores the state of the Originator object. It does not allow other objects to modify their state.
    \item \textbf{Caretaker}: This is responsible for storing mementos and restoring the Originator's state from the memento. It does not know the content of the memento, only that it holds the state of the Originator.
\end{itemize}

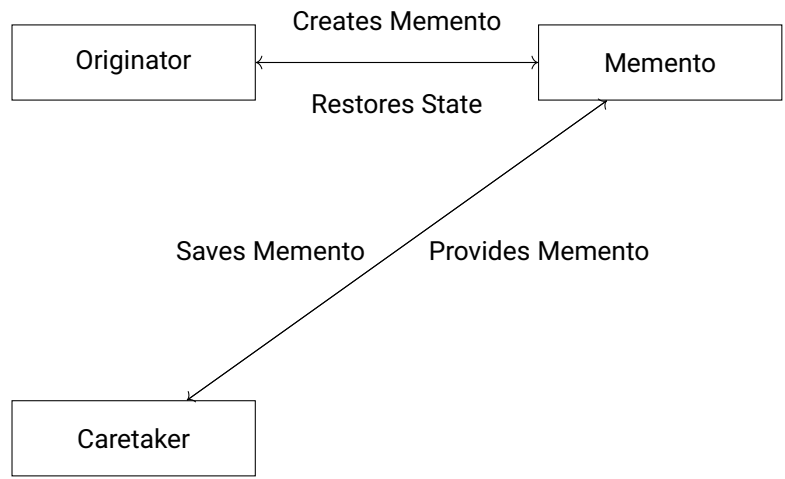
\begin{center}
\begin{tikzpicture}[node distance=4cm]
    \node (originator) [rectangle, draw, text centered, text width=3cm, minimum height=1cm] {Originator};
    \node (memento) [rectangle, draw, right of=originator, xshift=3cm, text centered, text width=3cm, minimum height=1cm] {Memento};
    \node (caretaker) [rectangle, draw, below of=originator, yshift=-1cm, text centered, text width=3cm, minimum height=1cm] {Caretaker};

    \draw[->] (originator) -- (memento) node[midway, above, yshift=0.3cm] {Creates Memento};  
    \draw[->] (memento) -- (originator) node[midway, below, yshift=-0.3cm] {Restores State};  
    \draw[->] (caretaker) -- (memento) node[midway, left, xshift=-0.3cm] {Saves Memento};  
    \draw[->] (memento) -- (caretaker) node[midway, right, xshift=0.3cm] {Provides Memento};  
\end{tikzpicture}
\end{center}

\subsubsection{Participants}
\begin{itemize}
    \item \textbf{Originator}: This class is responsible for creating and restoring its state. It uses a memento to store the state and restore it.
    \item \textbf{Memento}: A simple class that stores the state of the Originator. It does not provide methods to alter its content but only holds the information to be used by the Originator.
    \item \textbf{Caretaker}: This class is responsible for maintaining the history of mementos. It can store and retrieve mementos but does not alter or interpret the mementos' contents.
\end{itemize}

\subsubsection{How It Works}
The Memento Pattern works by allowing the Originator to save snapshots of its state at different moments in time. The process works as follows:
\begin{itemize}
    \item The Originator creates a Memento object that stores its internal state. 
    \item The Caretaker takes care of storing this Memento for future use, without knowing what is inside.
    \item When needed, the Caretaker provides the stored Memento back to the Originator, and the Originator restores its state based on the data in the Memento.
\end{itemize}

\textbf{Step-by-Step Example}:
\begin{enumerate}
    \item \textbf{Originator}: Let's say a drawing application allows a user to draw lines. The current drawing is saved as the state of the Originator.
    \item \textbf{Memento}: At certain points, the drawing is saved into a Memento object, which holds the state of the drawing.
    \item \textbf{Caretaker}: The Caretaker stores each Memento as the drawing progresses. If the user wishes to undo, the Caretaker gives the last saved Memento to the Originator to revert to the previous state.
\end{enumerate}

\subsubsection{Advantages and Disadvantages}

\textbf{Advantages}:
\begin{itemize}
    \item \textbf{Encapsulation}: The Memento Pattern keeps the state of the Originator hidden from the Caretaker. Only the Originator has access to the Memento’s content.
    \item \textbf{Undo and Redo functionality}: This pattern is very useful in providing undo functionality, allowing an object to restore a previous state.
    \item \textbf{Separation of Concerns}: It separates the responsibility of saving and restoring the state (handled by the Originator) from the responsibility of keeping track of these states (handled by the Caretaker).
\end{itemize}

\textbf{Disadvantages}:
\begin{itemize}
    \item \textbf{Memory Consumption}: Each saved state in the Memento consumes memory. If too many states are saved, it may lead to high memory consumption.
    \item \textbf{Complexity}: Implementing the Memento Pattern can add complexity to the system, especially when the object being saved has a complex state.
\end{itemize}

\subsubsection{Use Cases}
The Memento Pattern is most commonly used in situations where objects need to revert to previous states. Some typical use cases include:
\begin{itemize}
    \item \textbf{Undo/Redo functionality}: Applications like text editors, drawing tools, or IDEs often provide undo and redo capabilities. The Memento Pattern is ideal for implementing these features.
    \item \textbf{Game Saving}: Games often allow players to save their progress and reload from a previous state. The game's state can be saved as a Memento and restored later.
    \item \textbf{Transactional Systems}: In transactional systems, you might want to revert to a previous state if a transaction fails.
\end{itemize}

\subsubsection{Code Example}
Let’s look at a simple Python example of the Memento Pattern. In this example, we have a text editor that can save its state (text content) and undo changes by restoring a previous state.

\begin{lstlisting}[style=python]
class Memento:
    def __init__(self, state: str):
        self._state = state

    def get_state(self):
        return self._state

class TextEditor:
    def __init__(self):
        self._content = ""

    def type(self, words: str):
        self._content += words

    def save(self) -> Memento:
        return Memento(self._content)

    def restore(self, memento: Memento):
        self._content = memento.get_state()

    def get_content(self):
        return self._content

class Caretaker:
    def __init__(self):
        self._mementos = []

    def backup(self, memento: Memento):
        self._mementos.append(memento)

    def undo(self):
        if self._mementos:
            return self._mementos.pop()
        return None

# Usage
editor = TextEditor()
caretaker = Caretaker()

editor.type("Hello, ")
caretaker.backup(editor.save())

editor.type("World!")
caretaker.backup(editor.save())

editor.type(" How are you?")
print(editor.get_content())  # Output: "Hello, World! How are you?"

# Undoing changes
editor.restore(caretaker.undo())
print(editor.get_content())  # Output: "Hello, World!"

editor.restore(caretaker.undo())
print(editor.get_content())  # Output: "Hello, "
\end{lstlisting}

In this example:
\begin{itemize}
    \item The \texttt{Memento} class holds the state (the text content) of the \texttt{TextEditor}.
    \item The \texttt{TextEditor} (Originator) creates Mementos to save and restore its state.
    \item The \texttt{Caretaker} manages the Mementos, allowing the editor to undo the changes.
\end{itemize}

\subsubsection{Pattern Extensions}
There are some possible extensions or variations of the Memento Pattern:
\begin{itemize}
    \item \textbf{Multiple Undo Levels}: Instead of storing just one previous state, the Caretaker can keep a list of Mementos, enabling multiple undo levels.
    \item \textbf{State Difference Storage}: In cases where storing the entire state might be too memory-intensive, only the differences between states can be stored in each Memento.
    \item \textbf{Version Control}: The Memento Pattern can be extended to support version control, where every change creates a new version of the object’s state that can be reverted or reapplied.
\end{itemize}

\subsection{Observer Pattern}
The Observer Pattern \cite{pattern2021observer} is one of the most commonly used design patterns in software development. It is a behavioral pattern that defines a one-to-many dependency between objects so that when one object (called the subject) changes its state, all its dependents (called observers) are notified and updated automatically.

\subsubsection{Motivation}
The motivation behind the Observer Pattern arises when there is a need for multiple objects to react to changes in the state of another object. Consider a scenario where you have a data model, and multiple user interfaces need to be updated whenever the model changes. Instead of manually updating each user interface component, the Observer Pattern automates this process.

This pattern is especially useful in GUI frameworks, event-handling systems, and real-time applications where changes need to be propagated without tight coupling between components.

\subsubsection{Structure}
The Observer Pattern is typically composed of the following components:
\begin{itemize}
    \item \textbf{Subject}: The object that holds the state and notifies observers of any changes.
    \item \textbf{Observer}: The object that watches for changes in the subject and reacts accordingly.
    \item \textbf{ConcreteSubject}: A specific implementation of the subject, responsible for managing the state.
    \item \textbf{ConcreteObserver}: A specific implementation of an observer that updates its behavior based on the subject’s changes.
\end{itemize}

The relationship between these elements can be depicted using a tree diagram:

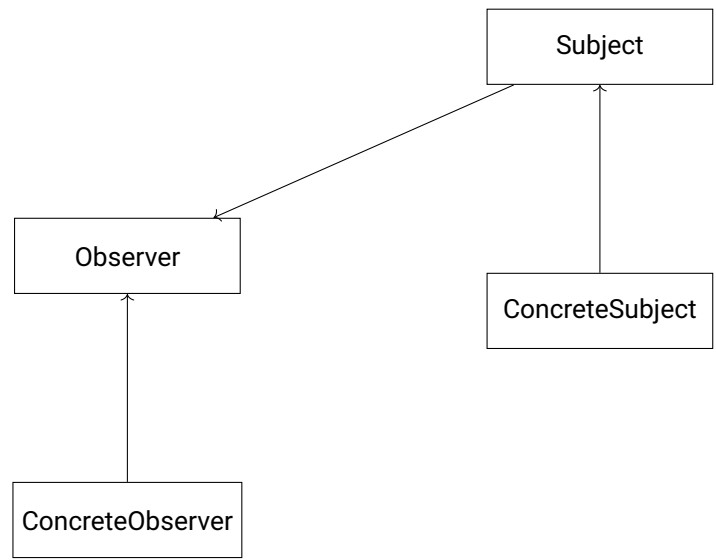
\begin{center}
\begin{tikzpicture}[node distance=2.5cm, auto]
    \node (Subject) [draw, rectangle, minimum width=3cm, minimum height=1cm] {Subject};
    \node (Observer) [draw, rectangle, below left=of Subject, xshift=-1.5cm, minimum width=3cm, minimum height=1cm] {Observer};
    \node (ConcreteSubject) [draw, rectangle, below=of Subject, minimum width=3cm, minimum height=1cm] {ConcreteSubject};
    \node (ConcreteObserver) [draw, rectangle, below=of Observer, minimum width=3cm, minimum height=1cm] {ConcreteObserver};

    \draw [->] (Subject) -- (Observer);
    \draw [->] (ConcreteSubject) -- (Subject);
    \draw [->] (ConcreteObserver) -- (Observer);
\end{tikzpicture}
\end{center}

\subsubsection{Participants}
The key participants in the Observer Pattern are:

\begin{itemize}
    \item \textbf{Subject}: The Subject knows its observers and can add, remove, or notify them of any state changes. The Subject maintains a list of observers.
    \item \textbf{Observer}: This interface defines a method, typically called \texttt{update()}, that gets called whenever the Subject changes.
    \item \textbf{ConcreteSubject}: The ConcreteSubject implements the Subject interface and stores the state that interests ConcreteObservers.
    \item \textbf{ConcreteObserver}: Each ConcreteObserver implements the Observer interface and updates its state to reflect changes in the ConcreteSubject.
\end{itemize}

\subsubsection{How It Works}
The Observer Pattern works by having the ConcreteSubject maintain a list of observers (ConcreteObservers). Whenever the subject’s state changes, it goes through its list of observers and calls the \texttt{update()} method on each one, passing along any necessary information.

Here is the step-by-step process:
\begin{enumerate}
    \item A ConcreteObserver is registered with the ConcreteSubject.
    \item When the state of the ConcreteSubject changes, it loops through the list of registered observers.
    \item The \texttt{update()} method is called on each observer, and the observer updates itself to reflect the change.
\end{enumerate}

\subsubsection{Advantages and Disadvantages}

\textbf{Advantages:}
\begin{itemize}
    \item Promotes loose coupling between subject and observers.
    \item Supports broadcast communication: one change can notify multiple observers.
    \item Easy to add new observers without modifying the subject.
\end{itemize}

\textbf{Disadvantages:}
\begin{itemize}
    \item Can lead to memory leaks if observers are not correctly removed.
    \item The order of notification is not guaranteed.
    \item Observers may become overloaded if they receive frequent updates.
\end{itemize}

\subsubsection{Use Cases}
The Observer Pattern is useful in scenarios where multiple objects need to be notified of changes in another object. Common use cases include:
\begin{itemize}
    \item Event systems: When an event occurs, multiple listeners or observers react to it.
    \item Data models: When the data model changes, the view is updated accordingly, typical in MVC (Model-View-Controller) frameworks.
    \item Real-time updates: Stock market applications, where many clients are notified of price changes.
\end{itemize}

\subsubsection{Code Example}
Let’s look at a Python implementation of the Observer Pattern. In this example, we simulate a weather station that broadcasts temperature updates to its observers.

\begin{lstlisting}[style=python]
# Subject (Observable)
class WeatherStation:
    def __init__(self):
        self._observers = []
        self._temperature = 0

    def add_observer(self, observer):
        self._observers.append(observer)

    def remove_observer(self, observer):
        self._observers.remove(observer)

    def notify_observers(self):
        for an observer in self._observers:
            observer.update(self._temperature)

    def set_temperature(self, temperature):
        self._temperature = temperature
        self.notify_observers()

# Observer Interface
class Observer:
    def update(self, temperature):
        raise NotImplementedError("Subclass must implement update method")

# Concrete Observer 1
class PhoneDisplay(Observer):
    def update(self, temperature: float) -> None:
        print(f"Phone Display: The current temperature is {temperature:.1f}\textdegree C")

# Concrete Observer 2
class WindowDisplay(Observer):
    def update(self, temperature: float) -> None:
        print(f"Window Display: The current temperature is {temperature:.1f}\textdegree C")


# Example usage
if __name__ == "__main__":
    # Create a weather station (subject)
    weather_station = WeatherStation()

    # Create observers
    phone_display = PhoneDisplay()
    window_display = WindowDisplay()

    # Register observers with the weather station
    weather_station.add_observer(phone_display)
    weather_station.add_observer(window_display)

    # Change temperature and notify observers
    weather_station.set_temperature(25)
    weather_station.set_temperature(30)
\end{lstlisting}

\textbf{Explanation:}
\begin{itemize}
    \item \texttt{WeatherStation} acts as the subject, and it holds a list of observers.
    \item \texttt{PhoneDisplay} and \texttt{WindowDisplay} are concrete observers that implement the \texttt{Observer} interface.
    \item When the temperature changes, the \texttt{set\_temperature()} method is called, which updates the internal state and notifies all registered observers.
\end{itemize}

Output from the code would be:
\begin{lstlisting}[style=cmd]
Phone Display: The current temperature is 25\textdegree C
Window Display: The current temperature is 25\textdegree C
Phone Display: The current temperature is 30\textdegree C
Window Display: The current temperature is 30\textdegree C
\end{lstlisting}

\subsubsection{Pattern Extensions}
The Observer Pattern can be extended in various ways:
\begin{itemize}
    \item \textbf{Push vs. Pull Models}: In a push model, the subject sends updated data to observers directly. In a pull model, the subject notifies the observers that something has changed, but the observer is responsible for retrieving the updated data.
    \item \textbf{Weak References}: In Python, it is possible to use weak references to avoid memory leaks, especially when observers forget to unregister themselves from the subject.
    \item \textbf{Event Channels}: In complex systems, you may introduce channels or event types, so observers only receive notifications they are interested in.
\end{itemize}

In conclusion, the Observer Pattern is a powerful and flexible way to handle event-driven programming. It helps in decoupling subjects from observers, making the system easier to maintain and extend.

\subsection{State Pattern}
The \textbf{State Pattern} \cite{dyson1996state} is a behavioral design pattern that allows an object to change its behavior when its internal state changes. The pattern encapsulates the varying behavior of the object based on its state and delegates state-specific behavior to the corresponding state classes. This is useful in scenarios where an object needs to change its behavior dynamically, depending on its current state.

\subsubsection{Motivation}
Imagine a scenario where you are building a simple media player application. The player can be in different states like "Playing", "Paused", and "Stopped". Depending on the current state, the actions of the media player (such as play, pause, and stop) behave differently. Without the State Pattern, managing this can lead to complex conditional logic and make the code difficult to maintain. By using the State Pattern, each state is encapsulated in its class, and the media player can delegate behavior changes by simply switching between states, making the design clean and modular.

\subsubsection{Structure}
The State Pattern consists of the following components:

\begin{itemize}
  \item \textbf{Context}: The object whose behavior varies depending on its state. It maintains a reference to the current state object and delegates behavior to it.
  \item \textbf{State}: This is an abstract class or interface that defines the behavior that is shared by all concrete states.
  \item \textbf{Concrete States}: These are specific state classes that implement the behavior associated with the state. Each state corresponds to a different behavior of the context object.
\end{itemize}

Below is the typical class diagram structure of the State Pattern.

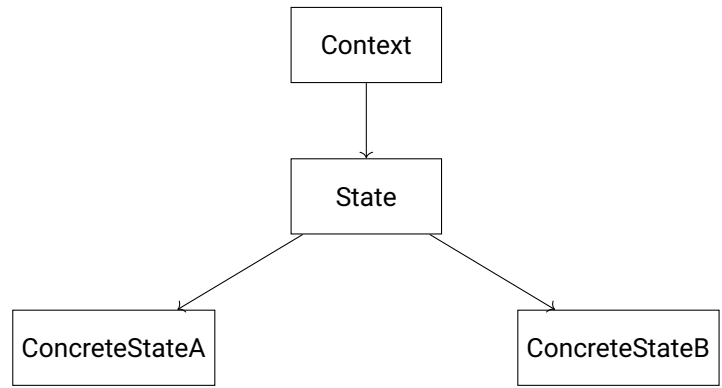
\begin{center}
\begin{tikzpicture}
    \node (context) [rectangle, draw, minimum width=2cm, minimum height=1cm] {Context};
    \node (state) [rectangle, draw, below=of context, minimum width=2cm, minimum height=1cm] {State};
    \node (concreteStateA) [rectangle, draw, below left=of state, minimum width=2cm, minimum height=1cm] {ConcreteStateA};
    \node (concreteStateB) [rectangle, draw, below right=of state, minimum width=2cm, minimum height=1cm] {ConcreteStateB};

    \draw[->] (context) -- (state);
    \draw[->] (state) -- (concreteStateA);
    \draw[->] (state) -- (concreteStateB);
\end{tikzpicture}
\end{center}

\subsubsection{Participants}
The participants in the State Pattern are:

\begin{itemize}
    \item \textbf{Context}: The object that has an internal state and delegates behavior to the state object.
    \item \textbf{State Interface}: Defines a common interface for all the possible states of the context.
    \item \textbf{Concrete State Classes}: Implements behavior specific to a particular state of the context.
\end{itemize}

\subsubsection{How It Works}
The Context holds a reference to a State object, and it delegates its behavior to the current state by invoking methods on the state object. Each concrete state class implements these methods differently, depending on the behavior it needs to exhibit in that particular state.

For example, a media player can be in different states such as "Playing", "Paused", or "Stopped". The context (media player) maintains a reference to the current state, and it switches between states when the user presses buttons like Play, Pause, or Stop.

The state objects can change the context's state, enabling the state transitions within the system. The context can ask its current state to handle an event, and the state can change the context's current state if necessary.

\subsubsection{Advantages and Disadvantages}

\textbf{Advantages}:
\begin{itemize}
    \item \textbf{Simplifies complex conditional logic}: The State Pattern helps in avoiding large, nested conditional statements (e.g., \texttt{if} or \texttt{switch} statements) by delegating the behavior to state classes.
    \item \textbf{Encapsulates state-specific behavior}: State-specific logic is kept within individual classes, making the code more modular and easier to maintain.
    \item \textbf{Easily extendable}: Adding a new state involves simply creating a new state class without altering the existing state classes or the context class.
\end{itemize}

\textbf{Disadvantages}:
\begin{itemize}
    \item \textbf{More classes}: The State Pattern can increase the number of classes as each state requires a separate class.
    \item \textbf{Increased complexity}: While it simplifies conditional logic, it introduces more complexity in terms of class management.
\end{itemize}

\subsubsection{Use Cases}
The State Pattern is ideal for scenarios where an object needs to change its behavior based on its state. Some common use cases include:

\begin{itemize}
    \item \textbf{UI components}: Widgets that have different states (e.g., enabled, disabled, focused, etc.) with different behaviors.
    \item \textbf{Media Players}: Media players can be in states like playing, paused, or stopped, and their behavior changes accordingly.
    \item \textbf{Network Connections}: Network connections may transition between states such as connecting, connected, and disconnected, with each state having different behaviors.
\end{itemize}

\subsubsection{Code Example}
Below is an example of the State Pattern in Python using a simple media player that can be in three states: Playing, Paused, and Stopped.

\begin{lstlisting}[style=python]
from abc import ABC, abstractmethod

# State Interface
class State(ABC):
    @abstractmethod
    def press_play(self, player):
        pass

    @abstractmethod
    def press_pause(self, player):
        pass

    @abstractmethod
    def press_stop(self, player):
        pass

# Concrete States
class PlayingState(State):
    def press_play(self, player):
        print("Already playing.")
    
    def press_pause(self, player):
        print("Pausing the player.")
        player.state = PausedState()
    
    def press_stop(self, player):
        print("Stopping the player.")
        player.state = StoppedState()

class PausedState(State):
    def press_play(self, player):
        print("Resuming playback.")
        player.state = PlayingState()
    
    def press_pause(self, player):
        print("Already paused.")
    
    def press_stop(self, player):
        print("Stopping the player.")
        player.state = StoppedState()

class StoppedState(State):
    def press_play(self, player):
        print("Starting playback.")
        player.state = PlayingState()
    
    def press_pause(self, player):
        print("Can't pause. The player is stopped.")
    
    def press_stop(self, player):
        print("Already stopped.")

# Context
class MediaPlayer:
    def __init__(self):
        self.state = StoppedState()
    
    def press_play(self):
        self.state.press_play(self)
    
    def press_pause(self):
        self.state.press_pause(self)
    
    def press_stop(self):
        self.state.press_stop(self)

# Usage
player = MediaPlayer()

player.press_play()   # Output: Starting playback.
player.press_pause()  # Output: Pausing the player.
player.press_play()   # Output: Resuming playback.
player.press_stop()   # Output: Stopping the player.
\end{lstlisting}

In this example, the `MediaPlayer' is the context that delegates its behavior to the current state (Playing, Paused, or Stopped). The behavior changes dynamically based on the state.

\subsubsection{Pattern Extensions}
The State Pattern can be extended in several ways:

\begin{itemize}
    \item \textbf{State transitions with conditions}: States can transition conditionally based on specific events or external conditions.
    \item \textbf{Shared behavior between states}: If there are common behaviors between states, they can be extracted into a shared parent class to avoid duplication.
    \item \textbf{State persistence}: In some cases, it might be necessary to persist the current state (e.g., saving to a file or database) and restore it later.
\end{itemize}

\subsection{Strategy Pattern}

The \textbf{Strategy Pattern} \cite{christopoulou2012automated} is a behavioral design pattern that allows you to define a family of algorithms, encapsulate each one, and make them interchangeable. The main idea is to separate the algorithm from the object that uses it, allowing the algorithm to change independently of the object. This pattern is especially useful when a class needs to switch between different behaviors or algorithms based on the context or user input.

\subsubsection{Motivation}
Imagine you're developing a program that calculates the price of a product. The pricing strategy may vary: sometimes you might apply a discount, sometimes a tax, and other times you might need to apply both. Hardcoding all possible strategies into one class would make the code difficult to maintain and extend.

With the \textbf{Strategy Pattern}, you can define different pricing strategies and swap them out as needed without altering the rest of your code. This improves flexibility and adheres to the \textbf{Open/Closed Principle} of SOLID design principles: objects should be open for extension but closed for modification.

\subsubsection{Structure}
The structure of the Strategy Pattern consists of three primary components:

\begin{itemize}
    \item \textbf{Context}: This is the class that requires the strategy to perform an operation. The context doesn't implement the algorithm but delegates it to a strategy object.
    \item \textbf{Strategy Interface}: An interface common to all concrete strategies, allowing the context to use them interchangeably.
    \item \textbf{Concrete Strategies}: These are the actual implementations of the algorithm or behavior that the context can choose from.
\end{itemize}

The following diagram illustrates the structure of the Strategy Pattern:

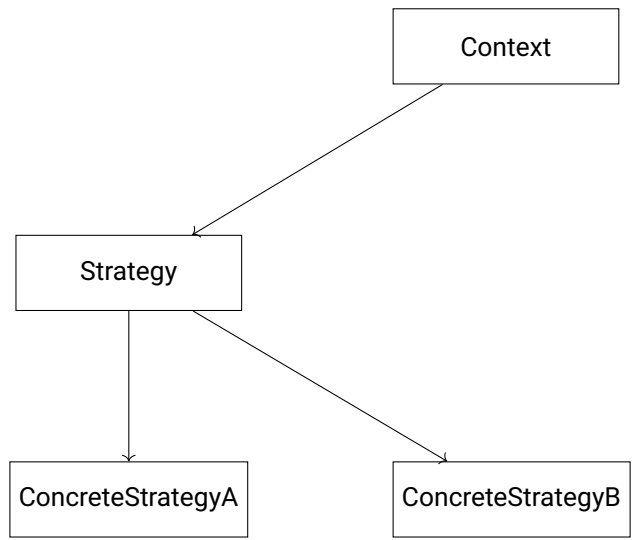
\begin{center}
\begin{tikzpicture}[node distance=2.5cm, auto]
    \node (context) [draw, rectangle, minimum width=3cm, minimum height=1cm] {Context};
    \node (strategy) [below left=2cm and 2cm of context, draw, rectangle, minimum width=3cm, minimum height=1cm] {Strategy};
    \node (concrete_strategy1) [below=2cm of strategy, draw, rectangle, minimum width=3cm, minimum height=1cm] {ConcreteStrategyA};
    \node (concrete_strategy2) [below right=2cm and 2cm of strategy, draw, rectangle, minimum width=3cm, minimum height=1cm] {ConcreteStrategyB};

    \draw[->] (context) -- (strategy);
    \draw[->] (strategy) -- (concrete_strategy1);
    \draw[->] (strategy) -- (concrete_strategy2);
\end{tikzpicture}
\end{center}

\subsubsection{Participants}
The main participants in the Strategy Pattern are:

\begin{itemize}
    \item \textbf{Context}: The class that contains a reference to the Strategy object and delegates the task to the strategy.
    \item \textbf{Strategy}: An interface that defines the contract for the algorithm. Concrete strategies will implement this interface.
    \item \textbf{ConcreteStrategy}: These are the specific implementations of the strategy that perform the actual algorithm. For example, you might have a \texttt{DiscountStrategy}, \texttt{TaxStrategy}, or a \texttt{CombinedStrategy}.
\end{itemize}

\subsubsection{How It Works}
The \textbf{Strategy Pattern} works by allowing the `Context' to hold a reference to a `Strategy' object. The context delegates work to the strategy rather than implementing it directly. This allows the strategy to be swapped out at runtime, and the `Context' doesn't need to know the specific details of the strategy's implementation.

For example, in a shopping cart system, you might want to apply different discount strategies based on the user's membership status or time of year. Instead of having a large `if-else' block to handle different discount rules, you define different discount strategies that implement a common interface. The `ShoppingCart' class will then be able to apply any discount strategy without being tightly coupled to the specific details.

\subsubsection{Advantages and Disadvantages}

\textbf{Advantages:}

\begin{itemize}
    \item \textbf{Flexibility}: You can switch strategies at runtime based on the context.
    \item \textbf{Clean Code}: It eliminates complex conditionals that determine which algorithm to use. This makes the code more maintainable.
    \item \textbf{Extensibility}: New strategies can be added without changing the context, adhering to the Open/Closed Principle.
\end{itemize}

\textbf{Disadvantages:}

\begin{itemize}
    \item \textbf{Increased Complexity}: Introducing strategy objects adds complexity because you need to manage different classes and strategies.
    \item \textbf{Overhead}: If the strategy needs to change infrequently, this pattern might add unnecessary overhead.
\end{itemize}

\subsubsection{Use Cases}
The Strategy Pattern is useful in situations where:

\begin{itemize}
    \item You have a family of algorithms or behaviors and need to switch between them at runtime.
    \item You want to eliminate conditionals that select different behaviors or algorithms.
    \item You need to be able to easily add new behaviors without changing the existing code.
\end{itemize}

Some practical examples include:

\begin{itemize}
    \item Different sorting algorithms: quicksort, mergesort, etc.
    \item Payment methods in an e-commerce application: credit card, PayPal, or bank transfer.
    \item Discount strategies in a shopping cart: percentage discount, fixed discount, or seasonal discount.
\end{itemize}

\subsubsection{Code Example}
Let’s illustrate the \textbf{Strategy Pattern} with a concrete Python example. In this case, we’ll implement a simple shopping cart system where different discount strategies can be applied.

\begin{lstlisting}[style=python]
from abc import ABC, abstractmethod

# Strategy Interface
class DiscountStrategy(ABC):
    @abstractmethod
    def apply_discount(self, price: float) -> float:
        pass

# Concrete Strategy 1: No Discount
class NoDiscountStrategy(DiscountStrategy):
    def apply_discount(self, price: float) -> float:
        return price

# Concrete Strategy 2: Percentage Discount
class PercentageDiscountStrategy(DiscountStrategy):
    def __init__(self, percentage: float):
        self.percentage = percentage

    def apply_discount(self, price: float) -> float:
        return price - (price * self.percentage / 100)

# Concrete Strategy 3: Fixed Discount
class FixedDiscountStrategy(DiscountStrategy):
    def __init__(self, discount: float):
        self.discount = discount

    def apply_discount(self, price: float) -> float:
        return price - self.discount

# Context
class ShoppingCart:
    def __init__(self, discount_strategy: DiscountStrategy):
        self.discount_strategy = discount_strategy

    def calculate_total(self, price: float) -> float:
        return self.discount_strategy.apply_discount(price)

# Client code
cart = ShoppingCart(PercentageDiscountStrategy(10))
total_price = cart.calculate_total(100)
print(f"Total price with discount: {total_price}")

cart = ShoppingCart(FixedDiscountStrategy(20))
total_price = cart.calculate_total(100)
print(f"Total price with fixed discount: {total_price}")

cart = ShoppingCart(NoDiscountStrategy())
total_price = cart.calculate_total(100)
print(f"Total price without discount: {total_price}")
\end{lstlisting}

In this example:
\begin{itemize}
    \item The `DiscountStrategy' is the strategy interface, which defines the `apply\_discount' method.
    \item `NoDiscountStrategy', `PercentageDiscountStrategy', and `FixedDiscountStrategy' are concrete implementations of the strategy.
    \item `ShoppingCart' is the context that uses a `DiscountStrategy' to calculate the total price.
\end{itemize}

\subsubsection{Pattern Extensions}
The Strategy Pattern can be extended in various ways:

\begin{itemize}
    \item \textbf{Dynamic Strategy}: You can modify the context to allow changing strategies at runtime, making the system even more flexible.
    \item \textbf{Composite Strategy}: Combine multiple strategies to execute a series of algorithms or behaviors.
    \item \textbf{Strategy with Parameters}: Pass parameters to the strategy methods to make them more dynamic and flexible.
\end{itemize}

For example, you can enhance the shopping cart example by adding the ability to choose a combination of discounts, such as a percentage discount followed by a fixed discount, thus implementing a \textbf{Composite Strategy}.

\subsection{Template Method Pattern}

The \textbf{Template Method Pattern} \cite{lyardet1997dynamic} is a behavioral design pattern that defines the skeleton of an algorithm in a method, deferring some steps to subclasses. It allows subclasses to redefine certain steps of the algorithm without changing its structure. This pattern is commonly used when a generic workflow or algorithm exists, but parts of it need to be customized by different clients or subclasses.

\subsubsection{Motivation}
The Template Method Pattern is useful when multiple classes share a common algorithm structure, but certain steps in the algorithm are meant to vary. Instead of duplicating the entire algorithm in each subclass, we define the overall algorithm in a base class and allow subclasses to modify only the necessary parts. This promotes code reuse and maintains the integrity of the overall structure of the algorithm.

Consider a scenario where we are developing a system that prepares different types of documents such as Word documents, PDFs, and web reports. Each document type has the same general steps: opening the file, writing content, formatting it, and saving it. However, the specific details for each of these steps vary depending on the type of document. The Template Method Pattern allows us to implement this generic flow once and leave the specific details to be defined by subclasses.

\subsubsection{Structure}

The Template Method Pattern has the following structure:

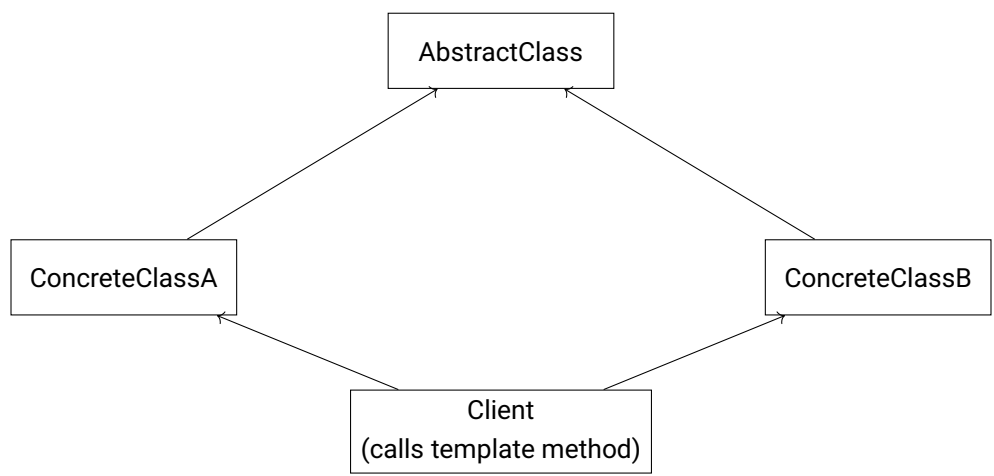
\tikzstyle{class} = [rectangle, draw, text centered, minimum height=2em, minimum width=3cm]

\begin{center}
\begin{tikzpicture}[node distance=2.5cm, auto]
    \node[draw, rectangle, minimum width=3cm, minimum height=1cm, text centered] (AbstractClass) {AbstractClass};
    \node[draw, rectangle, below left=2cm and 2cm of AbstractClass, minimum width=3cm, minimum height=1cm, text centered] (ConcreteClassA) {ConcreteClassA};
    \node[draw, rectangle, below right=2cm and 2cm of AbstractClass, minimum width=3cm, minimum height=1cm, text centered] (ConcreteClassB) {ConcreteClassB};

    \draw[->] (ConcreteClassA) -- (AbstractClass);
    \draw[->] (ConcreteClassB) -- (AbstractClass);

    \node[draw, rectangle, below=4cm of AbstractClass, align=center, minimum width=4cm, minimum height=1cm] (client) {Client\\(calls template method)};
    \draw[->] (client) -- (ConcreteClassA);
    \draw[->] (client) -- (ConcreteClassB);
\end{tikzpicture}
\end{center}

In this diagram:
\begin{itemize}
    \item \textbf{AbstractClass} defines the skeleton of the algorithm, represented by a method called the "template method." This method defines the steps of the algorithm, some of which are implemented in the abstract class and some of which are deferred to subclasses.
    \item \textbf{ConcreteClassA} and \textbf{ConcreteClassB} are subclasses that provide specific implementations of the steps defined by the abstract class.
\end{itemize}

\subsubsection{Participants}
The key participants in the Template Method Pattern are:
\begin{itemize}
    \item \textbf{AbstractClass}: Defines the template method and may also provide default implementations for some of the algorithm steps.
    \item \textbf{ConcreteClass}: Implements the abstract methods defined by AbstractClass to complete the specific details of the algorithm.
\end{itemize}

\subsubsection{How It Works}
1. The client calls the template method defined in the abstract class.
2. The template method executes the algorithm's skeleton, calling abstract methods that subclasses must implement.
3. Subclasses provide specific implementations of the abstract methods, which are integrated into the overall algorithm when the template method is executed.

\subsubsection{Advantages and Disadvantages}

\textbf{Advantages}:
\begin{itemize}
    \item \textbf{Code reuse}: The overall structure of the algorithm is defined once in the abstract class, and only specific steps are modified by subclasses. This avoids code duplication.
    \item \textbf{Ease of maintenance}: Changes to the overall algorithm need to be made in only one place (the abstract class), ensuring consistency across subclasses.
    \item \textbf{Flexibility}: Subclasses can override only the steps they need to change, leaving the rest of the algorithm intact.
\end{itemize}

\textbf{Disadvantages}:
\begin{itemize}
    \item \textbf{Rigid structure}: The overall structure of the algorithm is fixed, which might not be suitable for all cases where more flexibility is needed.
    \item \textbf{Complexity for beginners}: The pattern can be a bit abstract for beginners to grasp, especially with its reliance on inheritance and deferred implementation of methods.
\end{itemize}

\subsubsection{Use Cases}
The Template Method Pattern is commonly used in the following scenarios:

\begin{itemize}
    \item \textbf{Framework development}: When developing frameworks that require certain steps in a process to be standardized, but allow users of the framework to customize specific parts.
    \item \textbf{Algorithms with fixed structures}: When an algorithm has a clear structure but requires some flexibility in its implementation, such as generating reports or processing data in different formats.
    \item \textbf{Game development}: When creating a game where the core game loop is the same but the behavior of individual game elements can vary.
\end{itemize}

\subsubsection{Code Example}

Here’s a simple Python example of the Template Method Pattern where different types of documents (PDF and Word) follow a general workflow of preparation but implement their specific content formatting.

\begin{lstlisting}[style=python]
from abc import ABC, abstractmethod

class Document(ABC):
    def prepare_document(self):
        self.open_file()
        self.write_content()
        self.format_content()
        self.save_file()

    @abstractmethod
    def open_file(self):
        pass

    @abstractmethod
    def write_content(self):
        pass

    @abstractmethod
    def format_content(self):
        pass

    def save_file(self):
        print("Saving the document.")

class PDFDocument(Document):
    def open_file(self):
        print("Opening a PDF document.")

    def write_content(self):
        print("Writing content to the PDF document.")

    def format_content(self):
        print("Formatting the PDF document content.")

class WordDocument(Document):
    def open_file(self):
        print("Opening a Word document.")

    def write_content(self):
        print("Writing content to the Word document.")

    def format_content(self):
        print("Formatting the Word document content.")

# Client code
pdf_doc = PDFDocument()
pdf_doc.prepare_document()

print("\n")

word_doc = WordDocument()
word_doc.prepare_document()
\end{lstlisting}

In this code:
\begin{itemize}
    \item The \texttt{Document} class defines the template method \texttt{prepare\_document}, which outlines the workflow.
    \item \texttt{PDFDocument} and \texttt{WordDocument} provide specific implementations for the methods \texttt{open\_file}, \texttt{write\_content}, and \texttt{format\_content}.
    \item The \texttt{save\_file} method has a default implementation, but subclasses can override it if needed.
\end{itemize}

Output:

\begin{lstlisting}[style=cmd]
Opening a PDF document.
Writing content to the PDF document.
Formatting the PDF document content.
Saving the document.

Opening a Word document.
Writing content to the Word document.
Formatting the Word document content.
Saving the document.
\end{lstlisting}

\subsubsection{Pattern Extensions}
The Template Method Pattern can be extended and adapted in various ways:

\begin{itemize}
    \item \textbf{Hook methods}: Provide optional methods in the abstract class that can be overridden by subclasses if needed. These methods are often left empty in the abstract class.
    \item \textbf{Customization points}: Allow more customization by adding more abstract methods in the base class, giving subclasses finer control over the algorithm.
    \item \textbf{Combining with other patterns}: The Template Method Pattern can be combined with other design patterns like the Strategy or Factory patterns to add more flexibility and customization.
\end{itemize}

\subsection{Visitor Pattern}
The Visitor Pattern \cite{palsberg1998essence} is a behavioral design pattern that allows you to add further operations to objects without modifying their structure. It involves a visitor object that "visits" elements in a structure, enabling new functionality to be added while keeping the object’s code unchanged. This pattern is particularly useful when dealing with composite objects or collections of objects where you need to perform various operations.

\subsubsection{Motivation}
The main motivation for using the Visitor Pattern is when you want to define new operations without changing the classes of the elements on which it operates. In object-oriented design, adding new functionality often requires modifying classes, which can be cumbersome, especially if these classes are part of a complex hierarchy or if they are already in use in different parts of the system. 

The Visitor Pattern decouples operations from the objects, enabling you to add new operations without altering the objects themselves. This becomes beneficial when the object structure rarely changes, but the operations on these objects are frequently updated.

\textbf{Example Motivation:}
Suppose you have different shapes in a drawing application, such as circles, rectangles, and triangles. You might want to calculate their area, draw them, or export them to different formats. Instead of modifying the shape classes every time you want to add new functionality (e.g., exporting to a new format), you can use the Visitor Pattern to separate these operations from the shapes themselves.

\subsubsection{Structure}
The structure of the Visitor Pattern involves several key participants:

\tikzstyle{class} = [rectangle, draw, text centered, minimum height=3em]
\tikzstyle{arrow} = [thick,->,>=stealth]

The components of the Visitor Pattern include:

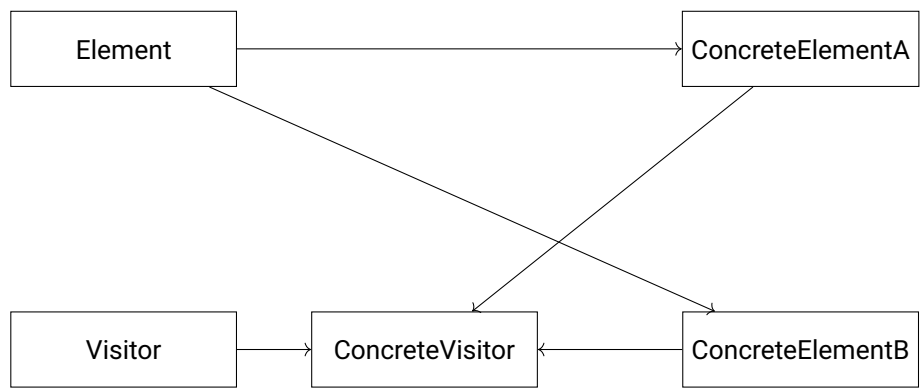
\begin{center}
\begin{tikzpicture}[node distance=3cm, auto]
    \node (element) [rectangle, draw, minimum width=3cm, minimum height=1cm, text centered] {Element};
    \node (visitor) [rectangle, draw, below of=element, yshift=-1cm, minimum width=3cm, minimum height=1cm, text centered] {Visitor};
    \node (concreteelementA) [rectangle, draw, right of=element, xshift=6cm, minimum width=3cm, minimum height=1cm, text centered] {ConcreteElementA};
    \node (concreteelementB) [rectangle, draw, below of=concreteelementA, yshift=-1cm, minimum width=3cm, minimum height=1cm, text centered] {ConcreteElementB};
    \node (concretevisitor) [rectangle, draw, right of=visitor, xshift=1cm, minimum width=3cm, minimum height=1cm, text centered] {ConcreteVisitor};
    
    \draw[->] (element) -- (concreteelementA);
    \draw[->] (element) -- (concreteelementB);
    \draw[->] (visitor) -- (concretevisitor);
    \draw[->] (concreteelementA) -- (concretevisitor);
    \draw[->] (concreteelementB) -- (concretevisitor);
\end{tikzpicture}
\end{center}

\subsubsection{Participants}

\begin{itemize}
    \item \textbf{Element (e.g., Shape)}: Declares an \texttt{accept} method that allows a visitor object to visit it. For instance, a \texttt{Shape} class might define an \texttt{accept} method for allowing various operations (like \texttt{DrawVisitor} or \texttt{ExportVisitor}).
    \item \textbf{ConcreteElement (e.g., Circle, Rectangle)}: Implements the \texttt{accept} method. Each concrete element allows the visitor to perform operations specific to that element.
    \item \textbf{Visitor (e.g., Visitor)}: An abstract class or interface that declares visiting methods for all types of concrete elements.
    \item \textbf{ConcreteVisitor (e.g., DrawVisitor, ExportVisitor)}: Implements the visitor interface and provides specific functionality for each type of element.
\end{itemize}

\subsubsection{How It Works}

The Visitor Pattern works by having the elements (such as objects in a collection or hierarchy) accept a visitor object. The visitor then "visits" the elements and performs operations on them. This process allows you to add new operations by simply creating new visitor classes without altering the existing elements.

\begin{enumerate}
    \item The object (e.g., a shape) calls its own \texttt{accept} method and passes the visitor to it.
    \item The visitor has a method for each concrete element and performs operations on the object.
\end{enumerate}

For example, consider a `Shape' interface with concrete implementations like `Circle' and `Rectangle'. Each shape will have an accept method to allow a visitor to "visit" and apply operations like drawing or exporting.

\begin{lstlisting}[style=python]
# Element Interface
class Shape:
    def accept(self, visitor):
        pass

# Concrete Element A
class Circle(Shape):
    def accept(self, visitor):
        visitor.visit_circle(self)

# Concrete Element B
class Rectangle(Shape):
    def accept(self, visitor):
        visitor.visit_rectangle(self)

# Visitor Interface
class ShapeVisitor:
    def visit_circle(self, circle):
        pass

    def visit_rectangle(self, rectangle):
        pass

# Concrete Visitor: Drawing shapes
class DrawVisitor(ShapeVisitor):
    def visit_circle(self, circle):
        print("Drawing a circle")

    def visit_rectangle(self, rectangle):
        print("Drawing a rectangle")

# Concrete Visitor: Exporting shapes
class ExportVisitor(ShapeVisitor):
    def visit_circle(self, circle):
        print("Exporting a circle to SVG")

    def visit_rectangle(self, rectangle):
        print("Exporting a rectangle to PNG")

# Client code
shapes = [Circle(), Rectangle()]

draw_visitor = DrawVisitor()
export_visitor = ExportVisitor()

for shape in shapes:
    shape.accept(draw_visitor)
    shape.accept(export_visitor)
\end{lstlisting}

In this example, you can easily extend the functionality to include new operations by creating new visitors, such as a `SaveVisitor' or `TransformVisitor', without modifying the `Shape` classes.

\subsubsection{Advantages and Disadvantages}
\textbf{Advantages:}

\begin{itemize}
    \item \textbf{Separation of Concerns}: The Visitor Pattern separates the algorithm from the object structure, making the code cleaner and easier to maintain.
    \item \textbf{Easy to Extend}: You can add new operations to existing object structures without modifying the structure.
    \item \textbf{Simplifies Complex Operations}: Complex operations on composite objects can be simplified by putting them in visitor classes.
\end{itemize}

\textbf{Disadvantages:}
\begin{itemize}
    \item \textbf{Difficult to Add New Element Types}: If you need to add new types of elements (like a \texttt{Triangle} in the shape example), you will have to modify all existing visitors to handle the new type, making the pattern less flexible in this regard.
    \item \textbf{Increased Complexity}: For simple object structures, the Visitor Pattern may introduce unnecessary complexity.
\end{itemize}

\subsubsection{Use Cases}
\begin{itemize}
    \item \textbf{When the structure rarely changes}: The pattern is most useful when the object structure (like classes and their relationships) remains stable but the operations performed on these objects change frequently.
    \item \textbf{When you need to add new behavior to a class hierarchy}: If adding new methods directly to the class hierarchy is impractical (e.g., due to design constraints or code ownership), the Visitor Pattern can help.
    \item \textbf{Working with Composite Structures}: For object structures like file systems, document trees, or graphics shapes, where multiple operations need to be performed.
\end{itemize}

Examples:
\begin{itemize}
    \item Compilers, where you need to perform different operations like type checking, optimization, or code generation on the abstract syntax tree (AST).
    \item GUI frameworks where you need to perform actions on various UI components, such as rendering, event handling, or applying themes.
\end{itemize}

\subsubsection{Code Example}
The following Python code demonstrates a simple implementation of the Visitor Pattern where we use visitors to draw and export shapes.

\begin{lstlisting}[style=python]
# Element Interface
class Shape:
    def accept(self, visitor):
        pass

# Concrete Element A
class Circle(Shape):
    def accept(self, visitor):
        visitor.visit_circle(self)

# Concrete Element B
class Rectangle(Shape):
    def accept(self, visitor):
        visitor.visit_rectangle(self)

# Visitor Interface
class ShapeVisitor:
    def visit_circle(self, circle):
        pass

    def visit_rectangle(self, rectangle):
        pass

# Concrete Visitor: Drawing shapes
class DrawVisitor(ShapeVisitor):
    def visit_circle(self, circle):
        print("Drawing a circle")

    def visit_rectangle(self, rectangle):
        print("Drawing a rectangle")

# Concrete Visitor: Exporting shapes
class ExportVisitor(ShapeVisitor):
    def visit_circle(self, circle):
        print("Exporting a circle to SVG")

    def visit_rectangle(self, rectangle):
        print("Exporting a rectangle to PNG")

# Client code
shapes = [Circle(), Rectangle()]

draw_visitor = DrawVisitor()
export_visitor = ExportVisitor()

for shape in shapes:
    shape.accept(draw_visitor)
    shape.accept(export_visitor)
\end{lstlisting}

\subsubsection{Pattern Extensions}

\begin{itemize}
    \item \textbf{Double Dispatch}: The Visitor Pattern is often used in languages like Python, where it achieves double dispatch—this means the operation executed depends on both the visitor and the element it is applied to.
    \item \textbf{Reflection}: In dynamic languages like Python, you can use reflection to avoid explicitly defining visit methods for each type of element, making it more flexible.
\end{itemize}

\section{Concurrency Patterns}
\subsection{Introduction}

Concurrency patterns \cite{soares2001concurrency} deal with the design challenges that arise when multiple processes or threads run in parallel within a system. These patterns are particularly important in modern computing environments where applications need to perform multiple tasks simultaneously, improving efficiency and performance. Concurrency in Python can be handled through threads, processes, or asynchronous programming techniques, but it also introduces complexity, such as race conditions, deadlocks, and synchronization issues. Concurrency patterns aim to provide solutions to these problems by structuring code in a way that allows safe and efficient parallel execution.

In Python, concurrency patterns help manage tasks that need to happen at the same time, whether across multiple CPU cores or on the same core by interleaving tasks. By using these patterns, we can avoid common pitfalls such as data corruption, inefficient resource usage, or sluggish performance.

\subsubsection{Key Concepts in Concurrency Patterns}
Before diving into specific concurrency patterns, there are a few important concepts to understand:
\begin{itemize}
    \item \textbf{Thread Safety:} Ensuring that shared resources are accessed and modified safely across multiple threads is crucial in concurrent programming. Concurrency patterns often include mechanisms to ensure that only one thread accesses shared data at a time, avoiding conflicts.
    \item \textbf{Synchronization:} This is the process of controlling the order in which tasks are executed, ensuring that certain tasks do not start until others have been completed. Synchronization helps prevent issues such as race conditions and inconsistent states.
    \item \textbf{Asynchronous Execution:} Asynchronous programming allows tasks to be performed without waiting for other tasks to complete, promoting more efficient use of resources, especially in I/O-bound programs.
    \item \textbf{Parallelism vs Concurrency:} While concurrency refers to the ability of a system to handle multiple tasks at once, parallelism means executing multiple tasks simultaneously. Concurrency patterns help balance and manage both concepts in your program.
\end{itemize}

\subsubsection{Common Concurrency Patterns}
Several concurrency patterns are commonly used in Python to manage multiple tasks running at the same time. These patterns help structure the execution of tasks so that the system can remain responsive and avoid the issues that arise with concurrent execution. Some of the most frequently used concurrency patterns include:
\begin{itemize}
    \item \textbf{Thread Pool Pattern:} This pattern uses a pool of threads to perform tasks concurrently, managing the threads automatically so that new threads do not need to be created for each task, which reduces overhead and improves performance.
    \item \textbf{Producer-Consumer Pattern:} This pattern involves a set of producer tasks that generate data and a set of consumer tasks that process this data. A common data structure, such as a queue, is used to coordinate between the producers and consumers.
    \item \textbf{Future Pattern:} This pattern allows the result of an operation to be computed asynchronously, returning a "future" object that can be used to retrieve the result once it is available. This is commonly used in asynchronous programming.
    \item \textbf{Active Object Pattern:} This pattern decouples method execution from method invocation by using a queue to store method calls and a separate thread to execute them, ensuring that tasks are processed in a thread-safe manner.
    \item \textbf{Reactor Pattern:} The reactor pattern is used in event-driven programming, where a single thread waits for events to occur and then dispatches them to appropriate event handlers, allowing for highly scalable, I/O-bound programs.
    \item \textbf{Barrier Pattern:} This pattern ensures that multiple threads or tasks reach a specific point before any of them can proceed, helping to synchronize complex parallel operations.
\end{itemize}

Each of these patterns addresses a specific challenge in concurrent programming, making it easier to manage parallel tasks safely and efficiently. Throughout this book, we will explore these patterns in detail with beginner-friendly Python examples, explaining how to apply them in real-world scenarios.

Concurrency patterns are essential for building responsive, scalable, and efficient applications in Python. By mastering these patterns, you'll be able to write programs that take full advantage of modern hardware and multi-core systems, while avoiding the common pitfalls that arise when multiple tasks run simultaneously.

\subsection{Thread Pool Pattern}

The Thread Pool Pattern \cite{birrell1989introduction} is a design pattern used to manage and control a group of reusable threads that perform tasks concurrently. Instead of creating a new thread every time a task needs to be executed, a pool of worker threads is maintained, which reduces the overhead associated with thread creation and termination. The thread pool manages the lifecycle of threads, allowing for better performance, especially when handling multiple tasks that do not require dedicated threads.

\subsubsection{Motivation}

The main motivation behind the Thread Pool Pattern is to improve performance by reusing threads instead of constantly creating and destroying them. Thread creation can be expensive in terms of time and system resources, particularly in environments where many threads are created and destroyed rapidly. In a thread pool, a predefined number of threads are created once and reused to execute tasks, minimizing the cost associated with frequent thread management. This is particularly useful for applications that need to handle numerous short-lived tasks or that experience high loads intermittently.

For example, in a web server, each incoming request can be processed by a thread. Rather than creating a new thread for every request, a thread pool can be used to efficiently manage the available threads and assign them to requests as they come in.

\subsubsection{Structure}

The structure of the Thread Pool Pattern includes:

\begin{itemize}
    \item \textbf{Thread Pool Manager}: The component responsible for managing the pool of threads and assigning tasks to available threads.
    \item \textbf{Worker Threads}: The threads that perform tasks assigned to them by the Thread Pool Manager.
    \item \item \textbf{Task Queue}: A queue where tasks are submitted and wait to be processed by available worker threads.
\end{itemize}

The structure of a basic thread pool can be visualized as follows:

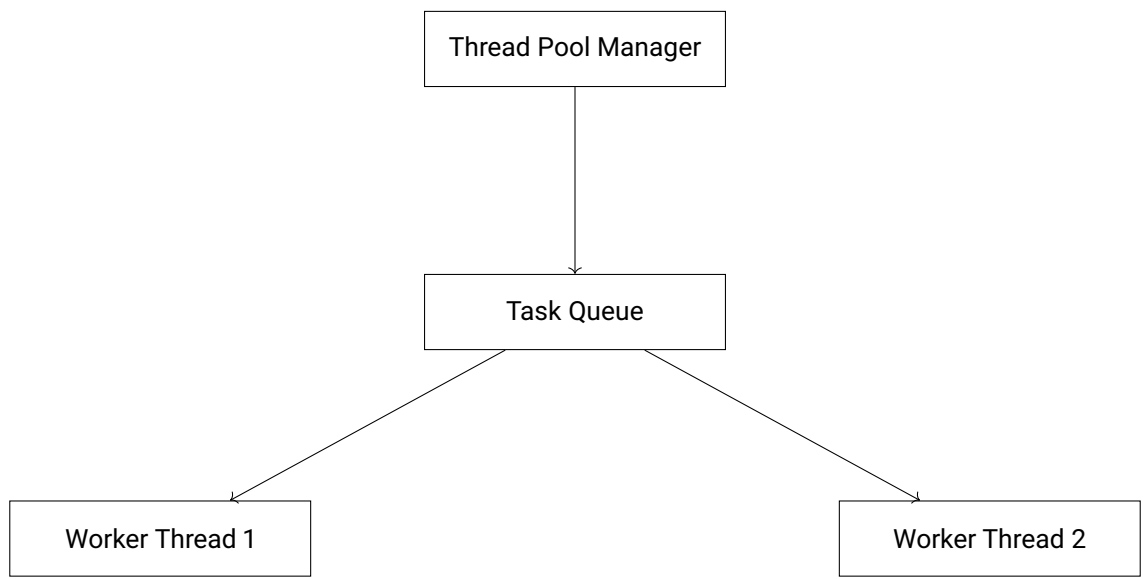
\begin{center}
\begin{tikzpicture}[node distance=3cm and 2cm, auto]
    \node[draw, rectangle, minimum width=4cm, minimum height=1cm, text centered] (manager) {Thread Pool Manager};
    \node[draw, rectangle, below of=manager, minimum width=4cm, minimum height=1cm, text centered, yshift=-0.5cm] (queue) {Task Queue};
    \node[draw, rectangle, below left=2cm and 1.5cm of queue, minimum width=4cm, minimum height=1cm, text centered] (worker1) {Worker Thread 1};
    \node[draw, rectangle, below right=2cm and 1.5cm of queue, minimum width=4cm, minimum height=1cm, text centered] (worker2) {Worker Thread 2};

    \draw[->] (queue) -- (worker1);
    \draw[->] (queue) -- (worker2);
    \draw[->] (manager) -- (queue);
\end{tikzpicture}
\end{center}

\subsubsection{Participants}

The key participants in the Thread Pool Pattern are:

\begin{itemize}
    \item \textbf{Thread Pool Manager}: This component oversees the creation and management of threads. It ensures that there are always a set number of threads available to process tasks and that the task queue is properly managed.
    \item \textbf{Worker Threads}: These are the actual threads that execute the tasks. Once a thread completes its assigned task, it becomes available to take on another task from the task queue.
    \item \textbf{Task Queue}: This is a queue where incoming tasks are placed. The Thread Pool Manager takes tasks from the queue and assigns them to worker threads.
\end{itemize}

\subsubsection{How It Works}

The Thread Pool Pattern works by keeping a fixed or dynamic number of threads in a pool. When a task is submitted to the thread pool, it is added to a task queue. The Thread Pool Manager assigns tasks to available worker threads as they become free. The number of threads can be adjusted dynamically, or it can remain constant depending on the implementation.

\begin{enumerate}
    \item \textbf{Initialization}: When the thread pool is created, a predefined number of threads are initialized and placed into the pool. These threads remain idle until a task is assigned to them.
    \item \textbf{Task Submission}: Tasks are submitted to the pool, usually through a task queue. Each task waits in the queue until a worker thread becomes available.
    \item \textbf{Task Execution}: Once a worker thread is free, the Thread Pool Manager assigns a task from the queue to the thread. The thread then executes the task.
    \item \textbf{Thread Reuse}: After the thread completes its task, it goes back to the pool and waits for another task to be assigned.
\end{enumerate}

\subsubsection{Advantages and Disadvantages}

\textbf{Advantages:}

\begin{itemize}
    \item \textbf{Efficiency}: Reduces the overhead of frequent thread creation and destruction by reusing existing threads.
    \item \textbf{Scalability}: Can efficiently manage multiple tasks concurrently, which is essential in high-load applications like web servers.
    \item \textbf{Control}: Provides better control over system resources by limiting the number of threads, preventing system overload.
\end{itemize}

\textbf{Disadvantages:}

\begin{itemize}
    \item \textbf{Resource Limitation}: If the number of worker threads is too small, tasks may spend too much time waiting in the queue.
    \item \textbf{Complexity}: Managing a thread pool can add complexity to an application, particularly when tasks vary significantly in execution time.
    \item \textbf{Starvation}: If certain tasks take too long, other tasks may get starved and not execute promptly.
\end{itemize}

\subsubsection{Use Cases}

Some common use cases for the Thread Pool Pattern include:

\begin{itemize}
    \item \textbf{Web Servers}: Handling multiple incoming requests concurrently without creating a new thread for each request.
    \item \textbf{Database Servers}: Managing multiple queries simultaneously, efficiently allocating threads to process different queries.
    \item \textbf{Background Tasks}: Running periodic background tasks or scheduled jobs in an application, where tasks are assigned to a limited set of threads.
\end{itemize}

\subsubsection{Code Example}

Here is a basic example of a thread pool implementation in Python using the `concurrent.futures.ThreadPoolExecutor' module:

\begin{lstlisting}[style=python]
import concurrent.futures
import time

# Define a task to be executed by the thread pool
def task(name):
    print(f"Task {name} is starting")
    time.sleep(2)  # Simulate a time-consuming operation
    print(f"Task {name} is completed")

# Create a thread pool with 3 worker threads
with concurrent.futures.ThreadPoolExecutor(max_workers=3) as executor:
    # Submit multiple tasks to the thread pool
    futures = [executor.submit(task, i) for i in range(5)]
    
    # Wait for all tasks to be completed
    for future in concurrent.futures.as_completed(futures):
        future.result()  # This will raise an exception if the task raised one
\end{lstlisting}

In this example:

\begin{itemize}
    \item We create a thread pool with 3 worker threads using \texttt{ThreadPoolExecutor}.
    \item Five tasks are submitted to the thread pool.
    \item Each task simulates a time-consuming operation by sleeping for 2 seconds.
    \item The pool assigns tasks to the threads, and tasks are executed concurrently.
\end{itemize}

\subsubsection{Pattern Extensions}

Some extensions to the basic Thread Pool Pattern include:

\begin{itemize}
    \item \textbf{Dynamic Thread Pools}: Adjust the number of worker threads based on the system load. For instance, if the queue grows too large, new threads can be created to handle the load.
    \item \textbf{Priority Task Queues}: In some applications, certain tasks may need to be processed before others. A priority queue can be introduced to ensure high-priority tasks are executed first.
    \item \textbf{Timeouts and Cancellation}: Tasks may need to be canceled or timeout if they take too long. Adding timeouts or cancellation capabilities to tasks within the thread pool can make the system more robust.
\end{itemize}

\subsection{Executor Pattern}
The Executor pattern \cite{rajan2010concurrency} is a design pattern that allows you to manage the execution of tasks, usually in the context of parallel or asynchronous programming. By encapsulating the creation and control of threads or processes, it simplifies task management and makes it easier to write scalable, maintainable code. This pattern is useful when you need to perform multiple tasks concurrently or in the background without manually handling threading or process management.

\subsubsection{Motivation}
When working with Python programs that require concurrent or parallel execution, managing threads or processes manually can become complex. You may need to handle thread creation, synchronization, task queuing, and resource management. The Executor pattern abstracts these details and provides a higher-level interface for task submission and execution. Python’s \texttt{concurrent.futures} module is a standard implementation of this pattern, providing an easy way to manage thread or process pools.

For example, imagine you are writing a program to download multiple files from the internet. You could create a separate thread for each download manually, but this approach can be error-prone, particularly as the number of tasks grows. The Executor pattern simplifies this by managing a pool of threads or processes and handling task submissions for you.

\subsubsection{Structure}
The Executor pattern typically involves the following components:

\begin{itemize}
    \item \textbf{Executor}: The central component responsible for managing task execution. It can be a thread pool, process pool, or another type of worker pool. In Python, this is often represented by \texttt{ThreadPoolExecutor} or \texttt{ProcessPoolExecutor}.
    \item \textbf{Future}: Represents the result of an asynchronous operation. A future object can be used to check if the task is complete and retrieve the result once it's available.
    \item \textbf{Task}: A function or callable that represents the work to be done.
\end{itemize}

Here is a visual representation of the Executor pattern using a tree structure:

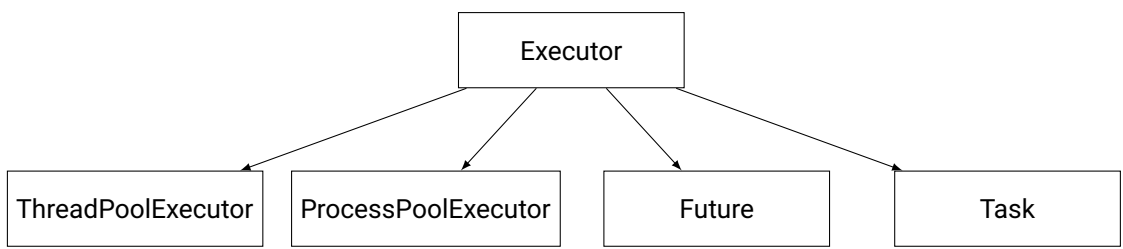
\begin{center}
\begin{tikzpicture}
    [grow=down, sibling distance=11em, level distance=6em, 
    edge from parent/.style={draw,-latex}, 
    every node/.style={draw, rectangle, minimum width=3cm, minimum height=1cm, text centered}]
    
    \node {Executor}
        child { node {ThreadPoolExecutor} }
        child { node {ProcessPoolExecutor} }
        child { node {Future} }
        child { node {Task} };
\end{tikzpicture}
\end{center}

\subsubsection{Participants}
\begin{itemize}
    \item \textbf{Executor}: Responsible for managing a pool of threads or processes and scheduling tasks for execution.
    \item \textbf{Task}: The unit of work submitted to the executor. This could be a function, method, or any callable object.
    \item \textbf{Future}: An object returned by the executor that represents the result of an asynchronous task. It provides methods for checking task completion and retrieving the result.
\end{itemize}

\subsubsection{How It Works}
The Executor pattern in Python works as follows:
\begin{enumerate}
    \item An executor (e.g., \texttt{ThreadPoolExecutor} or \texttt{ProcessPoolExecutor}) is created.
    \item Tasks (functions or callable objects) are submitted to the executor.
    \item The executor manages a pool of worker threads or processes to execute the tasks.
    \item Each task is run in the background, and a \texttt{Future} object is returned to represent the eventual result of the task.
    \item The \texttt{Future} object can be used to check the status of the task, wait for it to complete, or retrieve the result.
\end{enumerate}

\begin{lstlisting}[style=python]
from concurrent.futures import ThreadPoolExecutor

# Define a simple task
def task(name):
    print(f"Starting task {name}")
    return f"Task {name} completed"

# Create a ThreadPoolExecutor
with ThreadPoolExecutor(max_workers=3) as executor:
    # Submit tasks to the executor
    future1 = executor.submit(task, 'A')
    future2 = executor.submit(task, 'B')
    future3 = executor.submit(task, 'C')

    # Retrieve results from futures
    print(future1.result())
    print(future2.result())
    print(future3.result())
\end{lstlisting}

\textbf{Explanation}:
\begin{itemize}
    \item The \texttt{ThreadPoolExecutor} is created with a maximum of 3 worker threads.
    \item Three tasks are submitted to the executor using the \texttt{submit} method. Each task runs in a separate thread.
    \item \texttt{submit} returns a \texttt{Future} object, which can be used to check the status of the task and retrieve the result once it's completed.
    \item Finally, we call \texttt{result()} on each \texttt{Future} to block until the task is complete and get the result.
\end{itemize}

\subsubsection{Advantages and Disadvantages}

\textbf{Advantages}:
\begin{itemize}
    \item \textbf{Simplifies concurrency}: You don’t have to manually manage threads or processes. The executor handles the creation, scheduling, and termination of workers.
    \item \textbf{Resource management}: Executors can limit the number of concurrent tasks, preventing issues like thread exhaustion or CPU overload.
    \item \textbf{Scalability}: Executors can be used to distribute work across multiple CPU cores or machines in a cluster.
\end{itemize}

\textbf{Disadvantages}:
\begin{itemize}
    \item \textbf{Debugging difficulty}: Since tasks run in separate threads or processes, it can be harder to debug issues like race conditions or deadlocks.
    \item \textbf{Overhead}: Executors introduce some overhead due to task scheduling and worker management, which may not be ideal for very simple or lightweight tasks.
\end{itemize}

\subsubsection{Use Cases}
The Executor pattern is particularly useful in the following scenarios:
\begin{itemize}
    \item \textbf{Parallel processing}: When you need to distribute tasks across multiple CPU cores or machines to improve performance.
    \item \textbf{Background processing}: Performing tasks like file I/O, network requests, or database queries in the background without blocking the main thread.
    \item \textbf{Task scheduling}: Managing a queue of tasks that need to be executed asynchronously, such as handling user requests in a web server.
\end{itemize}

\subsubsection{Code Example}
Here's a more comprehensive example using both \texttt{ThreadPoolExecutor} and \texttt{ProcessPoolExecutor}:

\begin{lstlisting}[style=python]
from concurrent.futures import ThreadPoolExecutor, ProcessPoolExecutor
import time

def compute_task(value):
    time.sleep(2)  # Simulate a long-running task
    return value * 2

# Using ThreadPoolExecutor for I/O-bound tasks
with ThreadPoolExecutor(max_workers=4) as executor:
    futures = [executor.submit(compute_task, i) for i in range(5)]
    for future in futures:
        print(f"Thread result: {future.result()}")

# Using ProcessPoolExecutor for CPU-bound tasks
with ProcessPoolExecutor(max_workers=4) as executor:
    futures = [executor.submit(compute_task, i) for i in range(5)]
    for future in futures:
        print(f"Process result: {future.result()}")
\end{lstlisting}

\textbf{Explanation}:
\begin{itemize}
    \item We use \texttt{ThreadPoolExecutor} for I/O-bound tasks (which benefit from concurrent execution) and \texttt{ProcessPoolExecutor} for CPU-bound tasks (which benefit from parallel execution on multiple CPU cores).
    \item The example demonstrates how to submit multiple tasks to both thread and process pools and retrieve the results asynchronously.
\end{itemize}

\subsubsection{Pattern Extensions}
The Executor pattern can be extended or customized in several ways:
\begin{itemize}
    \item \textbf{Custom task scheduling}: You can build custom task queues or use priority queues to control the order in which tasks are executed.
    \item \textbf{Advanced error handling}: Executors can be extended to include detailed logging or retry mechanisms for failed tasks.
    \item \textbf{Distributed execution}: Executors can be extended to manage task distribution across a network of machines using libraries like \texttt{Dask} or \texttt{Ray}.
\end{itemize}

\subsection{Future Pattern}
The Future pattern \cite{lea2000concurrent} is a concurrency design pattern that is widely used for handling asynchronous operations. It represents the result of a computation that will be completed in the future, hence the name "Future." This pattern is useful for scenarios where you need to run operations in the background (such as downloading data or performing complex computations) without blocking the main thread of execution.

\subsubsection{Motivation}
The motivation behind the Future pattern is to enable concurrent execution of code, allowing programs to be more efficient and responsive. In real-world applications, tasks such as file I/O, network communication, or CPU-intensive calculations can take time. Instead of waiting for these tasks to be completed, the Future pattern allows you to continue executing other code while the task is being performed in the background.

For beginners, think of a Future as a placeholder for a result that you don’t have yet, but will have once a long-running operation finishes. This can make your program faster and smoother because it doesn't freeze while waiting for operations to complete.

\subsubsection{Structure}
The Future pattern typically consists of the following components:

\begin{itemize}
  \item \textbf{Future Object:} This object acts as a placeholder for the result of an asynchronous computation.
  \item \textbf{Executor:} Manages the execution of the tasks, often by maintaining a thread pool to run tasks in the background.
  \item \textbf{Task:} The actual function or computation that runs in the background.
\end{itemize}

The structure can be represented using a tree diagram:

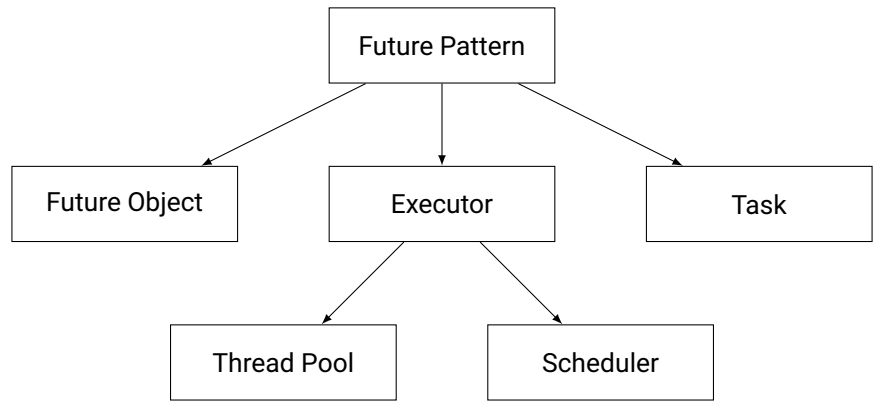
\begin{center}
\begin{tikzpicture}[
  grow=down, 
  sibling distance=12em, 
  level distance=6em, 
  edge from parent/.style={draw,-latex}, 
  every node/.style={draw, rectangle, minimum width=3cm, minimum height=1cm, text centered}] 
  
  \node {Future Pattern}
    child { node {Future Object} }
    child { node {Executor}
      child { node {Thread Pool} }
      child { node {Scheduler} }
    }
    child { node {Task} };
\end{tikzpicture}
\end{center}

\subsubsection{Participants}
There are several key participants in the Future pattern:

\begin{itemize}
  \item \textbf{Client:} The object or module that requests the asynchronous task.
  \item \textbf{Future Object:} Holds the result of the computation or task, which will become available in the future.
  \item \textbf{Executor:} Handles running tasks in the background, typically through a thread pool or event loop.
  \item \textbf{Task:} A callable or function that performs the actual work, which will return a result in the future.
\end{itemize}

\subsubsection{How It Works}
Here’s a step-by-step explanation of how the Future pattern works:

1. The \textbf{client} submits a task (a function or callable) to an \textbf{executor}.
2. The \textbf{executor} schedules the task for execution, often using a thread pool or event loop.
3. The executor returns a \textbf{Future object} to the client immediately. The Future object acts as a placeholder for the result of the task.
4. The task runs in the background, and when it completes, it sets the result in the Future object.
5. The client can later retrieve the result from the Future object, or check if the task is done.

A simple analogy is ordering food at a restaurant. You place your order (submit a task), receive a token (Future object), and then wait for your order (result) to be ready. Meanwhile, you can do other things.

\subsubsection{Advantages and Disadvantages}
\textbf{Advantages:}
\begin{itemize}
  \item \textbf{Non-blocking:} The client does not need to wait for the task to finish and can continue with other work.
  \item \textbf{Better resource utilization:} Background tasks can be run in parallel using thread pools, allowing for more efficient use of CPU and I/O resources.
  \item \textbf{Simplifies asynchronous code:} The Future object encapsulates the result of the task, making it easier to manage asynchronous operations.
\end{itemize}

\textbf{Disadvantages:}
\begin{itemize}
  \item \textbf{Complexity:} Introducing concurrency can make code harder to debug and reason about, especially for beginners.
  \item \textbf{Synchronization issues:} If multiple threads are involved, issues like race conditions or deadlocks can arise.
  \item \textbf{Resource overhead:} Managing threads or event loops can consume additional memory and CPU resources.
\end{itemize}

\subsubsection{Use Cases}
The Future pattern is commonly used in scenarios where you have long-running operations that you don't want to block the main thread. Some practical use cases include:

\begin{itemize}
  \item \textbf{Network operations:} Downloading files or fetching data from a server can take time. Using Futures allows the rest of the program to continue running while waiting for the data to arrive.
  \item \textbf{I/O-bound operations:} File reading or database querying often takes time, and using a Future allows your program to remain responsive during these operations.
  \item \textbf{CPU-bound tasks:} For tasks like image processing or data analysis, which can take a long time, you can run these tasks in parallel using the Future pattern.
\end{itemize}

\subsubsection{Code Example}
Here is a simple Python code example that demonstrates the use of the Future pattern with the \texttt{concurrent.futures} module:

\begin{lstlisting}[style=python]
import concurrent.futures
import time

# Simulate a long-running task
def slow_task(seconds):
    print(f"Sleeping for {seconds} second(s)...")
    time.sleep(seconds)
    return f"Task finished after {seconds} second(s)"

# Create an executor to manage threads
with concurrent.futures.ThreadPoolExecutor() as executor:
    # Submit tasks to be executed asynchronously
    future1 = executor.submit(slow_task, 2)
    future2 = executor.submit(slow_task, 3)

    # Do some other work while the tasks are running
    print("Doing some other work...")

    # Retrieve results from the futures
    result1 = future1.result()
    result2 = future2.result()

    print(result1)
    print(result2)
\end{lstlisting}

In this example:
\begin{itemize}
  \item We create an executor using a \texttt{ThreadPoolExecutor}.
  \item We submit two tasks, each running \texttt{slow\_task} with different durations.
  \item The \texttt{submit} method returns a Future object, which is used to retrieve the result later.
  \item Meanwhile, other work (a simple print statement) is done while the tasks are running.
  \item Once the tasks are complete, we retrieve their results using the \texttt{result()} method of the Future object.
\end{itemize}

\subsubsection{Pattern Extensions}
There are some ways to extend or modify the Future pattern for specific use cases:

\begin{itemize}
  \item \textbf{Combining futures:} Sometimes, you may want to wait for multiple futures to complete and then combine their results. Python’s \texttt{concurrent.futures.wait()} or \texttt{asyncio.gather()} can be used for this purpose.
  \item \textbf{Cancellation:} Futures can be extended to allow for task cancellation if the result is no longer needed. This is useful in cases where the result is time-sensitive.
  \item \textbf{Timeouts:} You can extend the pattern to handle timeouts by specifying a maximum time to wait for the result. If the task takes too long, an exception is raised.
\end{itemize}

\subsection{Producer-Consumer Pattern}
The Producer-Consumer pattern \cite{bjork2005games} is a classical synchronization design pattern, where one or more producers generate data (or tasks), and one or more consumers process them. This pattern is useful in concurrent programming where you want to decouple tasks, allowing producers and consumers to operate independently at their rates.

\subsubsection{Motivation}
The primary motivation for the Producer-Consumer pattern is to enable asynchronous processing, where producers generate data faster or slower than consumers can process them. This helps avoid bottlenecks, especially in multithreading environments where the tasks of generating and consuming data can be handled by separate threads.

\begin{itemize}
    \item \textbf{Producers:} These are components responsible for producing or generating tasks/data.
    \item \textbf{Consumers:} These are components responsible for consuming or processing the tasks/data generated by the producers.
\end{itemize}

In Python, this pattern is often used to handle tasks like reading and writing files, web scraping, handling requests, etc., where a producer generates work (e.g., fetching URLs), and a consumer processes the work (e.g., saving the content).

\subsubsection{Structure}
The basic structure of the Producer-Consumer pattern involves:
\begin{enumerate}
    \item A shared resource, often a queue, where producers place data and consumers retrieve data.
    \item Separate producer and consumer threads that work in parallel.
    \item Synchronization between the producer and consumer, often using locks or semaphores to manage access to the shared queue.
\end{enumerate}

Below is a simple visualization of the structure:

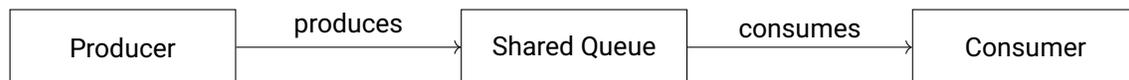
\begin{center}
\begin{tikzpicture}[node distance=6cm, auto]
    \node[draw, rectangle, minimum width=3cm, minimum height=1cm, text centered] (producer) {Producer};
    \node[draw, rectangle, right of=producer, minimum width=3cm, minimum height=1cm, text centered] (queue) {Shared Queue};
    \node[draw, rectangle, right of=queue, minimum width=3cm, minimum height=1cm, text centered] (consumer) {Consumer};

    \draw[->] (producer.east) -- (queue.west) node[midway, above] {produces};
    \draw[->] (queue.east) -- (consumer.west) node[midway, above] {consumes};
\end{tikzpicture}
\end{center}

\subsubsection{Participants}
\begin{itemize}
    \item \textbf{Producer:} The entity responsible for generating data. This could be a thread that reads from a database, scrapes web pages, or listens to user input.
    \item \textbf{Consumer:} The entity responsible for processing the data generated by the producer. This could be a thread that writes data to a file, displays it, or processes it in some other way.
    \item \textbf{Queue:} A shared buffer where producers place their data and consumers retrieve it. This is usually a thread-safe queue such as Python's \texttt{queue.Queue}.
\end{itemize}

\subsubsection{How It Works}
The Producer-Consumer pattern works through synchronization mechanisms that ensure smooth interaction between the producers and consumers. The producers and consumers typically work with a queue that acts as a buffer. Producers place their generated data into the queue, while consumers retrieve the data.

Here's how it works step by step:
\begin{enumerate}
    \item The producer generates data and places it in a queue.
    \item If the queue is full, the producer might need to wait until there's space available (blocking behavior).
    \item The consumer retrieves data from the queue.
    \item If the queue is empty, the consumer may need to wait until new data is available (blocking behavior).
\end{enumerate}

In Python, a typical way to implement this is with the \texttt{queue.Queue} class, which handles the locking and synchronization for you.

\subsubsection{Advantages and Disadvantages}
\textbf{Advantages:}
\begin{itemize}
    \item Decouples producers from consumers, allowing each to operate independently.
    \item Simplifies the architecture of multi-threaded applications.
    \item Reduces the risk of race conditions because the shared resource (queue) is thread-safe.
\end{itemize}

\textbf{Disadvantages:}
\begin{itemize}
    \item Managing the queue size can be tricky. A queue that's too small can lead to blocking, while a queue that's too large can consume too much memory.
    \item If not carefully managed, it can lead to deadlocks, especially if producers and consumers block each other indefinitely.
\end{itemize}

\subsubsection{Use Cases}
\begin{itemize}
    \item \textbf{Web Scraping:} A producer generates URLs, while consumers fetch the data from those URLs.
    \item \textbf{Task Processing:} A producer generates tasks (like files to process), and consumers perform those tasks (such as reading and writing to disk).
    \item \textbf{Logging:} A producer writes logs, and a consumer writes them to a file or database.
    \item \textbf{Image Processing:} A producer generates a set of images, and a consumer processes each image (resizing, applying filters, etc.).
\end{itemize}

\subsubsection{Code Example}
Here is a simple Python example of the Producer-Consumer pattern using the \texttt{queue.Queue} and \texttt{threading} modules:

\begin{lstlisting}[style=python]
import threading
import queue
import time
import random

# Define a thread-safe queue
q = queue.Queue()

# Producer function
def producer():
    for i in range(5):
        item = f"Item {i}"
        q.put(item)
        print(f"Produced {item}")
        time.sleep(random.random())  # Simulate production time

# Consumer function
def consumer():
    while True:
        item = q.get()
        if item is None:  # End condition
            break
        print(f"Consumed {item}")
        time.sleep(random.random())  # Simulate consumption time

# Create producer and consumer threads
producer_thread = threading.Thread(target=producer)
consumer_thread = threading.Thread(target=consumer)

# Start the threads
producer_thread.start()
consumer_thread.start()

# Wait for the producer to finish
producer_thread.join()

# Stop the consumer
q.put(None)  # Signal the consumer to stop
consumer_thread.join()
\end{lstlisting}

\textbf{Explanation:}
\begin{itemize}
    \item A thread-safe \texttt{queue.Queue} is used to share items between the producer and consumer.
    \item The \texttt{producer} function generates items and places them in the queue.
    \item The \texttt{consumer} function retrieves and processes items from the queue.
    \item The producer and consumer are run in separate threads to work concurrently.
    \item The \texttt{None} value is used to signal the consumer to stop processing when the producer is finished.
\end{itemize}

\subsubsection{Pattern Extensions}
There are several variations and extensions to the basic Producer-Consumer pattern, including:
\begin{itemize}
    \item \textbf{Multiple Producers, Multiple Consumers:} You can extend the pattern to have multiple producers and multiple consumers working concurrently, all sharing the same queue.
    \item \textbf{Bounded Queues:} In some cases, the queue may have a limited size, forcing the producers to wait if the queue is full.
    \item \textbf{Priority Queues:} You can use a priority queue (\texttt{queue.PriorityQueue}) where consumers process higher-priority tasks first.
    \item \textbf{Asynchronous Producers/Consumers:} With the \texttt{asyncio} library, you can build producers and consumers that operate asynchronously, without blocking threads.
\end{itemize}

\begin{lstlisting}[style=python]
import asyncio
import random

# Define an async producer
async def async_producer(q):
    for i in range(5):
        item = f"Async Item {i}"
        await q.put(item)
        print(f"Produced {item}")
        await asyncio.sleep(random.random())

# Define an async consumer
async def async_consumer(q):
    while True:
        item = await q.get()
        if item is None:
            break
        print(f"Consumed {item}")
        await asyncio.sleep(random.random())

# Main async function
async def main():
    q = asyncio.Queue()
    
    # Create producer and consumer coroutines
    producer_task = asyncio.create_task(async_producer(q))
    consumer_task = asyncio.create_task(async_consumer(q))
    
    # Wait for the producer to finish
    await producer_task
    
    # Stop the consumer
    await q.put(None)
    await consumer_task

# Run the async main function
asyncio.run(main())
\end{lstlisting}
This code shows how you can adapt the Producer-Consumer pattern to an asynchronous environment using Python's \texttt{asyncio} module.

\subsection{Reactor Pattern}

The Reactor Pattern \cite{mijavc2021reactor} is a design pattern primarily used in event-driven applications, where the flow of the program is determined by external events. It allows you to handle multiple event sources efficiently within a single thread, such as network connections, file descriptors, or timers. The reactor pattern is commonly used in server applications to manage I/O operations.

\subsubsection{Motivation}

The primary motivation behind the Reactor Pattern is to decouple event-handling logic from the main application logic, allowing the application to remain responsive and scalable when dealing with numerous events or I/O sources. In an event-driven environment, you want to avoid blocking the execution of your program while waiting for input/output (I/O) operations.

For instance, imagine you are building a web server in Python. Without the reactor pattern, the server might block every time it waits for a request, reducing its ability to handle multiple clients concurrently. The Reactor Pattern allows the server to remain responsive by non-blocking I/O and delegating the handling of each event to the appropriate handlers.

\subsubsection{Structure}

The Reactor Pattern is composed of several key components. Here's a diagram to show how the components interact:

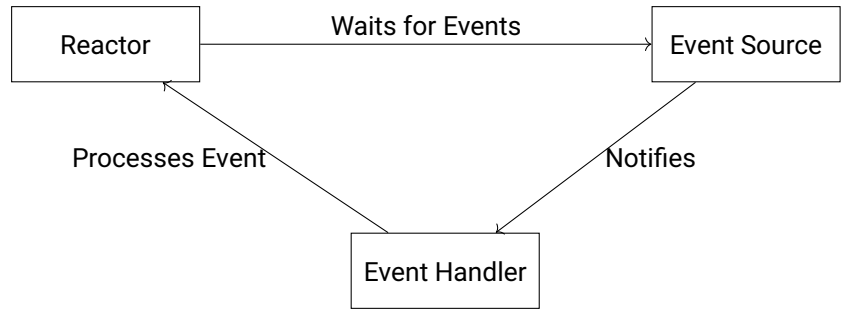
\begin{center}
\begin{tikzpicture}[node distance=3cm, auto]
  \node[draw, rectangle, minimum width=2.5cm, minimum height=1cm, text centered] (reactor) {Reactor};
  \node[draw, rectangle, minimum width=2.5cm, minimum height=1cm, text centered, right=6cm of reactor] (event) {Event Source};
  \node[draw, rectangle, minimum width=2.5cm, minimum height=1cm, text centered, below=2cm of event, xshift=-4cm] (handler) {Event Handler};

  \draw[->] (event) -- (handler) node[midway,right] {Notifies};
  \draw[->] (reactor) -- (event) node[midway,above] {Waits for Events};
  \draw[->] (handler) -- (reactor) node[midway,left] {Processes Event};

\end{tikzpicture}
\end{center}

\textbf{Components:}
\begin{itemize}
  \item \textbf{Reactor}: The core of the pattern, which waits for events to occur. It demultiplexes the events and dispatches them to the corresponding event handlers.
  \item \textbf{Event Source}: This is the source of events, such as a network socket or file descriptor. It produces events that need to be handled.
  \item \textbf{Event Handler}: Defines how to handle specific events (e.g., reading from a socket, processing user input). These handlers are called by the reactor when an event occurs.
\end{itemize}

\subsubsection{Participants}

1. \textbf{Reactor}:
   - This is responsible for listening to one or more event sources and dispatching the event to the appropriate handler.

2. \textbf{Event Source}:
   - Generates events that the reactor waits for, such as new client connections in a server or incoming data on a socket.

3. \textbf{Event Handler}:
   - Each type of event has a corresponding handler. The handler defines what happens when a specific event occurs. For example, a handler might read data from a socket and then process it.

4. \textbf{Dispatcher}:
   - Sometimes part of the reactor itself, the dispatcher is responsible for mapping events to the correct event handlers.

\subsubsection{How It Works}

1. \textbf{Event Registration}:
   - The application registers event sources (like sockets or file descriptors) with the reactor. Each event source is associated with an event handler that will process the event.

2. \textbf{Event Loop}:
   - The reactor enters an event loop where it waits for events to occur. This is typically done with non-blocking I/O using functions like \lstinline{select()} or \lstinline{epoll()}.

3. \textbf{Event Demultiplexing}:
   - When an event occurs, the reactor checks which event source it came from and looks up the corresponding event handler.

4. \textbf{Event Handling}:
   - The event handler is invoked to process the event. The handler might perform operations like reading data from a socket or writing a response to the client.

The event loop continues indefinitely, handling events as they occur, allowing the application to remain responsive to multiple I/O events simultaneously.

\subsubsection{Advantages and Disadvantages}

\textbf{Advantages}:
\begin{itemize}
  \item \textbf{Efficiency}: The reactor pattern allows handling multiple I/O operations within a single thread, reducing the need for context switching or thread synchronization.
  \item \textbf{Scalability}: With non-blocking I/O, applications can handle thousands of concurrent connections, making the pattern ideal for network servers and event-driven applications.
  \item \textbf{Modularity}: It provides a clear separation of concerns, with event-handling logic decoupled from the core application logic.
\end{itemize}

\textbf{Disadvantages}:
\begin{itemize}
  \item \textbf{Complexity}: Implementing the reactor pattern can be complex, especially when dealing with multiple event sources and handling errors.
  \item \textbf{Single Threaded}: By default, the reactor pattern operates in a single thread, which might become a bottleneck for CPU-bound tasks.
  \item \textbf{Limited Parallelism}: While the pattern handles I/O efficiently, CPU-bound operations might require a separate threading or multiprocessing approach to avoid blocking the reactor.
\end{itemize}

\subsubsection{Use Cases}

The Reactor Pattern is used in a variety of real-world applications, particularly where efficient I/O handling is crucial. Some common use cases include:

\begin{itemize}
  \item \textbf{Web Servers}: Servers like Nginx and Tornado in Python use the reactor pattern to handle multiple client requests simultaneously.
  \item \textbf{Networking Libraries}: Libraries such as Twisted and asyncio in Python are built around the reactor pattern for handling asynchronous I/O.
  \item \textbf{Event-Driven GUIs}: GUI frameworks often use the reactor pattern to process user input events like mouse clicks and keyboard input.
  \item \textbf{Message Queues}: Systems like RabbitMQ use event-driven architectures for efficient message passing between clients.
\end{itemize}

\subsubsection{Code Example}

Here’s an example of a simple Reactor Pattern implementation using Python’s \lstinline{selectors} module for non-blocking I/O:

\begin{lstlisting}[style=python]
import selectors
import socket

# Event handler
class EventHandler:
    def __init__(self, connection):
        self.connection = connection

    def handle_read(self):
        data = self.connection.recv(1024)
        if data:
            print("Received:", data.decode())
            self.connection.send(data)  # Echo back to the client
        else:
            print("Connection closed.")
            self.connection.close()

# Reactor
class Reactor:
    def __init__(self):
        self.selector = selectors.DefaultSelector()

    def register(self, connection, event_handler):
        # Register the connection for read events
        self.selector.register(connection, selectors.EVENT_READ, event_handler)

    def run(self):
        while True:
            events = self.selector.select()
            for key, mask in events:
                handler = key.data
                handler.handle_read()

# Simple server to test the reactor pattern
def run_server():
    reactor = Reactor()
    
    server_sock = socket.socket(socket.AF_INET, socket.SOCK_STREAM)
    server_sock.bind(('localhost', 65432))
    server_sock.listen()
    print("Server started on localhost:65432")
    
    while True:
        conn, addr = server_sock.accept()
        print(f"Accepted connection from {addr}")
        handler = EventHandler(conn)
        reactor.register(conn, handler)

        reactor.run()

if __name__ == "__main__":
    run_server()
\end{lstlisting}

This example demonstrates a simple echo server. When a client connects, the server registers the connection for read events. When the client sends data, the server echoes it back using the event handler.

\subsubsection{Pattern Extensions}

The Reactor Pattern can be extended or combined with other patterns for more sophisticated use cases:

\begin{itemize}
  \item \textbf{Proactor Pattern}: An alternative pattern that handles events after I/O completion, commonly used for asynchronous operations.
  \item \textbf{Multithreaded Reactor}: To overcome the limitation of single-threaded reactors, you can extend it to handle I/O events in one thread and process the events in another, allowing parallelism.
  \item \textbf{Chain of Responsibility}: The reactor can be combined with the Chain of Responsibility pattern to process events through a pipeline of handlers.
\end{itemize}

\section{Conclusion: Reflecting on Python Design Patterns}

As we reach the conclusion of this book, it's essential to reflect on the journey we've taken together through the world of Python design patterns. Throughout the book, we have explored the fundamental principles of design patterns, examined their relevance in software development, and dived deep into their practical application in Python. 

Design patterns offer reusable solutions to common software design problems. They promote clean, maintainable, and scalable code, which is especially important when working on larger projects. The goal of this book has been to equip you with the knowledge and tools needed to recognize when to apply these patterns, and more importantly, how to implement them effectively in Python.

In this final chapter, we'll summarize the key concepts we've covered, revisit the importance of design patterns, and encourage you to continue exploring and practicing these patterns as you advance in your Python programming journey.

\subsection{A Recap of Key Design Patterns}

In the previous chapters, we explored several essential design patterns grouped into three categories: Creational, Structural, and Behavioral. Each of these categories serves different purposes in structuring your code, and mastering them will give you the ability to build more complex, reliable software systems. Let's briefly revisit each category:

\subsubsection{Creational Patterns}
Creational patterns deal with object creation mechanisms, trying to create objects in a manner suitable for the situation. Some of the key patterns we've covered include:

\begin{itemize}
    \item \textbf{Singleton Pattern:} Ensures that a class has only one instance and provides a global point of access to it. We used the Singleton pattern to demonstrate scenarios where centralized management of resources is required, such as in logging systems.
    \item \textbf{Factory Pattern:} Defines an interface for creating objects, but lets subclasses alter the type of objects that will be created. This pattern is particularly useful for decoupling client code from the object creation process.
    \item \textbf{Builder Pattern:} A step-by-step construction of complex objects. We illustrated this pattern using examples like building different configurations of a complex car object, which required flexible construction options.
\end{itemize}

\subsubsection{Structural Patterns}
Structural patterns concern class and object composition. They help ensure that when one part of a system changes, the entire structure of the program doesn't need to change. These patterns help create large structures while keeping them flexible and efficient. Key patterns we discussed include:

\begin{itemize}
    \item \textbf{Adapter Pattern:} Allows incompatible interfaces to work together. We showed how to use this pattern to integrate two separate libraries with different APIs into a unified system without changing their underlying code.
    \item \textbf{Decorator Pattern:} Adds behavior to objects dynamically. This pattern was demonstrated with examples such as extending the functionality of a basic coffee object by dynamically adding sugar, milk, or other ingredients.
    \item \textbf{Facade Pattern:} Provides a simplified interface to a complex system. We used this pattern to illustrate how to hide complex subsystems behind a single, unified interface, making the system easier to use.
\end{itemize}

\subsubsection{Behavioral Patterns}
Behavioral patterns deal with communication between objects. They help in managing algorithms, relationships, and responsibilities between objects. Some patterns we studied include:

\begin{itemize}
    \item \textbf{Observer Pattern:} Allows an object (subject) to notify other objects (observers) of state changes. We used this pattern to demonstrate real-time notifications, such as a stock price tracking system that notifies users of changes in prices.
    \item \textbf{Command Pattern:} Encapsulates requests as objects, allowing for parameterization and queuing of requests. This pattern is perfect for undo functionality or task queues, which we illustrated through a command-line text editor with undo/redo operations.
    \item \textbf{Strategy Pattern:} Enables selecting an algorithm at runtime. We used this pattern to demonstrate different sorting strategies that could be swapped in and out depending on the input size and nature of the data.
\end{itemize}

\subsection{The Importance of Design Patterns in Software Development}

Why are design patterns so crucial in software development, especially in Python?

\begin{itemize}
    \item \textbf{Code Reusability:} One of the main benefits of design patterns is their ability to enhance reusability. By adhering to established patterns, you can write code that is reusable across different projects, reducing the need to reinvent the wheel.
    \item \textbf{Maintainability:} Patterns promote cleaner code that is easier to maintain over time. When working in teams or on large projects, adhering to design patterns ensures that the codebase remains understandable and manageable.
    \item \textbf{Scalability:} Well-architected code using design patterns scales better. By decoupling different components and managing responsibilities effectively, your software can grow in size and complexity without significant refactoring.
    \item \textbf{Flexibility:} Design patterns help you create flexible systems that can adapt to changing requirements. Instead of rigid structures, patterns encourage designs that can accommodate future growth and changes without extensive modifications.
\end{itemize}

\subsection{Applying Design Patterns in Python: Next Steps}

At this point, you've been introduced to the most commonly used design patterns in Python. However, learning design patterns is just the beginning. The next step is to practice applying these patterns in real-world projects. Here are some ways to continue your learning:

\begin{itemize}
    \item \textbf{Build Projects:} Start with small projects, such as creating a simple e-commerce system, a task manager, or a game, and try to incorporate design patterns where appropriate. Experiment with patterns you've learned to see how they enhance your design.
    \item \textbf{Contribute to Open Source:} Open source projects are an excellent way to see how design patterns are applied in real-world applications. Find a project you're passionate about, contribute to it, and observe how design patterns are used in different scenarios.
    \item \textbf{Review Codebases:} Look at other developers' code to see how they structure their applications. Pay attention to how design patterns are used, and think about how you might improve or refactor the code using what you've learned.
\end{itemize}

\subsection{Closing Thoughts}

Mastering design patterns is a long-term endeavor, but it is one of the most rewarding skills you can acquire as a developer. With this book, you've taken the first steps toward understanding the core principles behind these patterns and how to apply them in Python. Remember, the key to mastering design patterns is practice. The more you work with them, the more natural they will become.

As you continue your journey as a Python developer, always think critically about your designs. Use design patterns as tools, not rules. Sometimes, a pattern might not be the best fit for a particular problem, and that's okay. The real goal is to write code that is clear, maintainable, and scalable.

Thank you for joining me on this journey. I hope this book has been a helpful guide and that it has provided you with the tools you need to design better Python applications. Happy coding!

\begin{center}
\textit{— The Authors}
\end{center}

\bibliographystyle{ieeetr}
\bibliography{sample}

\end{document}